\documentclass[a4paper,12pt,makeidx]{book}

\usepackage{mathptmx}
\usepackage{helvet}
\usepackage{courier}
\usepackage{type1cm}

\usepackage{makeidx}         
\usepackage{graphicx}        
\usepackage{multicol}        
\usepackage[bottom]{footmisc}

\usepackage{amsmath, bm}
\usepackage{pstricks}
\usepackage{pst-node}
\usepackage[ansinew]{inputenc}
\usepackage{amssymb,amsmath}
\usepackage{amsfonts}
\usepackage{epsfig}
\usepackage[english]{babel}
\usepackage{graphics}
\usepackage{graphicx}
\usepackage{epigraph}

\usepackage{epic}
\usepackage{eepic}
\usepackage{algorithm}
\usepackage{algpseudocode}
\usepackage{listings}
\usepackage{url}

\makeindex             

\def\p{\partial}
\def\px{\partial_x}
\def\py{\partial_y}
\def\a{\alpha}
\def\b{\beta}
\def\g{\gamma}
\def\d{\delta}
\def\la{\lambda}
\def\o{\omega}

\def\k{{\bm k}}

\newcommand{\pa}{{\partial}}

\def\bse{\begin{subequations}}\def\ese{\end{subequations}}
\newcommand{\BE}[1]{\begin{equation}\label{#1}}
\newcommand{\BEA}[1]{\begin{eqnarray}\label{#1}}
\newcommand{\BSE}[1]{\begin{subequations}\label{#1}}

\let\*\cdot
\def\<{\left\langle} \def\>{\right\rangle} \def\({\left(} \def\){\right)}
\let\p\partial \let\~\widetilde \let\^\widehat 

 \let\~\widetilde
\def\BSE{\begin{subequations}}\def\ESE{\end{subequations}}
\let \= \equiv

\font\Sets=msbm10

\def\Integer {\hbox{\Sets Z}}    \def\Real {\hbox{\Sets R}}

\def\Complex {\hbox{\Sets C}}   \def\Natural {\hbox{\Sets N}}

\def\be{\begin{equation}}       \def\ba{\begin{array}}

\def\ee{\end{equation}}         \def\ea{\end{array}}

\def\bea {\begin{eqnarray}}      \def\eea {\end{eqnarray}}

\def\bean{\begin{eqnarray*}}    \def\eean{\end{eqnarray*}}

\def\pa  {\partial}             \def\ti  {\widetilde}

\def\la  {\lambda}

\def\eps{\varepsilon}           \def\ph{\varphi}

\def\const {\mathop{\rm const}\nolimits}

\def\res   {\mathop{\rm res}  \limits}

\def\im    {\mathop{\rm Im}   \nolimits}

\def\ker  {\mathop{\rm Ker} \nolimits}

\def\RA {\ \Rightarrow\ }         \def\LRA {\ \Leftrightarrow\ }

\def\qed   {\vrule height0.6em width0.3em depth0pt}

\def\<{\langle} \def\({\left(}  \def\>{\rangle} \def\){\right)}

\def\defeq {\stackrel{\mbox{\rm\small def}}{=}}

\def\a{\alpha}

\begin{document}

\author{Alexey Shabat and Elena Kartashova}
\title{Computable integrability}
\maketitle

\frontmatter

\include{dedic}
\include{foreword}

{\Large{Overview}}

 Already in 19th century Euler and
Lagrange established a mathematically satisfactory foundation of
Newtonian mechanics. Hamilton developed analogous formulation of
optics. Jacobi imported Hamilton's idea in mechanics, and
eventually arrived at a new formulation (now referred to as the
Hamilton-Jacobi formalism). The Hamilton-Jacobi formalism was a
crucial step towards Liouville's classical definition of the
notion of integrability. Moreover, Jacobi himself is also famous
for having discovered (1839) that the geodesic motion on an
ellipsoid is an integrable system which is solvable in terms of
hyper-elliptic functions \cite{Jaco1}. Jacobi's research on
dynamical started in 1837 and deeply motivated by Hamilton's
formulation of optics based on the least action. One may say that
Hamilton's work played the same role as Abel's work in Jacobi's
researches of elliptic functions.\\

 Liouville's definition of
integrability \cite{Liou1} is based on the notion of first
integrals, i.e. conservation laws. In his definition, a
(Hamiltonian) system is said to be integrable if it has
sufficiently many first integrals in involution. The same idea has
ever been inherited in many variants of the notion of
integrability. Liouville's definition of integrable Hamiltonian
systems naturally covered many classical examples. Among them are
the Kepler motion solved by Newton, harmonic oscillators solvable
by trigonometric functions, the rigid bodies (``spinning tops'')
of the Euler-Poisont type and the Lagrange type, and Jacobi's
example of geodesic motion on an ellipsoid. The spinning tops and
Jacobi's example were significant because they were known to be
solvable by elliptic functions. Soon after the work of Liouville,
C. Neumann discovered a new integrable Hamiltonian system, and
pointed out that this system can be solved by hyper-elliptic
functions. That was the beginning of subsequent discoveries of
many integrable systems. Neumann's Hamiltonian system \cite{Neum}
was a prototype of integrable Hamiltonian systems discovered in
the 19th century. They all are more or less connected with
hyper-elliptic
functions.\\

Some generalizations of Langrange's rigid body problem (spinning
top) has been done by G. Kirchhoff \cite{Kirch} who derived the
equation of motion of a rigid body in an ideal fluid and by  A.
Clebsch \cite{Cleb} and V. Steklov \cite{Stek} for some other
cases. Weber \cite{Web} solved the motion of a rigid body in an
ideal fluid in terms of genus-two hyper-elliptic functions.  Soon
afterwards Sophie Kowalevski \cite{Kow} discovered her famous the
third example of solvable rigid bodies problem  and solved it in
terms of genus-two hyper-elliptic functions. Triumphal procession
of classical theory of integrable systems has been crowned by
dissertation of  P. Stäckel \cite{Stae} where a systematic
classification of Hamiltoninan systems is presented that can be
solved by separation of variables.\\

All the researches, mentioned above, belong to the main stream of
studies on integrable systems in the 19th century. They are,
however, not the direct origin of the breakthrough in the
seventies of the 20th century. The breakthrough originates in
quite different sources in 19th century
--- study of solitary waves (or ``solitons'' in the modern language) and surface geometry.
Studies on solitons
 were initiated by J.
Scott Russell when he observed a solitary wave in a canal in 1834.
His report invoked a controversy. Sir G.B. Airy was suspicious,
but Lord Rayleigh advocated Scott Russel's observation by
calculation (1876). Finally, in 1895, D.J. Korteweg and G. de
Vries proposed a nonlinear partial differential equation (the KdV
equation) that well fits in the result of Lord Rayleigh. Note
that, unlike the equation of motion of point particles and rigid
bodies, the KdV equation is a partial differential equation. In
other words, it has an infinite degrees of freedom. This is a
common feature of many soliton equations discovered in the
breakthrough in the seventies. Surface geometry is also a subject
started and developed in the 19th century. The sine-Gordon
equation, which is also an important soliton equation, is a
partial differential equation that characterizes a special family
of surfaces. Furthermore, members of this family are connected by
the so called Bäcklund transformations. A good classical overview
of this subject is provided in G. Darboux's book \cite{Dar},
 in which one can even find the Toda lattice!\\

 The brilliant success of the
search for integrable Hamiltonian systems was rapidly fading on
the turn to the century. It was the beginning of a long blank that
continued until the sixties. Routes towards the breakthrough after
that blank, however, were already prepared in these days. A route
was opened by Paul Painlevé and the contemporaries. Actually, the
idea of Painlevé was conceptually very close to Kovalevskaya's
work on integrable rigid bodies. Kovalevskaya's method was to
search for the cases where the solutions of the equation of
motion, analytically continued to the complex plane, have no
singularity other than poles. Painlevé did the same for second
order nonlinear ordinary differential equations, slightly relaxing
the conditions on possible singularities of solutions.  It should
be noted that the celebrated six Painlevé equations (and
presumably their various generalizations) are NOT integrable in
the classical sense. Rather, apart from special cases, they are
not solvable by any abelian function, nor reducible to a linear
ordinary differential functions; this property is known as the
irreducibility of solutions of the Painlevé equations.
Nevertheless Garnier, one of the successors of Painlevé's
researches, pointed out a link with integrable systems \cite{Gar}.
In this paper Garnier discovered an integrable system solvable by
hyper-elliptic functions as a byproduct of his researches on a
generalization of Painlevés equations. This system was derived by
taking a special limit of his generalized Painlevé equations.
Remarkably, his calculations implicitly uses a ``Lax equation'',
which is a clue in the present approach to integrable systems.\\

Another route to the revival of integrable systems was discovered
in a quite different direction by Drach, Burchnall and Chaundy.
Drach considered \cite{Drach1},\cite{Drach2} the second order
linear ordinary equation
\be\frac{d^2}{dx^2} = [\phi(x) + h ]y\ee
from a Galois-theoretic point of view, and discovered the case
where the coefficient $\phi(x)$ and the solution $y(x)$ are both
related to hyper-elliptic abelian functions. J.L. Burchnall and
T.W. Chaundy considered \cite{Burch} the case where the same
linear ordinary operator $L$ as Drach's commutes with another
ordinary differential operator $M$, namely satisfies the
commutation relation $[L,M] = 0$, and got results similar to
Drach's work. In fact, Drach, Burchnall and Chaundy discovered or
finite gap operators, related to hyper-elliptic solutions of the
KdV equations. Their work is thus a precursor of
algebraic-geometric
studies of soliton equations in the seventies.\\

Theory of integrable systems resurrected as soliton theory in
 1965 when
M. Zabus-ky and M.D. Kruskal reported the celebrated numerical
computation of solutions of the KdV equation. Their research was
motivated by the work of E. Fermi, J. Pasta and S. Ulam  (FPU)
done in the fifties. Zabusky was re-examining this work,
approximating the nonlinear lattice of FPU by a continuous system.
The continuous system turned out to be described by the KdV
equation. He and Kruskal thus attempted to solve the KdV equation
numerically, and discovered numerical solutions in which many
solitary waves co-existed. The numerical solutions revealed
remarkable stability of the solitary waves, each of which behaved
like a ``particle''. Because of this behavior, they called the
solitary waves {\it solitons}. This observation stimulated
theoretical researches, and soon led to the discovery, by Gardner,
Greene, Kruskal and Miura \cite{Gardner} of exact multi-soliton
solutions and the {\it inverse scattering method} that produces
those solutions. The clue of their discovery was a relation
between the KdV equation and the stationary Schrödinger equation
in one spatial dimension. Using this relation, they could derive
the multi-soliton solutions, an finite number of conservation
laws, etc. systematically. Peter Lax \cite{Lax} soon proposed a
more convenient and universal reformulation of the work
\cite{Gardner} which is  now called the {\it Lax formalism}.\\

The Toda lattice was also discovered in these days. This
discovery, however, was done independently, \cite{Toda67a,Toda67b}.

Being unaware of the work of FPU, Toda was looking for an exact
model of heat conduction, and eventually arrived at his
exponential lattice. His method was to go back the usual way {\it
from equations to solutions}: starting from a nonlinear wave given
by an elliptic function, he attempted to derive a nonlinear
equation to be satisfied by the nonlinear wave, and eventually
arrived to his exponential lattice. Within less than a decade, the
method of \cite{Gardner}  and \cite{Lax}  was extended to many
other {\it soliton equations}, some of the most results in this direction are briefly presented below.\\

Ablowitz et al., \cite{AKNS74}, extended the
inverse scattering method to a matrix system. A wide range of new
equations, such as the modified KdV equation, the nonlinear
Schrödinger equation, and the classical sine-Gordon equations,
thus turned out to be solvable by the inverse scattering method, \cite{Fla74a,Fla74a,Ma74}
 Flaschka and
Manakov independently developed a Lax formalism of the Toda
lattice. The inverse scattering method was thus further extended
to a spatially discrete system.\\

Zakharov and Shabat, \cite{ZSh74}, extended
the inverse scattering method to systems with two
spatial-dimensions. A typical example was the
Kadomtsev-Petviashvili (KP) equation. While the inverse scattering
method was thus refined, several new techniques were also invented
in the seventies. In particular, the direct method
(bilinearization) of R. Hirota, the algebro-geometric method of
B.A. Durovin, V.B. Matveev, S.P. Novikov and I.M. Krichever, and
the group-theoretical (or Lie-algebraic) method due to M. Adler,
B. Kostant, W.W. Symes, A.G. Reyman and M.A. Semenov-Tian-Shansky
emerged in the second half of the seventies, and grew up to the
mainstream of the progress through the beginning of the
eighties.\\

Yet another route --- Calogero systems The world of integrable
Hamiltonian systems of the 19th century, too, came back soon after
the birth of the soliton theory, with a novel family of integrable
systems --- the Calogero systems. The Calogero systems are
Hamiltonian systems of interacting particle on a line. Calogero
discovered the simplest case of these systems as a quantum
integrable system in \cite{Cal71}.

 This paper deals
with the quantum integrability of the system with two-body
potential 1/x2. Sutherland soon proposed another version of
Calogero's systems in \cite{Su72} where he considered a variant of Calogero's system
with two-body potential 1/sin2x.

This system is nowadays called
the Sutherland system. Calogero conjectured the integrability of
classical analogues of these systems, and presented a partial
answer. Moser solved this probelm by constructing a Lax formalism
of these systems.

 Calogero and Marchioro proved in \cite{CaMa74} the classical
integrability of the three-body case by explicit calculation.  Moser, \cite{M075},
constructed a Lax representation for both the Calogero and
Sutherland systems, and proved the classical integrability (at
least for the Calogero system). In fact, this paper is also known
to be one of the papers in which Moser presented his method for
sovling the non-periodic finite Toda (and Kac-van Moerbeke) system
using Stiektjes' theory of continued fraction. Moser's work
suggested that the Lax formalism, originally developed in soliton
theory, would be also useful for finite-dimensional integrable
systems. Moser's idea was further extended by Olshanetsky and
Perelomov in \cite{OP76};
this work is also
one of the earliest attempts of the Lie algebraic methods for
constructing (and solving) integrable systems. A similar work was
done by Bogoyavlensky for the (finite) Toda lattice, \cite{Bo76}.

 After the aforementioned work, Moser
turned to classical integrable systems of the 19th century, i.e.,
Jacobi's geodesic flows on an ellipsoid, C. Neumann's system,
etc., and demonstrated that these systems, too, can be treated in
the Lax formalism, \cite{M080}.

(...)

\include{acknow}

\tableofcontents

\include{acronym}

\mainmatter
\chapter{General notions and ideas}

\section{Introduction}

In our first Chapter we are going to present  some general notions
and ideas of modern theory of integrability trying to outline its
computability aspects. The reason of this approach is that though
this theory was wide and deeply developed in the last few decades,
its results are almost unusable for non-specialists in the area
due to its complexity as well as due to some specific jargon
unknown to mathematicians working in other areas. On the other
hand, a lot of known results are completely algorithmic and can
be used as a base for developing some symbolic programm package
dealing with the problems of integrability. Creation of such a
package will be of a great help not only at the stage of
formulating of some new hypothesis but also as a tool to get new
systematization and classification results, for instance, to get
complete lists of integrable equations with given properties as it
was done already for PDEs
with known symmetries \cite{sh1987}.\\

Some results presented here are quite simple and can be obtained
by any student acquainted with the basics of calculus (in these
cases direct derivation is given) while some ideas and results
demand deep knowing of a great deal of modern mathematics (in this
cases only formulations and references are given). Our main idea
is not to present here the simplest subjects of integrability
theory but to give its general description in simplest possible
form in order to give a reader a feeling what has been done
already and what could/should be done further in this area. Most
of the subjects mentioned here
 will be discussed in  details in the next Chapters.\\

We will use the word "integrability" as a generalization of the
notion "exactly solvable" for differential equations. Possible
definitions of differential operator  will be discussed as well as
some definitions of integrability itself. Numerous examples
presented here are to show in particular that it is reasonable not
only to use different notions of integrability for different
differential equations but sometimes it proves to be very useful
to regard {\bf one equation} using  various definitions of
integrability, depending on what properties of the equation are
under the study. Two classical approaches to classification of
integrable equations - conservation laws and Lie symmetries  - are
also briefly presented and a few examples are given in order to
demonstrate deep difference between these two notions, specially
in case of PDEs. Two interesting semi-integrable systems are
introduced showing one more aspect of integrability theory - some
equations though not integrable in any strict sense, can be
treated  as "almost" integrable due to their intrinsic properties.

\section{Notion of differential operator}

There are many ways to define linear differential first-order
operator $D$, starting with Leibnitz formula for product
differentiating
\be
D(ab)=a^{'}b+ab^{'}.
\ee
This definition leads to
\be
D\cdot a= a^{'}+aD, \quad D^2a=aD^2+2a^{'}D+a^{''},....,
\ee
\be \label{dn} D^n \cdot a=\sum_{k} \left( \ba{c}n\\k \ea \right)
D^k(a)D^{n-k} \ee

where
\be
\left( \ba{c}n\\k \ea \right)=\frac{n(n-1)...(n-k+1)}{1\cdot 2
\cdots  k} \quad \mbox{with} \quad \left( \ba{c}n\\0 \ea \right)=1
\ee
are binomial coefficients. Thus, {\bf one} linear differential
first-order operator and its powers are defined. Trying to define
a composition of two linear operators, we get already quite
cumbersome formula

\be \label{d1d2}D_1D_2a=D_1(D_2(a)+aD_2)=D_1(a)D_2+D_1D_2(a) +
D_2(a)D_1+aD_1D_2, \ee

 which can be regarded as a {\bf definition}
of a factorizable linear differential second-order operator.
Notice that though each of $D_1$ and $D_2$ satisfies Leibnitz
rule, their composition $D_1D_2$ {\bf does not!} An important
notion of commutator of two operators $[D_1D_2-D_2D_1]$ plays a
 role of special multiplication (see Ex.1). In case of
{\bf non-factorizable} second-order operator an attempt to
generalize Leibnitz rule leads to a very complicated Bourbaki-like
constructions which we are not going to present here. All these
problems appear due to
coordinateness of this approach.\\

On the other hand, a linear differential operator being written in
coordinate form as
\be L=\sum_j f_j\p_j \ee
leads to
\be L=\sum_{|\a|\leq m} f_{\a}\p^{\a}, \quad \p^{\a}= \p^{\a_1}\cdots\p^{\a_n},   \quad |\a|=\a_1+...+\a_n \ee
which is  {\bf definition} of LPDO of order $m$ with $n$
independent variables. Composition of two operators $L$ and $M$ is
defined as
\be
L\circ M= \sum f_{\a}\p^{\a}\sum g_{\b}\p^{\b}=\sum h_{\g}\p^{\g}
\ee
and coefficients $h_{\g}$ are to be found from formula
(\ref{dn}).\\

Now,  notion of linear differential {\bf equation} (LDE) can be
introduced in terms of the kernel of differential operator, i.e.
\be
\ker (L):= \{\varphi| \sum_{|\a|\leq m} f_{\a}\p^{\a}\varphi =0\}.
\ee

Let us regard as illustrating example second-order LODO with one
independent variable $x$:
\be L= f_0+f_1\p+f_2\p^2, \quad \p:=\frac{d}{dx}. \ee

Notice that if two functions $\varphi_1$ and $\varphi_2$ belong to
its kernel $\ker (L)$, then
\be
c_1 \varphi_1 + c_2 \varphi_2 \in \ker (L),
\ee
i.e. $\ker (L)$ is a linear vector space over constants´ field
(normally, it is $\Real$ or $\Complex$).\\

Main theorem about ODEs states the existence and uniqueness of the
Cauchy problem for any ODE, i.e. one-to-one correspondence between
elements of the kernel and initial data. In our case Cauchy data
\be
\varphi |_{x=x_0}=\varphi^0, \quad \p \varphi |_{x=x_0}=\varphi^1
\ee
form a two-dimensional vector space and, correspondingly,
dimension of kernel is equal 2, $dim(\ker (L))=2$. Any two
functions $\varphi_1, \varphi_2 \in \ker (L)$ form its basis if
Wronskian $<\varphi_1, \varphi_2>$ is non-vanishing:
\be
<\varphi_1, \varphi_2>:= \left| \ba{cc}\varphi_1 \
\varphi_2\\\varphi_1^{'} \ \varphi_2^{'}\ea \right| \ne 0
\ee
while an arbitrary function $\psi \in \ker (L)$ has to satisfy
following condition:
\be
<\varphi_1, \varphi_2, \psi>:= \left| \ba{ccc}\varphi_1 \
\varphi_2 \ \psi \\ \varphi_1^{'} \ \varphi_2^{'}\ \psi^{'}\\
 \varphi_1^{''} \ \varphi_2^{''}\ \psi^{''}  \ea
\right| = 0.
\ee
Now we can construct immediately differential operator with a
given kernel as
\be
L(\psi)=<\varphi_1,\varphi_2, \psi>.
\ee
For instance, if we are looking for an LODE with solutions $\sin
{x}$ and $\sqrt{x}$, then
 corresponding LODE has form
\be
\psi^{''}(1-\frac 12 \tan {x} ) + \tan {x} \psi^{'} - \frac{1}{2x}
\psi - \frac 34 \frac{1}{x^2} \psi=0.
\ee

Coming back to LODO of order $m$, we re-write formula for the
kernel in the form

\be \label{kern} L(\psi)=\frac{<\varphi_1,..., \varphi_m,
\psi>}{<\varphi_1,...,\varphi_m>} \ee which provides that
high-order coefficient $f_m=1$. It is done just for our
convenience and we will use this form further.\\

Now, some constructive definition of linear differential operator
was given and importance of its kernel was demonstrated.
Corresponding differential equation was defined in terms of this
kernel and construction of operator with a given kernel was
described. All this is {\bf not possible} for nonlinear operator
because in this case manifold of solutions has much more
complicated structure then just linear vector space - simply
speaking, the reason of it is that in this case linear combination
of solutions is not a solution anymore.  Due to this reason some
other notions are to be used to study properties of nonlinear
operators - symmetries, conservation laws and, of course, as the
very first step - change of variables transforming a nonlinear
operator into a linear one. We will discuss all this in the next
sections.

\section{Notion of integrability}
\begin{itemize}

\item{}{\bf 1.3. Solution in elementary functions:}
\be {y''+y=0.} \ee
 General solution of this equation belongs to the class of
trigonometrical  functions, $y= a \sin( x + b)$, with arbitrary
const $a,b$. In order to find this solution one has to notice that
this equation is LODE with constant coefficients which possess
fundamental system of solutions, all of the form $e^{\lambda x}$
where $\lambda$ is a root of characteristic polynomial.

\item{}{\bf 1.2. Solution {\it modulo} class of functions:}
\begin{itemize}
\item[]{\bf 1.2.1.} {\bf Integrability in quadratures}\\
\be {y^{''}=f(y).}
\ee
In order to integrate this equation let us notice that
$y^{''}y'=f(y)y'$ which leads to
\be
\frac12 y'^2=\int f(y) dy + const = F(y)
\ee
and finally
\be
   {dx}=\frac {dy}{\sqrt{2F(y)}},
\ee
which describes differential equation with {\bf separable
variables}. In case when $F(y)$ is a polynomial of third or fourth
degree, this is {\bf definition of elliptic integral} and
therefore the initial nonlinear ODE is integrable in elliptic
functions. Particular case when polynomial $F(y)$  has multiple
roots might leads to a particular solution in elementary
functions. Let us regard, for instance, equation $y''= 2y^3$ and
put $const=0$, then $y'^2=y^4$ and $y=1/x$, i.e. we have ONE
solution in the class of {\it rational functions}. General
solution is written out in terms of {\it elliptic functions}.
Conclusion: equation is {\bf integrable} in the class of elliptic
functions and {\bf not integrable} in the class of
rational functions.\\

The case when  $F(y)$ is a not-specified smooth function, is
called {\bf integrability in quadratures}. This is collective name
was introduced in classic literature of 18th century and used in a
sense of "closed-form" or "explicit form" of integrability while
analytic theory of special functions was developed much later.\\

Notice that as a first step in finding of solution, the order $n$
of initial ODE was diminished to $n-1$, in our case $n=2$. Of
course, this is {\bf not possible} for any arbitrary differential
equation. This new ODE is called {\bf first integral} or {\bf
conservation law} due its physical meaning in applications, for
instance, our example can be reformulated as Newton second law of
mechanics and its first integral corresponds to energy
conservation law.\\

\item[]{\bf 1.2.2.} {\bf Painleve transcendent P1}\\
\be {y''=y^2+x.}\ee
This equation defines first {\bf Painleve transcendent}. About
this equation  it was proven that it has no solutions in classes
of elementary or special functions. On the other hand, it is also
proven that {\bf Painleve transcendent} is a meromorphic function
with known special qualitative  properties (\cite{p3}).\\

This example demonstrates us the intrinsical difficulties when
defining the notion of integrability. Scientific community has no
general opinion about integrability of  Painleve transcendent.
Those standing on the classical positions think about it as about
non-integrable equation. Those who are working on different
applicative problems of theoretical physics involving the use of
 Painleve transcendent look at it as at some new special
function (see also Appendix 1).
\end{itemize}

\item{}{\bf 1.3. Solution} {\it modulo} {\bf inexplicit function:}\\
 \be {u_t=2uu_x.} \ee

This equation describes so called {\bf shock waves} and its
solutions are expressed in terms of inexplicit function.  Indeed,
let us rewrite this equation in the new independent variables
$\tilde{t}=t$, $\tilde{x}=u$ and dependent one $\tilde{u}=x$, i.e.
now $x=\theta (t,u)$ is a function on $t,u$. Then

\be d\tilde{t}=dt, \quad dx= \theta_t dt + \theta_u
du, \quad u_t |_{dx=0}=-\frac{\theta_t}{\theta_u}, \quad u_x
|_{dt=0}=\frac{1}{\theta_u}\ee
 and
\be
-\frac{\theta_t}{\theta_u}=2 u \frac{1}{\theta_u} \Rightarrow
-\theta_t= 2u \Rightarrow -\theta=2ut-\varphi(u)
\ee
and finally  $ x+2tu= \varphi(u) $ where $\varphi(u)$ is {\bf
arbitrary function} on $u$. Now, we have finite answer but no
explicit form of dependence $u=u(x,t)$. Has the general solution
been found? The answer is that given some initial conditions, i.e.
$t=0$, we may define solution as $u=\varphi^{-1}(x)$ where is
$\varphi^{-1}$ denotes {\bf inverse function} for $\varphi$.

\item{}{\bf 1.4. Solution {\it modulo} change of variables
(C-Integrability):}
\be{\psi_{xy} +\alpha \psi_x + \beta \psi_y + \psi_x  \psi_y =0 .}\ee

This equation is called  {\bf Thomas equation} and it
 could be made linear with a change of variables. Indeed, let
$\psi= log \theta$ for some positively defined function $\theta$:
\bea
\psi_x  = (log \theta)_x = \frac{\theta _x}{\theta},\quad \psi_y =
(log \theta)_y = \frac{\theta _y}{\theta}, \\
 \psi_{xy}  = (
\frac{\theta_x}{\theta})_y = \frac {\theta_{xy} \theta  - \theta_y
\theta_x}{ \theta ^ 2}, \quad \psi_x  \psi _y = \frac {\theta_x
\theta_y}{\theta ^ 2}\eea
 and substituting this into Thomas equation
we get finally linear PDE
\be {\theta_{xy} + \alpha \theta_x + \beta \theta_y = 0.}\ee

Suppose for simplicity that $\b=0$ and  make once more change of
variables: $\theta=\phi e^{k_1y}$, then
\be
\theta_x= \phi_x e^{k_1y}, \quad \theta_{xy}= \phi_{xy}
e^{k_1y}+k_1\phi_{x} e^{k_1y} \quad  \mbox{and} \quad  \phi_{xy} +
(k_1+\a) \phi_x  = 0,
\ee
and finally
\be
\p_x (\phi_y + (k_1+\a) \phi)  = 0,
\ee
which yields to
\be
\phi_y -k_2 \phi=f(y) \quad  \mbox{with} \quad k_2=-(k_1+\a)
\ee
and arbitrary function $f(y)$. Now
 general solution can be obtained by the method
of variation of a constant. As a first step let us solve
homogeneous part of this equation, i.e. $ \phi_{y} - k_2\phi = 0$
and $\phi(x,y)= g(x)e^{k_2y}$ with arbitrary $g(x)$. As a second
step, suppose that $g(x)$ is function on $x,y$, i.e. $g(x,y)$,
then initial equation takes form
\bea (g(x,y)e^{k_2y})_{y} -
k_2g(x,y)e^{k_2y} = f(y),\\
g(x,y)_{y}e^{k_2y}+k_2g(x,y)e^{k_2y}- k_2g(x,y)e^{k_2y}=f(y),\\
g(x,y)_{y}e^{k_2y}=f(y), \quad g(x,y)= \int f(y)e^{-k_2y}dy +
h(x),
\eea
and finally the general solution of Thomas equation with $\a=0$
can be written out as
\be
\phi(x,y)= g(x,y)e^{k_2y}=e^{k_2y}(\int f(y)e^{-k_2y}dy + h(x))=
\hat{f}(y)+e^{k_2y}h(x)
\ee
with two arbitrary functions $\hat{f}(y)$ and $h(x)$.

 \item{}{\bf 1.5. Solution {\it modulo} Fourier transform
(F-Integrability):}
\be{u_t=2uu_x+ \varepsilon u_{xx}}\ee
where $\varepsilon$ is a constant. This equation is called {\bf
Burgers equation} and it differs from Thomas equation studied
above where change of variables was local in a sense that solution
in each point  does not depend on the solution in some other
points of definition domain, i.e. local in $(x,y)$-space. This
equation can be transformed into
\be u_t=2uu_x+ u_{xx} \ee
 by the
change of variables $\tilde{x}=\varepsilon x$, $u=\varepsilon
\tilde{u}$ and $\tilde{t}=\varepsilon^2 t$. We will use this form
of Burgers equation skipping tildes in order to simplify the
calculations below.
 To find solution of  Burgers equation one has to use Fourier transform
 which obviously
is nonlocal, i.e. here solution is local only in $k$-space. In
order to demonstrate it let us integrate it using notation
\be \int
u dx=v,\ee then after integration
\be v_t=v_x^2+ v_{xx}=e^{-v}(e^{v})_{xx}\ee and change of variables
$w=e^{v}$, Burgers equation is reduced to the {\bf heat equation}
\be w_t=w_{xx}\ee
which is linear. Therefore, solutions of
Burgers equation could be obtained from solutions of heat equation
by the change of variables
\be u=v_x=\frac{w_x}{w}\ee
as
\bea
w(x,0)=\int_{-\infty}^ {\infty}\exp(ikx)\hat{w}(k)dk \\
\Rightarrow
u(x,t)=\frac{\int \exp(ikx-k^2t)\hat{w}(k)ikdk}{\int
\exp(ikx-k^2t)\hat{w}(k)dk},\eea
where $w(x,0)$ is initial data  and $\hat{w}(k)$  is a function
called {\bf its Fourier transform} and it can be computed as
\be
\hat{w}(k)=\frac{1}{2\pi}\int_{-\infty}^
{\infty}\exp(-ikx)w(x,0)dx.
\ee
In fact, it is well-known that {\bf any} linear PDE with constant
coefficients on an infinite line can be solved using as standard
basis $\{e^{ikx}| k \in \Real\}$  because they are eigenfunctions
of these operators. Thus, Thomas equation with $\a=0$ where the
general solution was found explicitly, is {\bf an exception} while
heat equation
demonstrates the general situation. \\

 \item{}{\bf 1.6. Solution {\it modulo} IST
(S-Integrability):}  \be {u_t=6uu_x+ u_{xxx}.} \ee

This equation is called {\bf Korteveg-de Vries (KdV) equation} and
it is nonlinear PDE with nonconstant coefficients. In this case,
choosing set of functions $\{e^{ikx}\}$ as a basis is not helpful
anymore: Fourier transform does not simplify the initial equation
and only generates an infinite system of ODEs
on Fourier coefficients.\\

On the other hand, some new basis can be found which allows to
reduce KdV with rapidly decreasing initial data, $u\to 0,\,
x\to\pm\infty$, to the linear equation and to solve it. This new
basis can be constructed using solutions of {\bf linear
Schrödinger equation}
\be {\psi_{xx}+k^2\psi=u\psi} \ee
where function $u$ is called {\bf potential} due to its origin in
quantum mechanics. Solutions of linear Schrödinger equation are
called {\bf Jost functions}, $\psi^\pm(t,x,k)$, with asymptotic
boundary conditions:
\be
\psi^\pm(t,x,k; u(x,t))e^{-\pm i(kx+k^3t)}\to 1,\quad x\to\pm
\infty.
\ee
Jost function $ \phi (x,k)=\psi^+(t,x,k; u(x,t))e^{-i(kx+k^3t)}$
is defined by the integral equation
\be
\phi (x,k)=1+\int_x^\infty\frac{1-\exp[2k(x-x')]}{2k}
u(x')\phi(x',k)dx'
\ee
with $t$ playing role of a parameter. Second Jost function is
defined analogously with integration over $[-\infty,x]$. Notice
that asymptotically for $x\to\pm \infty$ linear Schrödinger
equation
\be \psi_{xx}+k^2\psi=u\psi \quad \mbox{is reduced to} \quad
\psi_{xx}+k^2\psi=0\ee
as in case of Fourier basis $\{e^{ikx}| k
\in \Real\} $. It means that asymptotically their solutions do
coincide and, for instance, any solution of linear  Schrödinger
equation
\be \psi\sim c_1e^{i(kx+k^3t)}+c_2 e^{-i(kx+k^3t)} \quad
\mbox{for} \quad x\to\infty.\ee

It turns out that solutions of KdV can be regarded as potentials
of linear Schrödinger equation, i.e. following system of equations
\be
\begin{cases}
u_t=6uu_x+ u_{xxx},\\
\psi_{xx}+k^2\psi=u\psi
\end{cases}
\ee
is consistent  and any solution of KdV can be written out as an
expansion of Jost functions which in a sense are playing role of
exponents $e^{ikx}$ in Fourier basis, \cite{sh1979}.\\

On the other hand, there exists a major difference between these
two basis: Fourier basis is written out in explicit form {\it via}
one function while Jost basis is written out in inexplicit form
{\it via} two functions with different asymptotic properties on
the different ends of a line. The crucial fact here is that two
Jost functions are connected by simple algebraic equation:
\be
\psi^{-}(x,k,t)=a(k)\psi^{+}(x,-k,t)+b(k)e^{ik^3t}\psi^{+}(x,k,t)
\ee
while it allows us to construct rational approximation of Jost
functions for given $a(k)$ and $b(k)$ and, correspondingly,
general  solution of KdV. The problem of reconstruction of
function $u$ according to $a(k), b(k)$ is called {\bf inverse
scattering problem} and this method, correspondingly, {\bf inverse
scattering transform (IST)}.\\

 \item{}{\bf 1.7. Solution {\it modulo} Dressing method
(D-Integrability):}  \be {iu_t=u_{xx} \pm |u|^2u.}\ee
This
equation is called {\bf nonlinear Schrödinger equation} (NLS) and
it is very important in many physical applications, for instance,
in nonlinear optics. Dressing method is generalization of IST and
in this case role of auxiliary linear equation (it was linear
Schrödinger equation, second order ODE, in the previous case)
plays a system of two linear first order ODEs \cite{zak1971}:
\be
\begin{cases}
\psi_x^{(1)}=\lambda\psi^{(1)}+u \psi^{(2)}\\
\psi_x^{(2)}=-\lambda\psi^{(2)}+v \psi^{(1)}
\end{cases}
\ee
where $v=\pm \bar{u}$. It turns out that system of equations
\be
\begin{cases}
iu_t=u_{xx} \pm |u|^2u\\
\psi_x^{(1)}=\lambda\psi^{(1)}+u \psi^{(2)}\\
\psi_x^{(2)}=-\lambda\psi^{(2)}+v \psi^{(1)}
\end{cases}
\ee
is consistent and is equivalent to Riemann-Hilbert problem.
Solutions of this last system are called {\bf matrix Jost
functions} and any solution of NLS can be written out as an
expansion of matrix Jost functions which in a sense are playing
role of
exponents $e^{ikx}$ in Fourier basis, \cite{sh1975}.
\end{itemize}

\section{Approach to classification}

Our list of definitions is neither full nor exhaustive, moreover
one equation can be regarded as integrable due to a few different
definitions of integrability. For instance, equation for shock
waves from § {\bf 1.3}, $u_t=2uu_x$, is a particular form of
Burgers equation from § {\bf 1.5}, $u_t=2uu_x+ \varepsilon u_{xx}$
with $\varepsilon=0$, and it can be linearized as above, i.e. it
is not only integrable in terms of   inexplicit function but also
C-integrable and  F-integrable, with general solution
\be u(x,t)=\frac{\int \exp(ikx-k^2t)\hat{u}(\varepsilon k)ikdk}{\int
\exp(ikx-k^2t)\hat{u}(\varepsilon k)dk}.\ee
 What form of
integrability is chosen for some specific equation depends on what
properties of it we are interested in. For instance, the answer in
the form of inexplicit function shows immediately dependence of
solution form on initial conditions - graphically presentation of
inverse function $u=\varphi^{-1}(x)$ can be obtained as mirrored
image of $x=\varphi (u)$. To get the same information  from the
formula above is a very nontrivial task. On the other hand, the
general formula is the only known tool to study solutions with
singularities. This shows that definitions of integrability do not
suit to serve as a basis for classification of integrable systems
and some more intrinsic ways should be used to classify and solve
them. Below we present briefly two possible classification bases -
 conservation laws and Lie symmetries.

\subsection{Conservation laws}

Some strict and reasonable  definition of a {\bf conservation law}
(also called {\bf first integral} for ODEs) is not easy
to give, even in case of ODEs. As most general definitions one
might regard
\be \label{clOd}
\frac{d}{dt}F(\vec{y})=0, \ee
for ODE
\be \frac{d}{dt}\vec{y}=f(\vec{y}), \quad  \vec{y}=(y_1,...,y_n),
\quad  f(\vec{y})=(f_1,...,f_n), \nonumber
\ee
 and
\be \label{clPa}
\frac{d}{dt}\int G(u,u_x,u_y,u_{xx},u_{xy},u_{yy}...)dxdy...=0, \ee
for PDE
\be \p_t u=
g(u,u_x,u_y,u_{xx},u_{xy},u_{yy},...). \nonumber
\ee
Obviously, without putting some restrictions on function $F$ or
$G$ these definitions are too general and do not even point out
some specific class of differential equations. For instance, let
us take {\bf any} second order ODE, due to well-known theorem on
ODEs solutions we can write its general solution in a form
\be
F(t,y,a,b)=0
\ee
where $a,b$ are two independent parameters (defined by initial
conditions). Theorem on inexplicit function gives immediately
\be
a=F_1(t,y,b) \quad \forall b\quad \mbox{and}\quad b=F_2(t,y,a)
\quad \forall a,
\ee
i.e. any  second order ODE has 2 independent conservation laws and
obviously, by the same way  $n$ independent conservation laws can
be constructed for ODE of order $n$. For instance, in the simplest
case of second order ODE with constant coefficients, general
solution and its first derivative have form
\be
\begin{cases}
y=c_1e^{\lambda_1x}+c_2e^{\lambda_2x} \\
y^{'}=c_1\lambda_1e^{\lambda_1x}+c_2\lambda_2e^{\lambda_2x}
\end{cases}
\ee
and multiplying $y$ by $\lambda_1$ and $\lambda_2$ we get
equations on $c_1$ and $c_2$ correspondingly:
\be
\begin{cases}
\lambda_2 y- y^{'}=c_1(\lambda_2-\lambda_1)e^{\lambda_1x} \\
\lambda_1 y- y^{'}=c_2(\lambda_1-\lambda_2)e^{\lambda_2x}
\end{cases} \Rightarrow \quad
\begin{cases}
\lambda_2x + \hat{c_2}=log(y^{'}-\lambda_1 y )\\
\lambda_1x + \hat{c_1}=log(y^{'}-\lambda_2 y )
\end{cases}
\ee
and two conservation laws are written out explicitly:
\be
\hat{c_2}=log(y^{'}-\lambda_1 y )-\lambda_2x, \quad
\hat{c_1}=log(y^{'}-\lambda_2 y )-\lambda_1x.
\ee
To find these conservation laws without knowing of solution is
more complicated task then to solve equation itself and therefore
they give no additional information about equation. This
 is the reason why often only polynomial or rational
conservation laws are regarded - they are easier to find and
mostly they describe qualitative  properties
of the equation which are very important for applications (conservation of energy, momentum, etc.) \\

On the other hand, conservation laws, when known, are used for
construction of ODEs solutions. Indeed, let us rewrite
Eq.(\ref{clOd}) as
\be
\frac{d}{dt}F(\vec{y})=(f_1\p_1 + f_2\p_2 + ... + f_n
\p_n)F=\mathcal{L}(F)=0,
\ee
i.e. as an equation in partial derivatives  $\mathcal{L}(F)=0$.
Such an equation has
 $(n-1)$ {\bf independent}
particular solutions $(\varphi_1, \varphi_2,...,\varphi_{n-1})$ if
its  Jacobian matrix has  maximal rank
 \be rank  \frac{\p(\varphi_1, \varphi_2,...,\varphi_{n-1})}{\p (y_1, y_2,...,y_{n-1})} =n-1,\ee
 with notation
\be
\frac{\p(\varphi_1, \varphi_2,...,\varphi_{n-1})}{\p (y_1,
y_2,...,y_{n-1})}= \left( \ba {cccc} \frac{\p \varphi_1}{\p y_1} \
\frac{\p
\varphi_1}{\p y_2} \ \cdots \ \frac{\p \varphi_1}{\p y_n}\\
\cdots \ \cdots \ \cdots \ \cdots \\
\frac{\p \varphi_n}{\p y_1} \ \frac{\p \varphi_n}{\p y_2} \ \cdots
\ \frac{\p \varphi_n}{\p y_n} \ea \right).
\ee

Now we can write out the general solution as $F(\varphi_1,
\varphi_2,...,\varphi_{n-1})$ with arbitrary function $F$ and
 initial ODE can be reduced to
\be
\frac{dz}{dt}=f(z), \quad \frac{dz}{f(z)}=dt
\ee
and solved explicitly in quadratures (see § {\bf 1.2}).

\subsection{Symmetry properties}
In order to give  definition of
 canonical form for $n$-order ODE
\be \label{canon}
 y^{(n)}=F(x,y,y^{'},...,y^{(n-1)})
\ee
 let us first introduce vector
\be \vec{y}=\begin{bmatrix}x\cr y\cr y^{'}\cr...\cr
y^{(n-1)}\end{bmatrix}
\ee
which is called {\bf vector of dynamical variables} with all its
coordinates regarded as independent, and its first derivative
\be
\frac{d\vec{y}}{dt}=\begin{bmatrix}1\cr y^{'}\cr y^{''}\cr...\cr
y^{(n)}\end{bmatrix} =\begin{bmatrix}1\cr y^{'}\cr y^{''}\cr...\cr
F\end{bmatrix}
\ee
with respect to some new independent variable $t$ such that
$dt=dx$, then the equation
\begin{equation}\label{dynamic}
\frac{d\vec{y}}{dt}=\begin{bmatrix}1\cr y^{'}\cr y^{''}\cr...\cr F
\end{bmatrix}
\end{equation}
is called {\bf canonical form} of an ODE. This canonical form is
also called {\bf dynamical system}, important fact is that
dimension of dynamical system is $ (n+1)$ for $n$-order ODE.\\

{\textbf{Definition 1.1.}} Dynamical system
\begin{equation}\label{symm}
\frac{d\vec{y}}{d \tau}=g(\vec{y})
\end{equation}
is called {\bf a symmetry} of  another dynamical system
\begin{equation*}
\frac{d\vec{y}}{d t}=f(\vec{y}),
\end{equation*}
if
\begin{equation}\label{cond}
\frac{d}{d \tau}(\frac{d\vec{y}}{d t})=\frac{d}{d
t}(\frac{d\vec{y}}{d \tau})\quad  \Leftrightarrow \quad \frac{d}{d
\tau}(f(\vec{y})=\frac{d}{d t}(g(\vec{y}))
\end{equation}
holds. Symmetry $g$ of dynamical system $f$ is called {\bf trivial} if $g=\const \cdot f$.\\

 Obviously, Eq.(\ref{cond}) gives necessary condition of
 compatibility of this two dynamical systems. It can be proven
 that this condition is also sufficient. Therefore, construction
 of each Lie symmetry with group parameter
$\tau$ is equivalent to a construction
 of some  ODE which have $\tau$
as independent variable and is consistent with a given ODE. System
of ODEs obtained this way, when being written in canonical form
 is called {\bf dynamical system
connected to the element of Lie symmetry algebra}. \\

 We will discuss it in all
details later (also see \cite{Olver}), now just pointing out the
fact that Lie algebra can be generated not only by Lie
transformation group (normally used for finding of solutions) but
also by the set of dynamical systems (\ref{dynamic}) (normally
used for classification purposes).\\

Two following elementary theorems show interesting
interconnections  between conservation laws and symmetries in case
of ODEs.\\

{\textbf{Theorem 1.1 }} {\it Let dynamical system
\begin{equation*}
\frac{d\vec{y}}{d \tau}=g(\vec{y})
\end{equation*}
is a symmetry of  another dynamical system
\begin{equation*}
\frac{d\vec{y}}{d t}=f(\vec{y})
\end{equation*}}
and
\be \frac{d}{dt}F(\vec{y})=(f_1\p_1 + f_2\p_2 + ... + f_n
\p_n)F=\mathcal{L}(F)=0
\ee
is a conservation law. Then $Fg$ is symmetry as well.\\

$\blacktriangleright$ Indeed, let us introduce
\be
\mathcal{M}=(g_1\p_1 + g_2\p_2 + ... + g_n \p_n),
\ee
then consistency condition Eq.(\ref{cond}) is written out as
\be \label{cons} \mathcal{L} \circ \mathcal{M}=\mathcal{M} \circ
\mathcal{L} \ee
 and after substituting $F\mathcal{M}$ instead of
$\mathcal{M}$ we get on the left hand of (\ref{cons})
\be
\mathcal{L}(F\mathcal{M})=\mathcal{L}(F)\mathcal{M}+F\mathcal{L}\mathcal{M}
=F\mathcal{L}\mathcal{M}=F\mathcal{M}\mathcal{L},
\ee
and on the right hand of
\be
(F\mathcal{M})\mathcal{L}=F\mathcal{M}\mathcal{L}.
\ee
\qed

{\textbf{Corollary 1.2 }} {\it  Let $n$-th order ODE  of the form (\ref{canon})} has a symmetry
\be
\frac{d\vec{g}}{dt}=\begin{bmatrix}g_0 \cr g_1\cr g_2\cr...\cr
g_n\end{bmatrix},
\ee
then $g_0$ is a conservation law and consequently without loss of
generality we may put $g_0=1$, if $g_0 \ne 0$.\\

{\textbf{Theorem 1.3 }} {\it An ODE of arbitrary order $n$ having  of $(n-1)$
independent conservation laws (i.e. complete set), has no
nontrivial symmetries consistent with conservation laws}.\\
$\blacktriangleright$ Indeed, using all conservation laws we can
reduce original $n$-order ODE into first order ODE
\be
\frac{da}{dt}=f(a)
\ee
and look for symmetries in the form
\be
 \frac{da}{d\tau}=g(a).
\ee
Then
\be
\frac{d}{d\tau}(\frac{da}{dt})=f^{'}(a)g(a), \quad
\frac{d}{dt}(\frac{da}{d\tau})=g^{'}(a)f(a)
\ee
and finally
\be
\frac{f^{'}}{f(a)}=\frac{g^{'}}{g(a)} \quad \Rightarrow \quad
ln(\frac{f}{g})= \const,
\ee
i.e. functions $f(a)$ and $g(a)$ are proportional. \qed \\

 Simply speaking, Eq.(\ref{symm}) defines a one-parameter
transformation group with parameter $\tau$ which conserves the
form of original equation.  The very important achievement of Lie
was his first theorem giving constructive procedure for obtaining
such a group. It allowed him to classify integrable differential
equations and to solve them. The simplest example of such a
classification
  for  second order ODEs with two
symmetries
is following: each of them can be transformed into one of the four types\\

{\bf (I)} $y^{''}=h(y^{'})$  \quad {\bf (II)} $y^{''}=h(x)$, \quad
{\bf  (III)} $y^{''}=\frac{1}{x}h(y^{'})$, \quad
{\bf (IV)} $y^{''}=h(x)y^{'}$,\\

where $h$ denotes an arbitrary smooth function while explicit form
of corresponding transformations was also
written out by Lie.  \\

Symmetry approach can also be also used for PDEs.  For instance,
first-order PDE for shock waves $u_t=2uu_x$ (§ {\bf 1.3}) has
following Lie symmetry algebra $u_{\tau}=\varphi(u) u_x$ where any
smooth function $\varphi=\varphi(u)$ defines one-parameter Lie
symmetry group with corresponding choice of parameter $\tau$.
Indeed, direct check gives immediately
\be
(u_t)_{\tau}=(2\varphi u_x)u_x + 2u(\varphi u_{xx} +
\varphi^{'}u_x^2)
\ee
and
\be
(u_{\tau})_t=\varphi^{'}u_x 2uu_x+\varphi(2u_x^2+2uu_{xx})
\ee
while relation $(u_t)_{\tau}= (u_{\tau})_t$ yields to the final
answer (here notation $\varphi^{'}\equiv \varphi_u$ was used). In
order to construct dynamic system for this equation, let us
introduce dynamical independent variables as
\be
u, \quad u_1=u_x, \quad u_2=u_{xx}, \quad ...
\ee
and  dynamical system as
\be
\frac{d}{dt}
\begin{bmatrix}u\cr u_1\cr u_2\cr ... \end{bmatrix}=
\begin{bmatrix}2uu_1\cr 2uu_2+2u_1^2 \cr 2uu_3+6u_1u_2\cr ... \end{bmatrix}.
\ee
This system can be transformed into finite-dimensional system
using characteristics method, \cite{sh1991}.\\

For PDE of order $n>1$  analogous dynamic system turns out to be
always infinite and only particular solutions are to be
constructed but no general solutions. Infinite-dimensional
dynamical systems of this sort are not an easy treat and
 also choice of dynamical variables presents sometimes a special
 problem to be solved, therefore even in such an exhaustive textbook as Olver's \cite{Olver}
 these systems are not even discussed. On the other hand,
 practically all known results on  classification of integrable
 nonlinear PDEs of two variables have been obtained
 using this approach (in this context the notion of F-Integrability is
 used as it was done to integrate Burgers equation, {\bf § 1.5.}) For instance, in \cite{zhi1979} for a PDE of
 the form
\be
u_{xy}=f(x,y)
\ee
with arbitrary smooth function $f$ on the right hand it was proven
that this PDE is integrable and has symmetries {\bf iff} right
part has one of the following forms: $e^u$, $\sin{u}$ or
$c_1e^u+c_2e^{-2u}$ with arbitrary constants $c_1,c_2$. Another
interesting result was presented in \cite{sh1991}  where all PDEs
of the form
\be
u_t=f(x, u , u_x, u_{xx})
\ee
have been classified. Namely, a PDE of this form is integrable and
has symmetries {\bf iff} if it can be linearized by some special
class of transformation. General form of transformation is written
out explicitly.

\subsection{Examples}
Few examples presented here demonstrate different
constellations of symmetries (SYM), conservation laws (CL) and solutions (SOL)
for a given equation(s).\\

\textbf{1.4.3.1}: {\bf  SYM +, CL +, SOL +}.\\

Let us regard a very simple equation
\be y^{''}=1,\ee
then its
dynamical system can be written out as
\be
 \frac{d\vec{y}}{dt} =\begin{bmatrix}1\cr
y^{'}\cr F\end{bmatrix}=
\begin{bmatrix} 1\cr y^{'}\cr 1\end{bmatrix} \quad \mbox{with}
\quad dt=dx
\ee
 and its general solution is
$y=\frac{1}{2}x^2+c_1x+c_2$ with two constants of integration. In
order to construct conservation laws, we need to resolve formula
for solution with respect to the constants $c_1, c_2$:
\be
c_1=y^{'}-x, \quad c_2=y+\frac{1}{2}x^2-xy{'}.
\ee
Now,  we look for solutions $F(\vec{y})$ of the equation
\be
\frac{d}{dt}F(\vec{y})=(\px  + y^{'} \py  + \p_{y^{'}}) F
=\mathcal{L}(F)=0
\ee
with  $\mathcal{L}=\px  + y^{'} \py  + \p_{y^{'}}$. Direct check
shows that $F=y^{'}-x$ and $F=y+\frac{1}{2}x^2-xy{'}$ are
functionally independent solutions of this equation. Moreover,
general solution is an {\bf arbitrary function} of two variables
\be F=F(y^{'}-x,y+\frac{1}{2}x^2-xy{'}),\ee
 for example,
\be F= (y^{'}-x)^2-2(y+\frac{1}{2}x^2-xy{'})= y^{'}-2y.\ee
On the other hand, if there are no restriction on the function
$F$, the conservation laws may take some quite complicated form,
for instance,
\be
F=Arcsin(y^{'}-x)/(y+\frac{1}{2}x^2-xy{'})^{0.93}.\ee
Now, that dynamical system, conservation laws and general solution
of the original equation have been constructed, let us look for
its  symmetry:
\be \quad g(\vec{y}): \quad
\frac{d\vec{y}}{d\tau}=g(\vec{y}),\quad g(\vec{y})=(g_1,g_2,g_3).
\ee
Demand
of compatibility
\be
\frac{d}{d \tau}(f(\vec{y})=\frac{d}{d t}(g(\vec{y}))\quad
\mbox{is equivalent to} \quad \mathcal{L}(g_1)=\mathcal{L}(g_3)=0,
\quad \mathcal{L}(g_2)=g_3,
\ee
and it can be proven that any linear combination of two vectors
$(1,0,0)$ and $(0,x,1)$ with (some) scalar coefficients provides
solution of compatibility problem. Thus, Lie symmetry group
corresponding to the vector $(1,0,0)$ is shift in $x$ while the
second vector $(0,x,1)$ corresponds to summing $y$ with particular
solutions of homogeneous equation.\\

{\bf 1.4.3.2: SYM --, CL +, SOL +}.\\

Let us regard as a system of ODEs
\bea \label{top}
n_1\frac{da_1}{dt}=(n_2-n_3)a_2a_3 \\
n_2\frac{da_2}{dt}=(n_3-n_1)a_1a_3 \\
n_3\frac{da_3}{dt}=(n_1-n_2)a_1a_2
\eea
with variables $a_i$ and constants $n_i$. This system is
well-known in physical applications - it describes three-wave
interactions of atmospheric planetary waves, dynamics of elastic
pendulum or swinging string, etc. Conservation laws for this
system are:
\begin{itemize}
\item{} Energy conservation law is obtained by multiplying the
$i$-th equation by $a_i$, $i=1,2,3$ and adding all three of them:
\be
n_1 a^{2}_{1}+n_2 a^{2}_{2}+n_3 a^{2}_{3}=\const.
\ee

\item{} Enstrophy conservation law is obtained by multiplying the
$\it i$-th equation by $n_i a_i$, $i=1,2,3$ and adding all three
of them:
\be
n_1^2 a^{2}_{1}+n_2^2 a^{2}_{2}+n_3^2 a^{2}_{3}=\const.
\ee
\end{itemize}

 Using these two conservation laws
one can easily obtain expressions for $a_2$ and $a_3$ in terms of
$a_1$. Substitution of these expressions into the first equation
of Sys.(\ref{top}) gives us
 differential equation  on $a_1$:
\be
 (y^{'})^2=f(y) \quad \mbox{with} \quad y=a_1
\ee
 whose explicit solution is one of Jacobian
 elliptic functions while $a_2$ and $a_3$ are two other Jacobian elliptic
 functions (e.g. \cite{CUP}). In fact, Sys.(\ref{top}) is often regarded as one
 of possible definitions of Jacobian elliptic
 functions.\\

Notice that from {\bf Theorem 1.3} one can conclude immediately
that all symmetries consistent with conservation laws, are trivial.\\

{\bf 1.4.3.3: SYM +, CL $\pm$, SOL +}.\\

Let us regard heat equation
\begin{equation}\label{noCL}
\p_t u = \p_{xx} u,
\end{equation}
which generates solutions of Burgers equation (see {\bf § 1.5}).
 Direct check shows that Eq.(\ref{noCL}) is invariant due to
transformations $x=\tilde{x}\tau$ and $t=\tilde{t}\tau^2$ with any
constant $\tau$, i.e. dilation transformations constitute Lie
symmetry group for Eq.(\ref{noCL}).  Moreover, Eq.(\ref{noCL}) is
integrable and its solution
\be
u=\int \exp(ikx-k^2t)\hat{u}(k)dk \quad \mbox{with}\quad
\hat{u}(k)=\frac{1}{2\pi}\int_{-\infty}^
{\infty}\exp(-ikx)u(x,0)dx
\ee
is obtained by Fourier transformation.\\

This example demonstrates also some very peculiar property - heat
equation (as well as Burgers equation) has {\bf only one
conservation law}:
\be
\frac{d}{dt}\int_{-\infty}^{\infty} u  dx =0.
\ee
Nonexistence of any other conservation laws is proven, for
instance, in \cite{sh1991}.

\section{Semi-integrability}
\subsection{Elements of integrability}

Dispersive evolution PDEs on compacts is our subject in this
subsection. In contrast to standard mathematical classification of
LPDO into hyperbolic, parabolic and elliptic operators there
exists some other classification - into dispersive and
non-dispersive operators - which is successfully used in
theoretical physics and {\bf is not complementary} to mathematical
one (for details see \cite{CUP}). Let regard LPDE with
constant coefficients in a form
\be
P(\frac{\partial}{\partial t},\frac{\partial}{\partial x})=0
\ee
where $t$ is time variable and $x$ is space variable, and suppose
that a {\bf linear wave}
\be
\psi (x)=\tilde{A}\exp{i(k x -\o t)}
\ee
with constant amplitude $\tilde{A}$, wave number $ k$ and
frequency $\o$ is its solution. After substituting a linear wave
into initial LPDE we get $ P(-i\o, ik)=0$, which means that  $k$
and $\o$ are connected in some way:
there exist some function $f$ such, that $ f(\o, k)=0.$\\

This connection is called {\bf dispersion relation} and  solution
of the dispersion relation is called {\bf dispersion function},
$\o = \o(k)$. If  condition
\be
\frac {\p^2\o}{\p k^2} \neq 0
\ee
holds, then initial LPDE is called {\bf evolution dispersive
equation} and its solutions obviously are completely defined by the form of dispersive
function.  All these definitions  can be easily reformulated for a
case of more space variables, namely $x_1, x_2,...,x_n$. In this
case linear wave takes form
\be
\psi (x)=\tilde{A}\exp{i(\vec{k}\vec{x} -\o t)}
\ee
with {\bf wave vector} $\vec{k}=(k_1,....k_n)$ and space-like
variable $\vec{x}=(x_1,...,x_n)$. Then
\be
P(\frac{\partial}{\partial t},\frac{\partial}{\partial
x_1},...,\frac{\partial}{\partial x_n})=0,
\ee
dispersion function can be computed from $ P(-i\omega,
ik_1,...,ik_n)=0 $ and the condition of non-zero second derivative
of the dispersion function takes a matrix form:
\be \arrowvert \frac {\partial^2\omega}{\partial k_i \partial k_j}
\arrowvert \neq 0.
\ee
Notice now that solutions of linear evolution dispersive PDE are
known {\bf by definition} and the reasonable question here is:
what can be found about solutions of {\bf nonlinear} PDE
\be
\mathcal{L}(\psi)=\mathcal{N}(\psi)
\ee
with dispersive linear
part $\mathcal{L}(\psi)$ and some nonlinearity $\mathcal{N}(\psi)$?\\

Nonlinear PDEs of this form play major role in the theory of wave
turbulence and in general there is no final answer to this
question. Case of {\bf weak turbulence}, i.e. when nonlinearity
$\mathcal{N}(\psi)$ is regarded small in a sense that wave
amplitudes $A$ are small enough (smallness of an amplitude can be
strictly defined), is investigated in much more details. Two
qualitatively different cases have to be regarded:
\begin{itemize}
\item[] {\bf 1.} coordinates of wave vector are real numbers,
$\{\vec{k}=(k_1,....k_n) | k_i \in \Real\}$ (corresponds to
infinite space domain);
\item[] {\bf 2.}  coordinates of wave vector are integer numbers,
$\{\vec{k}=(k_1,....k_n) | k_i \in \Integer\}$ (corresponds to
compact space domain).
\end{itemize}
 In the
first case method of {\bf wave kinetic equation} has been
developed in 60-th (see, for instance, \cite{has1962}) and applied
for many different types evolution PDEs. Kinetic equation is
approximately equivalent to initial nonlinear PDE but has more
simple form allowing direct numerical computations of each wave
amplitudes in a given domain of wave spectrum. Wave kinetic
equation is an averaged equation imposed on a certain set of
correlation functions and it is in fact one limiting case of the
quantum Bose-Einstein equation while the Boltzman kinetic equation
is its other limit.  Some statistical assumptions have been used
in order to obtain kinetic equations and limit of its
applicability then is a very complicated problem which should be
solved separately for
each specific equation, \cite{zak1999}.\\

 In the second case, it is proven, \cite{CUP}, that solving of initial nonlinear PDE {\bf
can be reduced} to solving a few small connected systems of ODEs of the form
\begin{eqnarray} \label{res}
\begin{cases}
\dot{A}_1= V A_2^* A_3, \\
\dot{A}_2= V A_1^* A_3, \\
\dot{A}_3= - V A_1A_2,
\end{cases}
\end{eqnarray}
in case of quadratic nonlinearity,
\begin{eqnarray}
\begin{cases}
\dot{A}_1= T A_2^* A_3 A_4, \nonumber\\
\dot{A}_2= T  A_1^* A_3A_4, \nonumber\\
\dot{A}_3= -T  A_1A_2A_4^*, \nonumber\\
\dot{A}_4= - T A_1A_2A_3^*, \nonumber
\end{cases}
\end{eqnarray}
in case of cubic nonlinearity and so on. Some of the resulting systems are proven to be integrable, e.g. \cite{BK09_1,Ver68a,Ver68b}.

Notice, that in contrast
to a linear wave with a constant amplitude
$\tilde{A}\neq\tilde{A}(t,\vec{x})$, waves in nonlinear PDE have
amplitudes $A_i$  {\bf depending on time}. It means that solutions
of initial nonlinear PDE have characteristic wave form as in
linear case but wave amplitudes can be expressed in  Jacobian elliptic functions on
time, $cn(T), dn(T)$ and $ sn(T)$. Notice that Sys.(\ref{res}) has
been studied in {\bf § 1.3.2} and its conservation laws were
found. Exact solutions of Sys.(\ref{res}) can be written out explicitly as
functions of initial conditions (see
\cite{BK09_1} for details).

\subsection{Levels of integrability}
 Let us formulate classical three-body problem whose
 integrability attracted attention of many investigators beginning
 with Lagrange. Computing the mutual gravitational interaction of
 three masses is surprisingly difficult to solve and only two integrable cases were found.
For simplicity we regard three-body problem with all masses
 equal, then equations of motion take form
\bea\label{3body}
\frac{d^2z_1}{d t^2}= z_{12}f_{12}+z_{13}f_{13}\\
\frac{d^2z_2}{d t^2}= z_{21}f_{12}+z_{23}f_{23}\\
\frac{d^2z_3}{d t^2}= z_{31}f_{13}+z_{32}f_{23}
\eea
where $z_j$ is a complex number, $z_j= x_j+iy_j$, describing
coordinates of $j$-th mass on a plane, $f_{jk}$ is a given
function depending on the distance between $j$-th and $k$-th
masses (physically it is attraction force) while following
notations are used: $z_{jk}=z_j-z_k$ and $f_{jk}=f(|z_{jk}|^2)$.\\

This system admits following conservation laws:
\begin{itemize}
\item{} {\bf Velocity of center of masses is constant}\\
Summing up all three equations, we get
\be
\frac{d^2}{d t^2}(z_1+z_2+z_3)=0.
\ee
This equality allows us to choose the origin of coordinate system
in such a way that  $$z_1+z_2+z_3=0$$ which simplifies all further
calculations significantly. That is the reason why till the end of
this section this coordinate system is used. Physically it means
that coordinate system is connected with masses center.

\item{} {\bf Conservation of energy}\\
Multiplying $j$-th equation by $\bar{z}^{'}_j$, summing up all
three equations and adding complex conjugate, we obtain on the
left
\be
\sum_{j=1}^{3}( \bar{z}^{'}_j z^{''}_j + z^{'}_j \bar{z}^{''}_j)=
\frac{d}{dt}\sum_{j=1}^{3} z^{'}_j \bar{z}^{'}_j.
\ee
i.e. left hand describes derivative of kinetic energy.\\

On the right we have derivative of potential energy $\mathcal{U}$:
\be
\frac{d}{dt}\mathcal{U}=\frac{d}{dt}(F(|z_{12}|^2)+F(|z_{13}|^2)+F(|z_{23}|^2))
\quad \mbox{with notation} \quad F^{'}=f
\ee
and finally energy conservation law takes form
\be
\sum_{j=1}^{3} z^{'}_j \bar{z}^{'}_j=
F(|z_{12}|^2)+F(|z_{13}|^2)+F(|z_{23}|^2) + \const
\ee
\item{}{\bf Conservation of angular momentum}

By differentiating of angular momentum
\be \im \sum_{j=1}^{3}
z^{'}_j \bar{z}_j \ee with respect to $t$,
we get
\be
\im\Big( \sum
|z^{'}_j|^2+f_{12}z_{12}\bar{z}_{12}+f_{13}z_{13}\bar{z}_{13}+f_{23}z_{23}\bar{z}_{23}\Big)=$$$$
\im\Big( f_{12}|z_{12}|^2 +f_{13}|z_{13}|^2+f_{23}|z_{23}|^2
\Big)=0,
\ee
\end{itemize}

while force $f$ is some real-valued function.\\

In general case there are no other conservation laws and the
problem is not integrable. On the other hand, one may look for
some  periodical solutions of Sys.(\ref{3body}) and try to deduce
the necessary conditions of periodicity. Importance of the
existence of periodical solutions was pointed out already by
Poincare and is sometimes even regarded as a {\bf definition} of
integrability - just as opposite case for a
chaos.\\

\textbf{Theorem 1.4.} {\it If $f_{ij} > 0, \quad \forall i,j$
(so-called repulsive case), then Sys.(\ref{3body}) has no
periodical solutions.}\\

$\blacktriangleright$ Indeed, in case of periodical solution,
magnitude of inertia momentum
\be
\mathcal{Z}:=|z_{12}|^2+|z_{13}|^2+|z_{23}|^2
\ee
should have minimums and maximums as sum of distances between
three masses. On the other hand,
\bea
\frac{1}{2}\frac{d^2}{dt^2}(|z_{12}|^2+|z_{13}|^2+|z_{23}|^2)
= \nonumber \\
|\dot{z}_{12}|^2+|\dot{z}_{13}|^2+|\dot{z}_{23}|^2 +
f_{12}|z_{12}|^2+f_{13}|z_{13}|^2+f_{23}|z_{23}|^2 > 0,
\eea
which contradicts to the fact that function $\mathcal{Z}$ has to
have different signs in the points of minimum and maximum. \qed \\

One interesting case - {\bf Poincare case} - though does not lead
to integrable reduction of Sys.(\ref{3body}), give quite
enlightening results and allows to regard this case as "almost"
integrable. In this case there exists one more conservation law -
conservation of inertia momentum
\be
|z_{12}|^2+|z_{13}|^2+|z_{23}|^2=\const
\ee
and it is possible due to a special choice of function
 $f_{jk}=1/|z_{jk}|^4$ which allows us to reduce initial system to the
ODE of the form

\be \label{per}
\mathcal{B}^{''}=a(\mathcal{B}^{'})^3+b(\mathcal{B}^{'})^2+c\mathcal{B}^{'}+d.
\ee

in new polar coordinates. This equation describes a geometrical
place of points  on the plane, i.e. some plane curve
$\mathcal{B}$, providing solutions of initial system. The curve
$\mathcal{B}$ is of figure-eight form and can not be described by
any known algebraic curve. On the other hand, it can be
approximated with desirable accuracy, for instance, by lemniscate
\be
x^4+\a x^2y^2+\b y^4=x^2-y^2.\ee Very comprehensive collection of
results and graphics one can find in \cite{fu}\\

There exists hypothesis that  {\bf the only periodical solution}
of Eq.(\ref{per}) is this eight-form curve (not proven). Existence
theorem for non-equidistant  periodical solutions is proven for a
wide class of functions $f$ (in variational
setting). \\

The simplest possible case of periodical solution can be obtained
if one of  $z_i$ is equal to zero (obviously,
the problem is reduced to two-body problem, see Ex.6). Two more complicated
classical integrable cases with periodical solutions for
particular choice of $z$-s are known:

\begin{itemize}
\item [--] {\bf Lagrange case}:
 $|z_{12}|=|z_{13}|=|z_{23}|$.\\
 It means that  distances between three masses are
 equal as well as all corresponding  attraction forces, and the masses are
 moving along a circle. Lagrange case is also called equidistance case. In this case
Sys.(\ref{3body}) can be reduced to ODE $y^{''}=f(y)$ and solved
in quadratures.
 \item [--] {\bf Calogero case}: all $z_j$ are real and
 $f_{jk}=1/|z_{jk}|^4$ for $j,k=1,...n$\\
It means that all masses are moving along a line (in fact, along a
real axes) and this is generalization of Euler case of three-body
problem which after appropriate change of variables
 Sys.(\ref{3body})  takes form
\bea
\frac{d^2x_1}{d t^2} &=& \frac{1}{(x_1-x_2)^3} +
\frac{1}{(x_1-x_3)^3} \\
\frac{d^2x_2}{d t^2}&=&  -\frac{1}{(x_1-x_2)^3} +
\frac{1}{(x_2-x_3)^3}\\
\frac{d^2x_3}{d t^2}&=&  -\frac{1}{(x_1-x_3)^3} -
\frac{1}{(x_2-x_3)^3}
\eea
Euler´s system was generalized in \cite{ca}, \cite{ca1} to
\be
\frac{d^2x_j}{d t^2}=\frac{\p \mathcal{U}}{\p x_j} \quad
\mbox{with} \quad \mathcal{U}=\sum_{i<j}\frac{1}{(x_i-x_j)^2}
\quad \mbox{and} \quad j=1,...,n.
\ee
All $n$ independent conservation laws were found and it was proven
that the system is integrable.
\end{itemize}

\section{Summary}
In our first Chapter we introduced a notion of differential
operator and gave few different definitions of integrable
differential equation. It was shown that some of them can be
equivalent for a given equation and it is reasonable to choose an
appropriate one depending on what properties of the equation are
under the study. Some interesting and physically important
examples of "almost" integrable systems were described. Very
intrinsic question on interconnections of conservation laws
and symmetries was also discussed. \\

Simplest possible example of nonlinear equation -  famous Riccati
equation which is ODE of first order with quadratic nonlinearity.
We will use this equation in the next Chapter for demonstrate many properties of
differential equations described above. Riccati equation will also
be very useful for introduction of some new notions like
singularities of solutions, integrability tests, etc.

\section{Exercises for Chapter 1}
\textbf{1.} Using (\ref{d1d2}) prove that $D_3=D_2D_1-D_1D_2$
satisfies Leibnitz rule.

\textbf{2.} Prove that
operator $$\mathcal{L}=\sum a_k x^k \p_x^k$$ can be transformed
into an operator with {\bf constant} coefficients by the following
change of variables: $x=e^t$. (Euler)

\textbf{3.}
Transform equation $\quad u_t=\varphi (u) u_x \quad$ into $ \quad
v_t=vv_x \quad$ by appropriate change of variables.

\textbf{4.}
Prove that LODE with constant coefficients has
conservation laws of the form
\be
\frac{y^{'}-\lambda_1 y}{\lambda_2}e^{-\lambda_3 x}=\const.
\ee

\textbf{5.} For potential energy $\mathcal{U}$ from {\bf § 5.2}
prove Lagrange-Jacobi identity:
\bea
x_1\frac{\p \mathcal{U}}{\p x_1} + y_1\frac{\p \mathcal{U}}{\p
y_1}+ x_2\frac{\p \mathcal{U}}{\p x_2}+y_2\frac{\p \mathcal{U}}{\p
y_2} + x_3\frac{\p \mathcal{U}}{\p x_3}+y_3\frac{\p
\mathcal{U}}{\p y_3}= \nonumber \\
2(f_{12}|z_{12}|^2+f_{13}|z_{13}|^2+f_{23}|z_{23}|^2).
\eea

\textbf{6.} For particular case $z_1=0$ and our choice of coordinate system solve
Sys.(\ref{3body}) explicitly.

\chapter{Riccati equation}

\section{Introduction}
Riccati equation (RE)
\be \label{ric}
\phi_x=a(x)\phi^2+b(x) \phi+c(x)
\ee

is one of the most simple nonlinear differential equations because
it is of {\bf first order} and with {\bf quadratic nonlinearity}.
Obviously, this was the reason that as soon as Newton invented
differential equations, RE was the first one to be investigated
extensively since the end of the 17th century \cite{Ric}. In 1726
Riccati considered the first order ODE
\be w_x=w^2+u(x) \ee
with polynomial in $x$ function $u(x).$ Evidently, the cases $\deg
u=1,\, 2$ correspond to the Airy and Hermite transcendent
functions, respectively. Below we show that Hermite transcendent
is integrable in quadratures.  As to Airy transcendent, it is only
F-integrable (See Ex.3)  though the corresponding equation
itself is at the first glance a simpler one.\\

Thus,  new transcendents  were introduced as solutions of the
first order ODE with the quadratic nonlinearity, i.e. as solutions
of REs. Some classes of REs are known to have general solutions,
for instance:
\be
y^{'} + ay^2=bx^{\alpha}
\ee
where all $a, b, \alpha$ are constant in respect to $x$. D.
Bernoulli discovered (1724-25) that this RE is integrable in
elementary functions if $\alpha=-2$ or $\alpha=-4k(2k-1), \
k=1,2,3,....$. Below some general results about RE are presented
which make it widely usable for numerous applications in different
branches of physics and mathematics.

\section{General solution of RE}
In order to show how to solve (\ref{ric}) in general form, let us
regard two cases.
\subsection{$a(x)=0$}
In case $a(x)=0$, RE takes particular form
\be \label{ric1}\phi_x=b(x) \phi+c(x), \ee
i.e. it is a first-order LODE and its general solution can be
expressed in quadratures. As a first step, one has to find a
solution $z(x)$ of its homogeneous part (see Ex.1), i.e.
\be
z(x): \quad z_x=b(x) z.
\ee
In order to find general solution of Eq.(\ref{ric1}) let us
introduce new variable $\tilde{\phi}(x)=\phi(x)/z(x)$, i.e.
$z(x)\tilde{\phi}(x)=\phi(x)$. Then
\be
(z(x)\tilde{\phi}(x))_x =b(x)z(x)\tilde{\phi}(x)+ c(x), \quad
\mbox{i.e.} \quad z(x)\tilde{\phi}(x)_x=c(x),
\ee
and it gives us general solution of Eq.(\ref{ric1}) in quadratures
\be \label{varcon}\phi(x)=z(x)\tilde{\phi}(x)=z(x)(\int
\frac{c(x)}{z(x)}dx +\const). \ee

This method is called {\bf method of variation of constants} and
can be easily generalized for a system of first-order LODEs
\be
\vec{y}^{'}=A(x)\vec{y}+\vec{f}(x).
\ee
Naturally, for the system of $n$ equations we need to know $n$
particular solutions of the corresponding homogeneous system in
order to use method  of variation of constants. And this is
exactly the bottle-neck of the procedure -
 in distinction with first-order LODEs which are all
integrable in quadratures, already second-order LODEs {\bf are not}.

\subsection{$a(x) \neq 0$}
In this case {\bf one known particular solution of a RE allows
to construct its general solution}. Indeed, suppose that $\varphi_1$ is a particular solution of
Eq.(\ref{ric}), then
\be
c= \varphi_{1,x}-a\varphi_1^2-b\varphi_1
\ee
and  substitution $\phi=y+\varphi_1$ annihilates free term $c$
yielding to an equation
 \be \label{help1} y_x=ay^2+\tilde{b}y \ee
with $\tilde{b}=b+2a\varphi_1.$ After re-writing Eq.(\ref{help1})
as
\be
\frac{y_x}{y^2}=a+ \frac{\tilde{b}}{y}
\ee
and making an obvious change of variables $\phi_1=1/y$, we get a
particular case of RE
\be
\phi_{1,x}+{\tilde{b}}\phi_{1}-a=0
\ee
and its general solution is written out explicitly in the previous
subsection.\\

{\textbf{Example 2.1}

As an important illustrative example leading to many applications
in mathematical physics, let us regard a particular RE in a form
\be \label{Hermit} y_x + y^2= x^2 + \a. \ee
For $\a=1$,  particular solution can be taken as $y=x$ and general
solution obtained as above yields to
\be
y=x+\frac{e^{-x^2}}{\int e^{-x^2} d x + \const},
\ee
i.e. in case (\ref{Hermit})  is integrable in quadratures.
Indefinite integral $\int e^{-x^2} d x$ though not expressed in
elementary functions, plays important role in many areas from
probability
theory till quantum mechanics.\\

For arbitrary $\a$,   Eq.(\ref{Hermit}) possess remarkable
property, namely, after an elementary fraction-rational
transformation
\be \label{Dirac}\hat{y}=x+\frac{\a}{y+x} \ee
it
takes form
\be
\hat{y}_x+\hat{y}^2=x^2+\hat{\a}, \ \  \hat{\a}=\a+2,
\ee
i.e. form of original Eq.(\ref{Hermit}) did not change while its
rhs increased by 2. In particular, after this transformation
Eq.(\ref{Hermit}) with $\a=1$ takes form
\be
\hat{y}_x+\hat{y}^2=x^2+3
\ee
and since $y=x$ is a particular solution of (\ref{Hermit}), then
$\hat{y}=x+1/x$ is a particular solution of the last equation. It
means that for any
\be \a=2k+1, \ \ k=0,1,2,...\ee
general solution
of
Eq.(\ref{Hermit}) can be found in quadratures as it was done for the case $\a=1$. \\

In fact, it means that Eq.(\ref{Hermit}) is {\bf form-invariant}
under the transformations  (\ref{Dirac}). Further we are going to
show that general RE possess similar property as well.

\subsection{Transformation group}
Let us check that general fraction-rational change of variables
\be\label{group}
\hat{\phi}=\frac{\a (x) \phi + \b (x)}{\g (x)\phi + \delta (x)}
\ee
transforms one Riccati equation into the another one similar to
Example 2.1. Notice that (\ref{group}) constitutes group of
transformations generated by
\be
\frac{1}{\phi} ,\quad  \a (x) \phi, \quad \phi +\b (x),
\ee
thus only actions of generators have to be checked:
\begin{itemize}
\item{} $\hat{\phi}=1/\phi$ transforms (\ref{ric})  into
\be \hat{\phi}_x+c(x)\hat{\phi}^2+b(x)\hat{\phi}+a(x)=0,\ee
\item{} $\hat{\phi}=\a (x) \phi$ transforms (\ref{ric})  into
\be \hat{\phi}_x-\frac{a(x)}{\a (x)}\hat{\phi}^2- [b(x)+(\log
\a(x))_x]\hat{\phi}-\a (x) c(x)=0,\ee
\item{} $\hat{\phi}=\phi +\b (x)$ transforms (\ref{ric})  into
\be \hat{\phi}_x-a(x)\hat{\phi}^2+[2\b (x) a(x)- b(x)]\hat{\phi} - \hat{c}= 0,\ee
where
\be \hat{c}=a(x)\b^2(x)- b(x)\b(x)+c(x)+\b(x)_x.\ee
\end{itemize}
Thus, having {\bf one solution} of a some Riccati equation we can
get immediately general solutions of the whole family of REs
obtained from the original one under the action of transformation
group
(\ref{group}).\\

It is interesting to notice that for Riccati equation knowing {\bf
any three solutions} $\phi_1,\phi_2,\phi_3$ we can construct all
other solutions $\phi$ using a very simple formula called {\bf
cross-ratio}:
\be
\label{cross}\frac{\phi-\phi_1}{\phi-\phi_2}=A\frac{\phi_3-\phi_1}{\phi_3-\phi_2}
\ee
with an arbitrary constant $A$, where choice of $A$ defines a
solution. In order to verify this formula let us notice that
system of equations
\be
\begin{cases}
\dot{\phi}=a(x)\phi^2+b(x)\phi+c(x)\\
\dot{\phi}_1=a(x)\phi_1^2+b(x)\phi_1+c(x)\\
\dot{\phi}_2=a(x)\phi_2^2+b(x)\phi_2+c(x)\\
\dot{\phi}_3=a(x)\phi_3^2+b(x)\phi_3+c(x)
\end{cases}
\ee
is consistent if
\be
\begin{bmatrix}
\dot{\phi}&\phi^2&\phi&1\cr \dot{\phi}_1&\phi_1^2&\phi_1&1\cr
\dot{\phi}_2&\phi_2^2&\phi_2&1\cr
\dot{\phi}_3&\phi_3^2&\phi_3&1\cr
\end{bmatrix}=0
\ee
and direct calculation shows that this condition is equivalent to
\be \label{Dratio}
\frac{d}{dx}\big(
\frac{\phi-\phi_1}{\phi-\phi_2} \cdot
\frac{\phi_3-\phi_1}{\phi_3-\phi_2}\big)=0.
\ee
 As it was shown, REs  are
not invariant under the action of (\ref{group}) while
(\ref{group}) conserves the form of equations but not form of the
coefficients. On the other hand, it is possible to construct  new
differential equations related to a given  RE which will be
invariant with respect to transformation group (\ref{group}) (see
next section).\\

At the end of this section we consider a very interesting example
\cite{Adler} showing  connection of Eq.(\ref{Dratio}) with first
integrals for
generalization of one of Kovalevskii  problems \cite{Kov}.\\

\textbf{{Adler´s example.}} System of equations
\be \label{AdlerKov} y_{j,x}+2y_j^2=sy_j, \quad s=\sum_{j=1}^n
y_j, \quad j=1,2,...n
 \ee
was studied by Kovalevskii in case $n=3$ and it was shown
that there exist two quadratic first integrals
\be
F_1=(y_1-y_2)y_3, \quad F_2=(y_2-y_3)y_1
\ee
and therefore  Kovalevskii problem  is integrable in
quadratures.\\

In case $n \ge 4$ the use of (\ref{Dratio}) gives us
immediately following some first integrals
\be
\frac{y_l-y_i}{y_l-y_j} \frac{y_k-y_i}{y_k-y_j},
\ee
i.e. Sys.(\ref{AdlerKov}) has nontrivial first integrals for
arbitrary $n$.\\

 It is interesting that for this example solution of
Sys.(\ref{AdlerKov}) is easier to construct without using its
first integrals. Indeed, each equation of this system is a Riccati
equation if $a$ is regarded as given, substitution
$y_i=\phi_{i,x}/2\phi_{i}$ gives
\be \phi_{i,xx}=s\phi_{i,xx}, \ \phi_{i,xx}=a(x)+c_i, \
s=a_{xx}/a_x \ee
and equation for $a$ has form
\be
\frac{a_{xx}}{a_x}=\frac{a_{x}}{2}(\frac{1}{a-c_1}+...+\frac{1}{a-c_n}).
\ee
After integration 
\be a_x^2=\const(a-c_1)...(a-c_n),\ee
 i.e. problem is
integrable in quadratures (more precisely, in hyper-elliptic
functions).\\

In fact, one more generalization of Kovalevskii problem can be
treated along the same lines - case when function $s$ {\bf is not}
sum of $y_j$ but some arbitrary function $s=s(x_1,...,x_n)$. Then
equation on $a$ takes form
\be
\frac{a_{xx}}{a_x}=a_{x}s(\frac{a_{x}}{a-c_1}+...+\frac{a_{x}}{a-c_n})
\ee
which concludes Adler´s example.

\subsection{Singularities of solutions}
All the properties of Riccati equations which have been studied
till now, are in the frame of local theory of differential
equations.  We just ignored possible existence of singularities of
solutions regarding all its properties locally, in a neighborhood
of a point. On the other hand, in order to study analytical
 properties of solutions, one needs to know character of
 singularities, behavior of solutions at infinity,
 etc.

 One can distinguish between two main types of singularities -
 singularities,
 not depending on initial conditions (they are called {\bf fixed}) and
depending on initial conditions (they are called {\bf movable}).
Simplest possible singularity is a pole, and that was the reason
why first attempt of classification of the ordinary nonlinear
differential equations of the first and second order, suggested by
Painleve, used this type of singularities as criterium. Namely,
list of all equations was written out, having only {\bf poles as
movable singularities} (see example of P1 in Chapter 1), and nice
analytic properties of their solutions have been found. It turned
out that, in particular, Painleve equations  describe self-similar
solutions of solitonic equations (i.e. equations in partial
derivatives): P2 corresponds to KdV (Korteweg-de Vries equation),
P4 corresponds to NLS (nonlinear Schrödinger equation)
and so on. 

Using cross-ratio formula (\ref{cross}), it is easy to demonstrate
for a Riccati equation that {\bf all  singularities} of the
solution $\phi$, with an exception  of singularities of particular
solutions $\phi_1,\phi_2,\phi_3$, {\bf are movable poles}
described as following:
\be
\phi_3=\frac{1}{1-A}(\phi_2-A\phi_1)
\ee
where $A$ is a parameter defining the solution $\phi$. Let us
construct a solution with poles for Eq.(\ref{Hermit}) from Example
2.1. We take a solution in a form
\be
\label{series}y=\frac{1}{x+\varepsilon}
+a_0+a_1(x+\varepsilon)+a_2(x+\varepsilon)^2+... \, \ee
with
indefinite coefficients $a_i$, substitute it into (\ref{Hermit})
and make equal terms corresponding to the same power of
$(x+\varepsilon)$. The final system of equations takes form
\be
\begin{cases}
a_0  =  0, \\
3a_1-\a - \varepsilon^2=0, \\
4a_2+2\varepsilon=0, \\
5a_3-1+a_1^2=0, \\
6a_4+2a_1a_2=0, \\
7a_5+2a_1a_3+a_2^2=0 \\
 ...
\end{cases}
\ee
and in particular for $\a=3, \ \varepsilon=0$ the coefficients are
\be a_1=1, \ a_2=a_3=...=0\ee
 which corresponds to the solution
\be
y=x+\frac{1}{x}
\ee
which was found already in Example 2.1.

 This way we have also
learned that each pole of solutions have order 1. In general case,
it is possible to prove that series (\ref{series}) converges  for
arbitrary pair $(\varepsilon, \ \a)$ using the connection of RE
with the theory of linear equations (see next section). In
particular for fixed complex $\a$, it means that for any point
$x_0=-\varepsilon$ {\bf there exist the only solution of
(\ref{Hermit})} with a pole
in this point.

As to nonlinear first order differential equations (with
non-quadratic nonlinearity), they have more complicated
singularities. For instance, in a simple example
\be
y_x=y^3+1
\ee
if looking for a solution of the form $y=ax^k+....$ one gets
immediately
\be
akx^{k-1}+...=a^3x^{3k}+... \ \ \RA k-1=3k  \ \ \RA 2k=-1
\ee
which implies that singularity here is a branch point, not a pole
(also see \cite{Reid}). It make RE also very important while
studying degenerations of Painleve transcendents. For instance,
(\ref{Hermit}) describes particular solutions of P4 (for more
details see Appendix).

\section{Differential equations related to RE}
\subsection{Linear equations of second order}
One of the most spectacular properties of RE is that its theory is
in fact {\bf equivalent} to the theory of second order homogeneous
LODEs
\be \label{hom}
 \psi_{xx}=b(x)\psi_x+c(x)\psi
\ee
because it can easily be shown that these equations can be
transformed into Riccati form and {\it viceversa}. Of course, this
statement is only valid if
Eq.(\ref{ric}) has non-zero coefficient $a(x)$, $a(x) \neq 0$.\\

$\blacktriangleright$ Indeed, let us regard second-order
homogenous LODE (\ref{hom}) and make change of variables
\be
\phi=\frac{\psi_x}{\psi}, \quad \mbox{then}
 \quad \phi_x=\frac{\psi_{xx}}{\psi} -
 \frac{\psi_{x}^2}{\psi^2},\ee
which implies
\be
 \frac{\psi_{xx}}{\psi}=\phi_x +  \frac{\psi_{x}^2}{\psi^2}  =\phi_x +  \phi^2
\ee
and after substituting the results above into initial LODE,  it
takes  form
\be
\phi_x=\phi^2+b(x) \phi+c(x).
\ee
which is particular case of RE. \qed \\

$\blacktriangleleft$ On the other hand, let us regard general RE
\be
\phi_x=a(x)\phi^2+b(x) \phi+c(x)
\ee
and suppose that $a(x)$ is not $\equiv 0$ while condition of
$a(x)\equiv 0$ transforms RE into first order linear ODE which can
be solved in quadratures analogously to Thomas equation (see
Chapter 1). Now, following change of variables
\be
\phi = -\frac{\psi_x}{a(x)\psi}
\ee
transforms RE into
\be
 -\frac{\psi_{xx}}{a(x)\psi}+
 \frac{1}{a(x)}\Big(\frac{\psi_{x}}{\psi}\Big)^2+
 \frac{a(x)_x}{a(x)^2}\frac{\psi_{x}}{\psi}=
 a(x)\Big(\frac{\psi_{x}}{a(x)\psi}\Big)^2
- \frac{b(x)}{a(x)}\frac{\psi_{x}}{\psi}+
 c(x)
\ee
and it  can finally be reduced to
\be
a(x)\psi_{xx}-\big[ a(x)_x + a(x)b(x)\big]\psi_x+ c(x)a(x)^2\psi=0
\ee
which is second order homogeneous LODE. \qed\\

Now, analog of the result of Section 2.2 for second order
equations can be proved.

\textbf{Proposition 2.1}  {\it Using one solution of a second order
homogeneous
LODE, we can construct  general solution as well.}\\

$\blacktriangleright$ First of all, let us prove that without loss
of generality we can put $b(x)=0$ in $
\psi_{xx}+b(x)\psi_x+c(x)\psi=0$. Indeed, change of variables
\be
\psi(x)=e^{-\frac{1}{2}\int b(x)dx} \hat{\psi}(x) \quad
\Rightarrow
\quad\psi_x=(\hat{\psi}_x-\frac{1}{2}b\hat{\psi})e^{-\frac{1}{2}\int
b(x)dx}
\ee
and finally
\be \label{canon2}
\hat{\psi}_{xx}+\hat{c}\hat{\psi}=0, \quad
\hat{c}=c-\frac{1}{4}b^2-\frac12 b_x.
 \ee
Now, if we know one particular solution $\hat\psi_1$ of
Eq.(\ref{canon2}), then it follows from {the considerations above
that RE
\be
\phi_x+\phi^2+\hat{c}(x)=0
\ee
has a solution $\phi_1=\hat{\psi}_{1,x}/\hat{\psi}_1.$ The change
of variables $\hat{\phi} = \phi-\phi_1$ annihilates the
coefficient $\hat{c}(x)$:
\be
(\hat{\phi}+\phi_1)^{'}+\big( \hat{\phi}+\phi_1\big)^2)+
\hat{c}(x)=0  \quad \Rightarrow \quad \hat{\phi}_x+\hat{\phi}^2
+2\phi_1 \hat{\phi}=0\quad \Rightarrow
\ee
\be \label{temp} (\frac{1}{\hat{\phi}})_x = 1
+2\phi_1\frac{1}{\hat{\phi}}, \ee
i.e. we reduced our RE to the particular case Eq.(\ref{ric1})
which is integrable in quadratures. Particular solution
$z=1/\hat{\phi}$ of homogeneous part of Eq.(\ref{temp}) can be
found from
\be
z_x=2z\frac{\hat{\psi}_{1,x}}{\hat{\psi}_1} \quad \mbox{as} \quad
z=\hat{\psi}_1^2
\ee
and Eq.(\ref{varcon}) yields to
\be \label{psi2} \hat{\psi}_2(x)=\hat{\psi}_1\int
\frac{dx}{\hat{\psi}_1^2(x)}. \ee
Obviously, two solutions $\hat{\psi}_1$ and $\hat{\psi}_2$ are
linearly independent since Wronskian $<\hat{\psi}_1,
\hat{\psi}_2>$ is non-vanishing (see Ex.2):
\be <\hat{\psi}_1, \hat{\psi}_2>:= \left| \ba{cc}\hat{\psi}_1 \
\hat{\psi}_2\\\hat{\psi}_1^{'} \ \hat{\psi}_2^{'}\ea \right|=
\hat{\psi}_1 \hat{\psi}_2^{'}-\hat{\psi}_2 \hat{\psi}_1^{'}= 1\ne
0.
\ee
Thus their linear combination gives general solution of  Eq.(\ref{hom}).\qed\\

\textbf{Proposition 2.2} Wronskian $<\psi_1, \psi_2>$ is
constant {\bf iff} $\psi_1$ and $\psi_2$ are solutions of
\be
\psi_{xx}=c(x)\psi .\ee

$\blacktriangleright$ Indeed, if $\psi_1$ and $\psi_2$ are
solutions, then
\be
(\psi_1 \psi_2^{'}-\psi_2 \psi_1^{'})^{'}=\psi_1
\psi_2^{''}-\psi_2 \psi_1^{''}=c(x)(\psi_1 \psi_2-\psi_1 \psi_2)=0
\ \ \RA $$$$\RA \ \ \psi_1 \psi_2^{'}-\psi_2 \psi_1^{'}=\const.
\ee \qed

$\blacktriangleleft$ if Wronskian of two functions $\psi_1$ and
$\psi_2$ is a constant,
\be
\psi_1 \psi_2^{'}-\psi_2 \psi_1^{'}=\const \ \ \RA \ \ \psi_1
\psi_2^{''}-\psi_2 \psi_1^{''}=0$$$$ \RA \ \
\frac{\psi_2^{''}}{\psi_2}=\frac{\psi_1{''}}{\psi_1}.
\ee
\qed

{\bf Conservation of the Wronskian} is one of the most important
characteristics of second order differential equations and will be
used further for construction of modified Schwarzian equation.

To illustrate  procedure described in Proposition 2.1, let us take
{\bf Hermite equation} \be \label{H2}\o_{xx}-2x\o_x+2 \lambda
\o=0. \ee Change of variables $z=\o_x/\o$ yields to
\be
z_x=\frac{\o_{xx}}{\o}-z^2, \ \ z_x+z^2-2xz+2\lambda=0
\ee
and with $y=z-x$ we get finally
\be
y_x+y^2=x^2-2\lambda-1,
\ee
i.e. we got the equation studied in Example 2.1 with
$\a=-2\lambda-1$. It means that all solutions of Hermite equation
with positive integer $\lambda$,  $ \ \lambda=n, \ \ n \in
\Natural$ can easily be found while for negative integer $\lambda$
one needs change of variables inverse to (\ref{Dirac}):
\be
y=-x + \frac{\g}{\hat{y}-x}, \ \ (\hat{y}-x)(y+x)=\g, \ \
\g=\hat{\a}-1, \ \ \hat{\a}=\a+2.
\ee
It gives us {\bf Hermite polynomials}
\be
\begin{cases}
\lambda=0, \ \ y=-x, \ \ \o=1 \\
\lambda=1, \ \ y=-x+\frac{1}{x}, \ \ \o=2x \\
\lambda=2, \ \ y=-x+\frac{4x}{2x^2-1)} \ \ \o=4x^2-2 \\
..........\\
\lambda=n, \ \ y=-x+\frac{\o_x}{\o}, \ \ \o= H_n(x)=(-1)^n
e^{x^2}\frac{d^n}{dx^n}(e^{-x^2}) \\
........
\end{cases}
\ee
as solutions.\\

Notice that the same change of variables
\be \label{log} \phi=\frac{{\psi}_{x}}{\psi} \ee
which linearized original RE, was also used for linearization of
Thomas equation and Burgers equation in Chapter 1. This change of
variables is called {\bf log-derivative} of function $\psi$ or
{\bf Dlog}$(\psi)$ and plays important role in many different
aspects of integrability theory, for instance, when solving
factorization
problem. \\

\textbf{Theorem 2.3.} Linear ordinary differential operator $L$ of order $n$ could be
 factorized with factor of first order ,
 i.e. $L=M\circ(\pa_x - a)$
  for some operator $M$, {\bf iff}
  \be \label{ker}
  a = \frac{\psi_x }{\psi}, \quad \mbox{where} \quad  \psi \in \ker(L).
 \ee

$\blacktriangleright$  $L=M\circ(\pa_x - a), \ a = \psi_x/ \psi$
implies $(\pa_x - a)\psi=0$, i.e. $\psi \in \ker(L)$. \qed \\

$\blacktriangleleft$
 Suppose that $\psi_1=1$ is an element of
the $\ker(L)$, i.e. $\psi_1 \in \ker(L)$. It leads to $a=0$ and
operator $L$ has zero free term  and is therefore divisible by
$\pa_x$.\\

If constant function $\psi_1=1 $ does not belong to the kernel of
initial operator, following change of variables
\be
\hat{\psi}= \frac{\psi}{\psi_1}
\ee
lead us to a new operator \be \label{sim} \hat{L}=f^{-1}L \circ f
\ee which has a constant as a particular solution $\hat{\psi_1}$
for $f=\psi_1$.
\qed \\

\textbf{Remark.} Operators $L$ and $\hat{L}$ given by (\ref{sim}), are
called {\bf equivalent  operators} and their properties will be
studied
in detailed in the next Chapter.\\

 Notice that Theorem 2.3 is analogous to the Bezout´s theorem on
divisibility criterium of a polynomial: A polynomial $P(z)=0$ is
divisible on the linear factor, $P(z)=P_1(z)(z-a)$,  iff $a$ is a
root of a given polynomial, i.e. $P(a)=0$. Thus, in fact this
classical theorem constructs one to one correspondence between
factorizability and solvability of $L(\psi)=0$. \\

The factorization of differential operators is in itself a very
interesting problem which we are going to discuss in details in
 Chapter 3. Here we will only regard one very simple
example - LODO with constant coefficients
\be
L(\psi):=\frac{d^n \psi}{dx^n}+a_1\frac{d^{n-1}
\psi}{dx^{n-1}}+\ldots+a_n\psi =0.
\ee
In this case each root $\lambda_i$ of a {\bf characteristic
polynomial}
\be
\lambda^n+a_1\lambda^{n-1}+\ldots+a_n=0
\ee
generates a corresponding first order factor with
\be
\lambda_i= \frac{\psi_x }{\psi}
\ee
ant it yields to
\be
\psi_x =\lambda_i \psi \quad \Rightarrow \quad
\psi=c_ie^{\lambda_i x}
\ee
and finally
\be
L=\frac{d^n }{dx^n}+a_1\frac{d^{n-1}
}{dx^{n-1}}+\ldots+a_n=(\frac{d}{dx}-\lambda_1)\cdots(\frac{d}{dx}-\lambda_n).
\ee
This formula allows us to construct general solution for
$L(\psi)=0$, i.e. for $\psi \in \ker(L)$, of the form
\be
\psi=\sum c_i e^{\lambda_i x}
\ee
in the case of all distinct roots of characteristic polynomial.\\

In case of  double roots $\lambda_k$ with  multiplicity $m_k$ it
can be shown that
\be \label{multiple} \psi=\sum P_k(x)
e^{\lambda_k x} \ee
where degree of a polynomial $P_k(x)$ depends
on the multiplicity of a root, $deg P_k(x) \leq m_k-1$ (cf. Ex.4)

\subsection{Schwarzian type equations}
Let us regard again second-order LODE
\be\label{second}
\psi_{xx}+b(x)\psi_x+c(x)\psi=0
\ee
and suppose we have two solutions $\psi_1, \psi_2$ of
(\ref{second}). Let us introduce new function $\varphi=
\psi_1/\psi_2$, then
\be
 \varphi_x =
\frac{\psi_{1x}\psi_2-\psi_{2x}\psi_1}{\psi_2^2}, \ee
\be
\varphi_{xx} =
b(x)\frac{\psi_{1x}\psi_2-\psi_{2x}\psi_1}{\psi_2^2}+
2\frac{\psi_{1x}\psi_2-\psi_{2x}\psi_1}{\psi_2^3}\psi_{2x},
\ee
which yields
\be
\frac{\varphi_{xx}}{\varphi_x}= - b(x) - 2
\frac{\psi_{2x}}{\psi_2}
\ee
and substituting $\phi =\frac{\psi_{2x}}{\psi_2}= (\log \psi_2)_x$
into (\ref{ric}) related to  (\ref{second}) we get finally
\be\label{schwarz}
\frac 34 \left(\frac{\varphi_{xx}}{\varphi_x}\right)^2 - \frac 12
\frac{\varphi_{xxx}}{\varphi_x} = c(x).
\ee
Left hand of (\ref{schwarz}) is called {\bf Schwarz derivative} or
just {\bf Schwarzian} and is invariant in respect to
transformation group (\ref{group}) with constant coefficients $\a,
\b, \g, \d$:
\be
\hat{\varphi}= \frac{\a\varphi+\b}{\g\varphi+\d}.
\ee
It is sufficient to check only two cases:
\be
\hat{\varphi}=\frac{1}{\varphi} \ \mbox{and} \ \hat{\varphi}=\a
\varphi + \b.
\ee
which can be done directly.\\

\textbf{Remark.}
The schwarzian equation (\ref{schwarz}) with
\be
c(x)=\sum_1^N\frac{\a_k}{(x-a_k)^2}+\sum_1^N\frac{\b_k}{(x-a_k)}
\ee
plays major role in the theory of conformal transformations of
polygons \cite{Gol}. In this case $\ph(x)$ maps polygon with
vertexes $a_1,\dots a_n$ on the complex $x-$plane into unit
circle. \vspace{3mm}

Notice now that for linear second order ODE in the canonical form
\be\label{can2} \psi_{xx}=v(x)\psi \ee
 we have
\be
 \varphi_x =
\frac{\psi_{1x}\psi_2-\psi_{2x}\psi_1}{\psi_2^2}=\frac{\const}{\psi_2^2}
\ee
and it gives rise the question what should be equation for the
$A=\psi_1 \psi_2$ instead $\varphi_x=\psi_2^{-2}.$ This question
leads us to equation which generalize the classical Schwarzian
equation (\ref{schwarz}).

\textbf{Theorem 2.4} Let $\psi_1, \ \psi_2$ are two linear
independent solutions of Eq.(\ref{can2})
 Then functions
\be
\psi_1^2,\ \ \psi_2^2, \ \ \psi_1 \psi_2
\ee
constitute a basis in the solution space of the following third
order equation: \be \label{***}A_{xxx}=4v(x)A_{x}+2v_x(x)A. \ee
Moreover function $\ph=\psi_1 /\psi_2$ satisfies Schwarzian
equation (\ref{schwarz}) as well as $A=\psi_1 \psi_2$ is a
solution of the equation
\be\label{**}
  4v(x)A^2+A_x^2-2AA_{xx}= w^2
\ee
where $w=\psi_{1x}\psi_2-\psi_{2x}\psi_1.$\\

 $\blacktriangleright$ Using notations
\be
A_1=\psi_1^2, \ A_2=\psi_2^2, \ A_3=\psi_1 \psi_2,
\ee
we can compute Wronskian $W$ of these three functions
\be W=<A_1,A_2,A_3>=(\psi_1 \psi_{2,x}- \psi_2
\psi_{1,x})^3=< \psi_1,\psi_2>^3 \ee
 and use Proposition 3.2 to
demonstrate that $$W=\const\neq 0,$$ i.e. functions $A_i$ are
linearly independent.

After introducing notations
\be w= < \psi_1,\psi_2> \ \ \mbox{and} \
\ f_j=\frac{\psi_{j,x}}{\psi_j} \ee
it is easy to obtain
\be
 \frac{w}{A_3}=f_2-f_1,\quad \frac{A_{3,x}}{A_3}=f_2+f_1 \ee
which yields to
\be \label{ff} f_1= \frac{A_{3,x}-w}{2A_3}, \quad
f_2= \frac{A_{3,x}+w}{2A_3}. \ee

Substitution of these $f_j$ into
\be
f_{j,x}+f_j^2=v(x)
\ee
gives rise  Eq.(\ref{**}) with $A=A_3$ and it's differentiation of
with respect to $x$ gives Eq.(\ref{***}) and it easy to see that
{\bf equations (\ref{**}) and (\ref{***}) are equivalent}.
Analogous reasoning shows that $A_1, \ A_2$  are also
solutions of Eq.(\ref{***}) but in this case $w=0$ in Eq.(\ref{**}).

At last, rewrite  Eq.(\ref{**}) as
\be
  4v(x)+\frac{A_x^2}{A^2}-\frac{2 A_{xx}}{A}=
\frac{w^2}{A^2}
 \ee
and introduce notation $H=1/A$, then
\be \label{modSchwarz}
  v(x)=\frac 34 \frac{H_x^2}{H^2}-\frac 12 \frac{H_{xx}}{H}+
w^2 H^2
 \ee
and compare this equation with Schwarzian Eq.(\ref{schwarz}) one
can see immediately that it corresponds to Eq.(\ref{modSchwarz})
with $w=0$ and $H=\ph_x.$
\qed \\

 Substitution $\varphi=
\psi_1 /\psi_2$ allowed us to get invariant form of the initial
Eq.(\ref{second}), and this is the
reason why we call it {\bf modified Schwarzian equation}. This
form of Schwarzian equation turns out to be useful for a
construction of approximate solutions of Riccati equations with
parameter (see next section). is called {\bf
modified Schwarzian equation}.\\

Notice that after the substitution $a=e^{2b}$,  rhs of modified
Schwarzian equation, i.e. {\bf modified Schwarzian derivative,
Dmod}, takes a very simple form
\be Dmod(a):=\frac 34 \frac{a_x^2}{a^2}-\frac 12 \frac{a_{xx}}{a}=
b_{xx}+b_x^2
 \ee
which is in a sense similar to {\bf Dlog}. Indeed, for
$\psi=e^\varphi$,
\be Dlog(\psi)=\frac{\psi_x}{\psi}=\varphi_x=e^{-\varphi}\frac{d}{dx}e^{\varphi},\ee
while
\be Dmod(e^{2\varphi})=e^{-\varphi}\frac{d^2}{dx^2}e^{\varphi}.\ee

At the end of this section let us stress the following  basic
fact: we have shown that from some very logical point of view,
first order  nonlinear Riccati equation, second order linear
equation and third order nonlinear Schwarzian equation {\bf are
equivalent}! It gives us freedom to choose the form of equation
which is most adequate for specific problem to be solved.

\section{Series in spectral parameter}
In our previous sections we have studied Riccati equation and its
modifications as classical ordinary differential equations, with
one independent variable. But many important applications of
second order differential equations consist some additional
parameter $\la$, for instance one of the most significant
equations of one-dimensional quantum mechanics takes one of two
forms
\bea
\psi_{xx} & = & (\lambda+u)\psi \label{sch} \\
\psi_{xx} & = & (\lambda^2+u_1\lambda+u_2)\psi \label{zs} \eea
where Eq.(\ref{sch}) is called Schrödinger equation and
Eq.(\ref{zs}) can be considered as  modified Dirac equation in
quantum mechanics while in applications to solitonic hierarchies
it is called Zakharov-Shabat equation.

 Notice that Schrödinger equation, yet not  (\ref{zs}), can be readily rewritten as the equation
 for eigenfunctions of the operator $L=D^2-u:$
(That fact will be used in next chapter).
\be \psi_{xx} =  (\lambda+u)\psi\LRA L\psi=\lambda\psi
\ee
Correspondingly, we call $L=D^2-u$ as {\bf Schrödinger operator}
and $\la$ is called {\bf spectral parameter}. In order to unify
terminology it is convenient to name the whole coefficient before
$\psi$ as {\bf generalized potential} allowing it sometimes to be
a polynomial in $\la$ of {\bf any finite degree}. Thus, coming
back to Eq.(\ref{can2}), the generalized potential here is just
the function $c(x)$ but in Example 2.1 the parametric dependence
has been important and as matter of fact this example corresponds
to Schrödinger equation (\ref{sch}) with the harmonic potential
$u(x)=x^2.$

 Now, with the equation
having a parameter, problem of its integrability became, of
course, more complicated and different approaches can be used to
solve it. If we are interested in a solution for all possible
values of a parameter $\la$, asymptotic solution presented by a
formal series can always be obtained (section 4.1) while for some
specific exact solutions can be constructed (sections 4.2, 4.3) in
a case of truncated series. It becomes possible while existence of
a  parameter gives us one more degree of freedom to play with. Cf.
with Example 4.2 where exact solution has been obtained also as a
series and its convergence resulted from the main theorem of the
theory of differential equations on solvability of Cauchy problem.
On the other hand, this solution is valid only for some restricted
set of parameter´ values, namely for integer odd $\a$.

\subsection{RE with a parameter $\la$}
Let us  show first that the RE with a parameter $\la$
corresponding to Eq.(\ref{zs}), namely
\be \label{zsR}
f_x+f^2=\la^2+u_1\la+u_2, \quad \mbox{with} \quad f=D_x\log(\psi),
\ee
has a solution being represented as a formal series.\\

\textbf{Lemma 2.4} Eq.(\ref{zsR}) has a solution
\be \label{fSeries}f=\la + f_0 +\frac{f_1}{\la}+... \ee where
coefficients $f_j$ are differential polynomials in $u_1$ and
$u_2$.\\

$\blacktriangleright$ After direct  substituting the series
(\ref{fSeries}) into the equation for $f$ and making equal
corresponding coefficients in front of the same powers of $\la$,
we get
\be
\begin{cases}
2f_0=u_1 \\
2f_1+f_{0,x}+f_{0}^2=u_2\\
2f_2+f_{1,x}+2f_{0}f_{1}=0\\
2f_3+f_{2,x}+2f_{0}f_{2}+f_{1}^2=0\\
.....
\end{cases}
\ee
and therefore, coefficients of (\ref{fSeries}) are differential
polynomials of potentials $u_1$ and $u_2$. \qed\\

Notice that taking a series
\be \label{fSeries-}g=-\la + g_0
+\frac{g_1}{\la}+\frac{g_2}{\la^2}+\frac{g_3}{\la^3} \quad ... \ee
as a form of solution , we will get a different system of
equations on its coefficients $g_i$:
\be
\begin{cases}
-2g_0=u_1 \\
-2g_1+g_{0,x}+g_{0}^2=u_2\\
-2g_2+g_{1,x}+2g_{0}g_{1}=0\\
-2g_3+g_{2,x}+2g_{0}g_{2}+g_{1}^2=0\\
 .....
\end{cases}
\ee
Solution of Eq.(\ref{zsR}) constructed in Lemma 4.1 yields to the
solution of original Zakharov-Shabat equation (\ref{zs}) of the
form
\be \label{psi1W}\psi_1 (x,\la)=e^{\int f(x,\la) dx}=e^{\la
x}(\eta_0(x)+\frac{\eta_1(x)}{\la}+
\frac{\eta_2(x)}{\la^2}+\frac{\eta_3(x)}{\la^3}+....) \ee
 and
analogously, the second solution is
\be \label{psi2W}\psi_2
(x,\la)=e^{\int g(x,\la) dx}=e^{-\la
x}(\xi_0(x)+\frac{\xi_1(x)}{\la}+
\frac{\xi_2(x)}{\la^2}+\frac{\xi_3(x)}{\la^3}+....) \ee
In fact,
it can be proven that Wronskian $<\psi_1,\psi_2>$ is a power
series on $\lambda$ (see Ex.8) with constant coefficients. Notice
that existence of these two solutions is not enough to construct
general solution of initial Eq.(\ref{zs}) because linear
combination of these formal series is not defined, also
convergence problem has to be considered. On the other hand,
existence of Wronskian in a convenient form allows us to construct
family of potentials giving convergent series for (\ref{psi1W})
and (\ref{psi2W}). We
 demonstrate it at the more simple example,
namely Schrödinger equation (\ref{sch}).

Let us regard Schrödinger equation (\ref{sch}), its RE has form
\be \label{schR} f_x+f^2=\la+u, \quad \mbox{with} \quad
f=Dlog(\psi), \ee and it can be regarded as particular case of
(\ref{zs}), i.e. the series for $f$ yields to
\be
f=k + f_0 +\frac{f_1}{k}+... , \quad \la=k^2,\ee
and
$g(x,k)=f(x,-k)$. We see that in case of (\ref{sch}) there exists
a simple way to calculate function $g$ knowing function $f$ and it
allows us to construct two  solutions of Schrödinger equation
(\ref{sch}):
\be \label{psi1} \psi_1 (x,k)=e^{\int f(x,k) dx}=e^{k
x}(1+\frac{\zeta_1(x)}{k}+
\frac{\zeta_2(x)}{k^2}+\frac{\zeta_3(x)}{k^3}+....)\ee
and
\be \psi_2 (x,k)= \psi_1 (x,-k). \ee
 Substitution of say $\psi_1$ into
(\ref{sch}) gives a recurrent relation between coefficients
$\zeta_i$:
\be \label{recur} \zeta_{j+1,x}=\frac 12 (u\zeta_j-\zeta_{j,xx}),
\quad \zeta_0=1. \ee
In particular,
\be \label{recur1}  u= 2 \zeta_{1,x}\ee
which means that in order to compute potential $u$  it is enough
to know only {\bf one coefficient $\zeta_{1}$} of the formal
series! Below we demonstrate how this recurrent relation helps us
to define potentials corresponding to a given solution.\\

\textbf{Example 2.5.} Let us regard truncated series corresponding
to the solutions of (\ref{sch})
\be \psi_1=e^{kx}(1+\frac{\zeta_1}{k}), \ \
\psi_2=e^{-kx}(1-\frac{\zeta_1}{k}),\ee then due to (\ref{recur})
\be
u=2\zeta_{1,x}, \ \ \zeta_{1,xx}=2\zeta_{1,x}\zeta_{1}
\ee
and Wronskian $W$ of these two functions has form
\be
\label{wron1} W=<\psi_1, \psi_2> =
-2k+\frac{1}{k}(\zeta_1^2-\zeta_{1,x}). \ee
Notice that
\be (\zeta_1^2-\zeta_{1,x})_x=0 \ee and it means that $W$
does not depend on $x$, $W=W(k)$.
 Introducing notation $k_1$ for a zero of the Wronskian,
$\mathcal{W}(k_1)=0$, it is easy to see that
\be \zeta_1^2-\zeta_{1,x}=k_1^2\ee which implies that $\psi_1$ and
$\psi_2$ {are solutions} of (\ref{sch}) with
\be
\zeta_1 = k_1- \frac{2k_1}{1+e^{-2k_1(x-x_0)}} =
-k_1\tanh{k_1(x-x_0)}
\ee
and potential
\be \label{1soliton}
u=-2\frac{(2k_1)^2}{(e^{k_1(x-x_0)}+e^{-k_1(x-x_0)})^2}=
\frac{-2k_1^2}{\cosh^2 (k_1(x-x_0))}, \ee
where $x_0$ is a
constant of integration. \\

It is important to understand here that general solution of
Schrödinger equation (\ref{sch}) can be now found as a linear
combination of $\psi_1$ and $\psi_2$ for all  values of a
parameter $\la=k^2$ {\bf with exception} of two  special cases:
$k=0$  and $k=k_1$ which implies
functions $\psi_1$ and $\psi_2$ are {\bf linearly dependent} in these points.  \\

\begin{figure}[h]
\begin{center}
\end{center}
\end{figure}

At the Fig. 1  graph of potential $u$ is shown and it is easy to
see that magnitude of the potential in the point of extremum is
defined by zeros of the Wronskian $\mathcal{W}$. At the end of
this Chapter it will be shown that this potential represents a
solitonic solution of stationary KdV equation, i.e. {\bf solution
of a Riccati equation generates solitons!}

\subsection{Soliton-like potentials}
It this section we regard only Schrödinger equation (\ref{sch})
and demonstrate that generalization of the Example 4.2 allows us
to describe a very important special class of potentials having
solutions in a form of truncated series.\\

\textbf{Definition 2.5}  Smooth real-valued function  $u(x)$ such that
\be \ u(x) \to  0 \quad \mbox{for} \quad x \to \pm \infty,\ee
is called {\bf transparent potential} if there exist solutions of
Schrödinger equation (\ref{sch}) in a form of truncated series (\ref{psi1})\\

Another name for a transparent potential is {\bf soliton-like} or
{\bf solitonic} potential due to many reasons. The simplest of
them is just its form  which is a bell-like one and "wave" of this
form was called a soliton by \cite{zabu} and this notion became
one of the most important in the modern nonlinear physics, in
particular while many nonlinear equations have solitonic
solutions.

Notice that truncated series for $\psi_1$ and $\psi_2$ can be
regarded as polynomials in $k$ of some degree $N$ multiplied by
exponent $e^{\pm kx}$ (in Example 4.2 we had $N=1$). In
particular, it means that Wronskian $W=<\psi_1, \psi_2>$ is odd
function, $W(-k)=-W(k)$, vanishing at $k=0$ and also it is a {\bf
polynomial} in $k$  of degree $2N+1$:
\be \label{wronn}
W(k)=-2k\prod_1^N (k^2-k_j^2). \ee

As in Example 2.4., functions $\psi_1$ and $\psi_2$ are linearly
dependent at the points $k_j$, i.e.
\be \label{interp}
\psi_2(x,k_j)=(-1)^{j+1}A_j\psi_1(x,k_j), \quad j=1,2,...,N\ee
with
some constant proportionality coefficients which we write down as  $(-1)^{j+1}A_j$.\\

In order to find the coefficients of the polynomial
\be
e^{-kx}\psi_1(x,k)=k^N+a_1k^{N-1}+...+a_N
\ee
it is enough to write out explicitly $\psi_1$ and $\psi_2$ in
roots $k= k_j$ of polynomial (\ref{wronn}) and substitute these in
equations  (\ref{interp}). Thus we find that equations
(\ref{interp}) are equivalent to the following
 system of $N$ linear equations on $N$ unknowns $ a_1,\dots,a_N:$
 \be \label{adler}
\begin{cases}
\frac{a_1}{k_1}+E_1\frac{a_2}{k_1^2}+\frac{a_3}{k_1^3}+\dots+ E_1=0\\
E_2\frac{a_1}{k_2}+\frac{a_2}{k_2^2}+E_2\frac{a_3}{k_2^3}+\dots+ 1=0\\
\frac{a_1}{k_3}+E_3\frac{a_2}{k_3^2}+\frac{a_3}{k_3^3}+\dots+ E_3=0\\
\quad \dots \qquad \dots   \  \ \qquad \dots
\end{cases} \ee
where following notations has been used:
 \be \label{Ej}
 E_j=\frac{e^{\tau_j}
-e^{-\tau_j}}{e^{\tau_j} +e^{-\tau_j}}= \tanh \tau_j, \quad
\tau_j=k_jx+\b_j, \quad A_j=e^{2\b_j}.\ee

\textbf{Lemma 2.6.}  Let
numbers $k_j$ are ordered in following way
\be \label{ordN}  \ee
 and the values of variables $E_j$ in
coefficients of system (\ref{adler}) obey conditions
\be
\label{adler1} -1< E_j<1, \qquad \forall \ j. \ee
 Then the
determinant $\Delta(E_1,\dots, E_N)$ of system (\ref{adler}) of
linear equations on $N$ unknowns $a_1,\dots,a_N$ does not vanish
and satisfies inequality as follows
 \be \label{adler2}
\Delta(E_1,\dots, E_N)>\frac{1}{k_1\cdots
k_N}\prod_{j<i}\left(\frac{1}{k_i}-\frac{1}{k_j}\right) \ee

$\blacktriangleright$  Obviously  determinant
$\Delta(E_1,\dots,E_N)$ is linear functions in $E_j$ and as such
takes minimal value on the boundary of the set (\ref{adler1}) at
some sequence $E_j=\eps_j, \ \eps_j=\pm 1.$ Thus
\be \Delta>\det\left( \ba{cccr}
 k_1^{-1}& \eps_1k_1^{-2}&k_1^{-3}&\dots \\
\eps_2k_2^{-1}& k_2^{-2}&\eps_2k_2^{-3}&\dots \\
k_3^{-1}& \eps_3k_3^{-2}&k_3^{-3}&\dots \\
\quad &\dots \qquad &\dots   \  \ \qquad &\dots
\ea\right)\defeq\Delta_\eps.
\ee
 Since $\eps_j^2=1$ latter determinant can be computed in the closed form. Indeed
\be
\Delta_\eps= \frac{\eps_2\eps_4\cdots}{k_1\cdots k_N}\det\left(
\ba{cccr}
 1& (\eps_1k_1)^{-1}&(\eps_1k_1)^{-2}&\dots \\
1& (\eps_2k_2)^{-1}&(\eps_2k_2)^{-2}&\dots \\
1& (\eps_3k_3)^{-1}&(\eps_3k_3)^{-2}&\dots \\
\quad &\dots \qquad &\dots   \  \ \qquad &\dots \ea\right).\ee

Thus, use formula for Vandermond determinant we obtain
\be
\Delta_\eps=\frac{\eps_2\eps_4\cdots}{k_1\cdots k_N}
\prod_{j<i}\left(\frac{1}{\eps_ik_i}-\frac{1}{\eps_jk_j}\right)=
\frac{\eps_2^2\eps_4^2\cdots}{k_1\cdots k_N}
\prod_{j<i}\left(\frac{1}{k_i}-\frac{1}{\eps_i\eps_jk_j}\right)\ge
\ee
\be
\frac{1}{k_1\cdots
k_N}\prod_{j<i}\left(\frac{1}{k_i}-\frac{1}{k_j}\right)\ee
\qed \\

\textbf{Theorem 2.7.}
Suppose we have two sets of real numbers
\be \{k_j\}, \quad \{\beta_j\}, \quad j=1,2,...,N,  \quad k_j, \beta_j \in \Real \ee
and numbers $k_j$ are ordered as it shown in (\ref{ordN}). Then
conditions (\ref{adler1}) are fulfilled and defined by the linear
system (\ref{adler}) the functions $\psi_1(x,k), \psi_2(x,k)$
 \be \label{psi12}\psi_1(x,k)=e^{kx}(k^N+a_1k^{N-1}+...+a_N), \quad
\psi_2(x,k)=(-1)^N\psi_1(x,-k) \ee
are such that

 i) Wronskian
$W(k)=<\psi_1, \psi_2>=-2k\prod_1^N (k^2-k_j^2)$

ii) Schrödinger equation (\ref{sch}) is satisfied with transparent
potential $u$ such that
 \be \label{k1u}-2k_1^2<u<0.\ee \\

{\it Proof} of i).

 In order to compute the Wronskian $ W$ of functions  (\ref{psi12}), notice first
 that $W$ is a polynomial with leading  term $-2k^{2N+1}$.  Condition of proportionality
(\ref{interp}) for functions $\psi_1(x,k_j), \ \psi_2(x,k_j)$
provides that $k_j$ are zeros of the Wronskian and that
$\mathcal{W}$ is an odd function on $k$, i.e. (\ref{wronn}) is
proven.

{\it Formal proof of ii)} is a bit technical and we postpone it up
to next chapter. In any case in order to prove
that functions (\ref{psi12}) satisfy Schrödinger equation
\be \frac{\psi_{j,xx}}{\psi_j}=k^2+u,\quad  u= 2 a_{1,x} \ee
it is sufficient to verify equations (\ref{recur})
\be
a_{j+1,x}=\frac 12 (u a_j-a_{j,xx}), \quad a_0=1 \ee
for solution
$a_1,\dots,a_N$ of the linear system (\ref{adler}).
In order to illustrate  how to use this theorem for construction
of exact solutions with transparent potentials let us address  two
cases:  $N=1$ and $N=2$.\\

\textbf{Example 2.8.}  In case $N=1$ explicit form of functions
\be \psi_1=e^{kx}(k+a_1), \quad \psi_2=e^{-kx}(k-a_1)\ee
allows us to find $a_1$ immediately:
\be a_1=-k_1E_1=-k_1 \tanh y_1= -k_1\tanh{(k_1x+\b_1)}\ee
which coincides with formula for a solution of the same equation
obtained in Example 2.5. Notice that using this approach we have
found solution of Schrödinger equation by {\bf pure algebraic
means} while in Example 4.2  we had to solve Riccati equation in
order to compute coefficients of the corresponding truncated
series.  \\

The system (\ref{adler}) for case $N=2$ takes form
\be
\begin{cases}
k_1a_1+E_1a_2+k_1^2E_1=0\\
E_2k_2a_1+a_2+k_2^2=0
\end{cases} \RA a_1=\frac{(k_2^2-k_1^2)E_1}{k_1-k_2E_1E_2}\ee
which yields
\be \label{2soliton}
a_1=D_x\log\left((k_2-k_1)\cosh(\tau_1+\tau_2)+
(k_2+k_1)\cosh(\tau_1-\tau_2)\right) \ee
and corresponding
potential  $u=2a_{1,x}$ has explicit form
 \be\label{u2} u=
2 D_x^2\log\left((k_2-k_1)\cosh(\tau_1+\tau_2)+
(k_2+k_1)\cosh(\tau_1-\tau_2) \right) \ee
where
\be x \to \pm \infty \quad \RA a_1 \to \pm
(k_1+k_2),\ee
 i.e. $u$ is a smooth function such that
\be u \to 0 \quad \mbox{for} \quad x \to \pm \infty. \ee

\textbf{Definition 2.9.} The point $\la\in\Complex$ is called {\bf eigenvalue}
of Schrödinger operator $L=D^2-u$ if there exist solution $\ph$
{\bf eigenfunction} of equation (\ref{sch}) such that
\be \ \ph(x) \to  0 \quad \mbox{for} \quad x \to \pm \infty.\ee

\textbf{Remark 2.10} In spectral theory
(see classical textbook \cite{Codd}) it is proven that for smooth
real-valued potentials $u(x)$ vanishing rapidly as $x \to \pm
\infty$ the set of eigenvalues of Schrödinger operator $L=D^2-u$
is finite and they are all real positive numbers 
$
\la=\la_j, \
j=1,\dots, N.$ 
Moreover, in the case of standard numeration
\be L\ph_j=\la_j\ph_j,\quad \la_1>\la_2,\dots,\la_N>0 \ee
the eigenfunction $\ph_j$ has $j-1$ change of sign on the $x-$
axis. \\

In application to transparent potentials from Theorem 4.6 we find
that
\be x\to\pm\infty\RA E_j\to\pm 1 \ee
and, therefore, in these limits
\be \psi_1(k,x)e^{-kx}\to A^\pm(k) \ \mbox{for} \
x\to\pm\infty \ee
 where $A^\pm(k)$ are polynomials in $k$ of $N-$th
degree. Come back to Lemma 4.5 and the linear system of equations
(\ref{adler}) for $a_1,\dots,a_N$ one finds that
 \be\label{apm}
A^{-}(k)=\prod_1^N(k+k_j) \  \ \mbox{and} \
A^{+}(k)=\prod_1^N(k-k_j)\ee
It is easy to see now that for the
our case the eigenvalues $\la_j=k_j^2$ correspond to zeros of
Wronskian (\ref{wronn}) and eigenfunctions  $\ph_j$ are just
$\psi_1(k_j,x)$ with positive $k_j$ numbered as in
Eq.(\ref{ordN}). Indeed, for any positive $k$ the solutions
$\psi_1(k,x)$ and $\psi_2(k,x)$ vanish when $x\to \mp\infty,$
respectively, and are proportional each other at $k=k_j$ by
Eq.(\ref{interp}).

 To proceed further and apply more deep results of the spectral theory
 it is crucial to introduce {\bf roots variables} i.e. zeros
$k=r_j(x), \ j=1,\dots, N$ of the polynomial $e^{-kx}
\psi_1(k,x)$:
\be \psi_1(k,x)=e^{kx}\prod_1^N(k-r_j(x)). \ee
Consider the function $\psi_1(k,x)$ upon the interval
$-\infty<x<T$ one should notice that the positive roots $r_j(T)$
are eigenvalues of boundary value problem for Eq. (\ref{sch}) on
interval $-\infty<x<T$ with zero boundary conditions. Similarly,
the negative roots $r_j(T)$ are eigenvalues of boundary value
problem on interval $T<x<\infty<x<T$ for the same equation. Apply
the spectral theory one can prove that roots variables $r_j(T)$
increase monotone (Cf Exercise 8) along with $T$ and
\be T\to\pm\infty\RA r_j\to\pm k_j. \ee

Furthermore, the $2N-$th degree polynomial in $k$
 \be\label{rojki} \Phi=\psi_1(k,x)\psi_2(k,x)=\prod_1^N(k^2-r_j^2) \ee
due Eq.(\ref{interp}) change the sign on ends of intervals
\be [k_j,k_{j+1}],\quad k_1>k_2>...>k_N>0 \ee
and it imply that
 \be\label{zeroe}  k_1^2\ge r_1^2\ge k_2^2\ge...\ge k_N^2 \ge
r_N^2\ge 0 \ee
Substitute the expansion of the polynomial $\Phi$ in $\la=k^2$
into Eq.(\ref{**}) from Theorem 3.4 and equate coefficients on
$\la^{N-1}$ one can get the important formula
\be\label{trace1}
 u(x)=2\sum_1^N(r_j^2-k_j^2)
 \ee
of transparent potentials in terms of roots variables. Thus
reality conditions and inequalities (\ref{zeroe}) yields the
estimate (\ref{k1u}) formulated in Theorem 4.5.

Summing up we notice that the fraction
\be
\ph=\frac{\psi_1(k,x)}{\psi_2(k,x)}=e^{2kx}
\prod_1^N\left(\frac{k-r_j}{k+r_j}\right)
\ee
is totally defined by it values at the zeros $k=k_n$ of the
Wronskian  and due (\ref{interp}) we have
 \be\label{pmr}
\log|\prod_1^N\frac{k_n+r_j}{k_n-r_j}|=2(k_nx+\beta_n) \ee
It is
solution Eq.(\ref{schwarz}) and as can be shown  Ansatz
$\beta_n=k_n^3t+\d_n$ gives rise \be \label{kdv}
u_t+6uu_x+u_{xxx}=0 \ee which is important model equation in the
theory of surface waves.
 Formulae (\ref{2soliton}) and (\ref{u2})
have been generalized for the case of arbitrary $N$ by Hirota
whose work gave a rise to a huge amount of papers dealing with
construction of soliton-like solutions for  {\bf nonlinear
differential equations} because some simple trick allows to add
new variables in these formulae (see \cite{kodama} and
bibliography herein).

\subsection{Finite-gap potentials}
In previous section it was shown how to construct integrable cases
of Schrödinger equation with soliton-like potentials vanishing at
infinity. Obvious - but not at all a trivial - next step is to
generalize these results for construction of integrable cases for
Schrödinger equation with {\bf periodic potentials}. In the
pioneering work \cite{Novikov1} the finite-gap potentials were
introduced and described in terms of their spectral properties but
deep discussion of spectral theory lays beyond  the scope of this
book (for exhaustive review see, for instance, \cite{Dubr1}). The
bottleneck of present theory of  finite-gap potentials is
following: spectral properties  formulated by Novikov´s school
provide only
 almost periodic potentials but do not guarantee periodic ones in
all the cases.

We are going to present here some simple introductory results
about finite-gap potentials and discuss a couple of examples. For
this purpose, most of the technique demonstrated in the previous
section can be used though as an auxiliary equation we will use
not Riccati equation but its
equivalent form, modified Schwarzian (\ref{modSchwarz}).\\
In particular, it allows us to generalize Lemma 2.1 for the case
of arbitrary polynomial $U(x,\la)$ in $\la$. Namely we have\\

\textbf{Lemma 2.11.} Equation for modified Schwarzian
 \be \label{modSchwarz1}
  \frac 34 \frac{h_x^2}{h^2}-\frac 12 \frac{h_{xx}}{h}+
\la^m h^2=U(x,\la):=\la^m+u_1 \la^{m-1}+...+ u_m
 \ee
with any polynomial generalized potential $U(x,\la)$ has unique
asymptotic solution represented by formal Laurant series such
that:
\be \label{series1}
h(x,\la)=1+\sum_{k=1}^{\infty}\la^{-k}h_k(x) \ee
where
coefficients $h_j$ are differential polynomials in all
$u_1,...,u_m$.\\

$\blacktriangleright$ The proof can be carried
out directly along the same lines as for Lemma 2.1. \qed

In particular, for $m=1$ which corresponds to Schrödinger equation
(\ref{sch}) with potential $U=\la + u$ we find
\be \label{h1} h_1=\frac 12 u, \quad 2h_2= \frac 12
h_{1,xx}-h_1^2, ... \ee

We remind that for the function $\Phi=1/h$ we have (see Theorem
2.4)
\be
  4U(x,\la)\Phi^2+\Phi_x^2-2\Phi \Phi_{xx}=\la^m\RA
\Phi_{xxx}=4U(x,\la)\Phi_{x}+2U_x(x,\la)\Phi.\ee

\textbf{Definition 2.12.} Generalized potential
\be U(x,\la)=\la^m+u_1 \la^{m-1}+...+ u_m \ee of an equation
\be
\psi_{xx}=U(x,\la)\psi
\ee
is called  {\bf $N$-phase potential} if Eq. (\ref{***}),
\be \varphi_{xxx}=4U(x,\la)\varphi_{x}+2U_x(x,\la)\varphi,\ee
 has a
solution $\Phi$ which is a polynomial in $\la$ of degree $N$:
\be \label{gamma} \Phi(x,\la)=\la^N+\varphi_1(x) \la^{N-1}+...+
\varphi_N(x)=\prod_{j=1}^N (\la-\rho_j(x)).\ee
Roots $\rho_j(x)$ of the solution $\Phi(x,\la)$ are called {\bf
root variables}.

In particular case of Schrödinger equation this potential is also
called {\bf finite-gap potential}. As it follows from
\cite{Dubr1}, original "spectral" definition of a finite-gap
potential is equivalent to our Def. 2.12 which is more convenient
due to its applicability not only for Schrödinger equation but
also for arbitrary equation of the second order.

\textbf{Example 2.13.} Let a solution $\Phi(x, \la)=\la-\rho (x)$ is a
polynomial of first degree and potential is also linear, i.e.
$m=1, \ N=1$. Then after integrating the equation from definition
above, we get
\be \label{**1}
  4(\la + u)\Phi^2+\Phi_x^2-2\Phi \Phi_{xx}=\Omega(\la).
\ee
Obviously the "constant of integration" $\Omega(\la)$ in left
part of (\ref{**1}) is a polynomial in $\la$ of degree 3
\be
\label{polC}\Omega(\la)=4\la^3+c_1\la^2+c_2\la+c_3=4(\la-\la_1)(\la-\la_2)(\la-\la_3),
 \ee
where $\la_i$ are all roots of the polynomial $\Omega(\la)$.
Eq.(\ref{**1}) is identity on $\la$ and therefore without loss of
generality we write further $C(\la)$ for both sides of it. This
identity has to keep true for all values of $\la$, in particular,
also  for $\la=\g (x)$ which gives
\be
\label{ellipt}\g^2_x=C(\g)=4(\g-\la_1)(\g-\la_2)(\g-\la_3).\ee
Now instead of solving Eq.(\ref{**1}), we have to solve
Eq.(\ref{ellipt}) which is
 integrable in elliptic functions.

If we are interested in real solutions without singularities, we
have to think about initial data for Eq.(\ref{ellipt}). For
instance, supposing that all $\la_j$ are real,  without loss of
generality
\be  \la_1 > \la_2 > \la_3, \ee and for  initial data $(x_0,\g_0)$ satisfying
\be  \forall (x_0,\g_0) : \ \la_3<\g_0<\la_2,\ee
Eq.(\ref{ellipt}) has real smooth periodic solution expressed in
elliptic functions
\be
\label{elliptPoten}u=2\g-\la_1-\la_2-\la_3\ee
 It is our finite-gap potential
(1-phase potential) and its period can be computed explicitly as
\be
T=\int_{\la_3}^{\la_2}\frac{d\la}{\sqrt{(\la-\la_1)(\la-\la_2)(\la-\la_3)}}.
\ee

We have regarded in Example 4.8 particular case $mN=1$. Notice
that in general case of $mN>1$  following the same reasoning,
after integration we get  polynomial $C(\la)$ of degree $2N+m$
\be
\label{**2}
  4U(x,\la)\varphi^2+\varphi_x^2-2\varphi \varphi_{xx}=
C(\la):=4\la^{2N+m}+... \ee
and correspondingly a system of
$2N+m-1$ equations on $N$ functions, i.e. the system will be
over-determined.  On the other hand, choice of $\la=\g_j$ makes it
possible to get a closed subsystem of $N$ equations for $N$
functions as above:
\be \label{gammmN} \g^2_{j,x}=C(\g_j)/\prod_{j
\neq k} (\g_j-\g_k)^2. \ee
Following Lemma shows that this
over-determined system of equations has unique solution which is
defined by Sys.(\ref{gammmN}).

\textbf{Dubrovin´s Lemma.} Let system of differential equations
(\ref{gammmN}) on root variables $\rho_j$ defined by (\ref{gamma})
 with
\be
\Omega(\la) =
4(\la^{2N+m}+c_1\la^{2N+m-1}+\dots+c_{2N+m}\defeq\omega^2
\ee and $\Phi=\prod(\la - \rho_k)$. Then following keeps true:\\
1. $\Omega(\la)|_{\la=\rho_j}=\varphi_x^2(x,\la)|_{\la=\rho_j}, \
j=1,...,N$,\\
2. expression
\be \Phi^{-1}(2\Phi_{xx}+\frac{\Omega(\la)-\Phi^2_x}{\Phi}) \ee
is a polynomial in $\la$ of degree $m$ and leading coefficient
1.

$\blacktriangleleft$ Differentiate the product we get
\be
-\Phi_{x}= \rho_{1,x}\prod_{k\ne
1}(\la-\rho_k)+\rho_{2,x}\prod_{k\ne 2}(\la - \rho_k)+\dots \ee
 which implies by $\rho_{jk}\defeq \rho_{j}-\rho_{k}$
\be
\Phi_x|_{\la=\rho_j}= -\rho_{j,x} \prod_{j\neq k }\rho_{jk} \RA
\Omega(\la)|_{\la=\rho_j}=\Phi_x^2(x,\la)|_{\la=\rho_j},
\ee
i.e. first statement of the lemma is proven.\\

The second differentiation yields
\be \Phi_{xx}|_{\la=\rho_j}=
-\rho_{j,xx}\prod_{k\ne j}\rho_{jk}+2\rho_{j,x}\sum_{i\ne
j}^N\prod_{k\ne i,k}\rho_{i,x}\rho_{jk}\ee
 On the other hand
differentiate $\omega(\rho_j)$ we find
\be
\rho_{j,x}\frac{d\omega}{d\la}|_{\la=\rho_j}=\rho_{j,xx}\prod_{k\ne
j}\rho_{jk}+\dots
\ee
It remains only to apply l'Hopital's rule to the fraction from (ii). \qed

Let a function $V(x,\la)$ is defined by Eq.(\ref{**2}) as
\be
4V(x,\la)=\varphi^{-1}(2\varphi_{xx}+\frac{C(\la)-\varphi_x^2}{\varphi})
\ee
(....)
 Below it will be shown that also {\bf
transparent potentials themselves} can be computed algebraically.

The idea of constructing also transparent potentials by pure
algebraic means is very appealing because (....) In order to show
how to do it, we  concentrate further not on the coefficients
$a_j$ of the polynomials $\psi_1$ and $\psi_2$ but on their zeros.
More precisely, let us study properties a polynomial
\be
\label{roots} P=\psi_1 \psi_2 \ee
and its roots. (....)

\section{Summary}
In this Chapter, using Riccati equation as  our main example,
 we tried to demonstrate at least some of
the ideas and notions introduced in Chapter 1. Regarding transformation
group and singularities of solutions for RE we  constructed some
equivalent forms of Riccati equation. We also compared three
different approaches to the solutions of Riccati equation and its
equivalent forms. The classical form of RE allowed us to construct
easily asymptotic solutions represented by formal series. Linear
equation of the second order turned out to be more convenient to
describe finite-gap potentials for exact solitonic solutions which
would be a much more complicated task for a RE itself while
generalization of soliton-like potentials to finite-gap potentials
demanded modified Schwarzian equation.

In our next Chapter we will show that modified Schwarzian equation
also plays important role in the construction of a differential
operator commuting with a given one while existence of commuting
operators allows us to obtain examples of hierarchies for
solitonic equations using Lemma 4.6. In particular, for $m=1$
coefficients $h_k(x)$ of Eq.(\ref{series1}) describe a set of
conservation laws for KdV equations.

\section{Exercises for Chapter 2}
\textbf{1.} Prove that general solution of $z^{'}=a(x)z$ has a
form \be z(x)=e^{\int a(x)dx}.\ee

\textbf{2.} Deduce formula (\ref{psi2}) regarding $<\hat{\psi}_1,
\hat{\psi}_2>=1$ as a linear first order equation on
$\hat{\psi}_2$.

\textbf{3.} Check that Airy function $...$ is a solution of
Eq.(\ref{Airy}).

\textbf{4.} Prove that for $L=\frac{d^m}{dx^m}$ its kernel
consists of all polynomials of degree $\leq m-1$.

\textbf{5.} Let functions $A_1$ and $A_2$ are two solutions of
(\ref{***}).  Prove that the Wronskian $<A_1,\,
A_2>=A_1A_2'-A_2A_1'$ is solution as well.

\textbf{6.} Let the function $A$ satisfies (\ref{**}). Prove that the
functions
\be  f_\pm=\frac12D\log A\pm\frac{\sqrt z}{A}  \ee
satisfies the Riccati equations (\ref{ric}).

\textbf{7.} Proof that
\be \frac 34 \frac{a_x^2}{a^2}-\frac 12 \frac{a_{xx}}{a}=k^2
\LRA a=(\eps_1e^{kx}+\eps_2e^{-kx})^{-2}. \ee

\textbf{8.} (Duke Math.J, v.104(2),2000) , (Nuovo Cimento, v.43,1978) Prove for roots variables
\be
r_{j,xx}+2r_{j,x}r_j=\sum_{1\le i\le N, i\ne
j}\frac{2r_{j,x}r_{i,x}}{r_j-r_i}, \ j=1,\dots, N.
\ee
\textbf{9.} Proof that Wronskian of two functions (\ref{psi1W}) and (\ref{psi2W})
is well-defined and has a form
\be
.....
\ee

\chapter{Factorization of linear operators}

\section{Introduction}
Let us notice first that different definitions of integrability,
as a rule, use linearization of initial equation and/or expansion
on some basic functions which are themselves solutions of some
linear differential equation. Important fact here is that
linearization of some differential equation is its simplification
but not solving yet. For instance, in case of linear Schrödinger
equation, $\psi_{xx}+k^2\psi=u\psi$, we are not able  {\bf to
find} its solutions explicitly but only  {\bf to name} them  Jost
functions and to exploit their useful
properties (see previous Chapters).\\

On the other hand, well-known fact is that for LODE with constant
coefficients operator itself can always be factorized into
first-order factors and thus the problem is reduced to the solving
of a few first-order LODEs:
$$\frac{d}{dx}\psi + \lambda
\psi=f(x)$$
which are solvable in quadratures.\\

In case of differential operators with variable coefficients
factorization is not always possible but for the great number of
operators BK-factorization gives factorization conditions
explicitly which we are going to demonstrate in the next Section.

\section{BK-factorization}

Speaking generally, BK-factorization produces following result: in
case of LPDO with characteristic polynomial having at least one
distinct root, factorization is constructed algebraically for an
operator of arbitrary order $n$ while in case of some multiple
roots of characteristic polynomial of LPDO, factorization is
formulated in terms of Riccati equation(s). Factorization of  LODO
is always equivalent to solving some Riccati equation(s). Below
explicit procedure for order 2 and 3 is briefly described.

\subsection{LPDO of order 2}
Let us  outline here  BK-factorization procedure \cite{bk2005} for
the simplest case of bivariate LPDO of second order. Consider an
operator
\begin{equation}\label{A2}
A_2=\sum_{j+k\le2}a_{jk}\px^j\py^k
=a_{20}\px^2+a_{11}\px\py+a_{02}\py^2+a_{10}\px+a_{01}\py+a_{00}.
\end{equation}
with smooth coefficients and seek for  factorization
$$
A_2=(p_1\px+p_2\py+p_3)(p_4\px+p_5\py+p_6).
$$

Let us write down the equations on $p_i$ explicitly, keeping in
mind the rule of left composition, i.e. that $ \px (\alpha \py) =
\px (\alpha) \py +
\alpha \partial_{xy}.$\\

Then in all cases

$$
 \begin{cases}
  a_{20} &= \ p_1p_4\\
  a_{11} &= \ p_2p_4+p_1p_5\\
  a_{02} &= \ p_2p_5\\
  a_{10} &= \ \mathcal{L}(p_4) + p_3p_4+p_1p_6\\
  a_{01} &= \ \mathcal{L}(p_5) + p_3p_5+p_2p_6\\
  a_{00} &= \ \mathcal{L}(p_6) + p_3p_6
  \end{cases}
  \eqno (\it{2SysP})
 $$\\
where we use the notation $\mathcal{L} = p_1 \px + p_2 \py $. In
generic case we assume that (after a linear change of variables if
necessary)
$$
a_{20}\ne 0 \quad \mbox{and} \quad p_1=1.
$$

Then  first three equations of
 {\it 2SysP}, describing the highest order terms are
equations in the variables $p_2, p_4, p_5$ and to find them we
have to find roots of a quadratic polynomial
$$
\mathcal{P}_2(\o):=  a_{20}(\o)^2 +a_{11}(\o) +a_{02} = 0
$$

and this leads to a linear system for $p_4$, $p_5$ with $\o$ as
parameter:
$$
\begin{bmatrix} 1&0\cr -\o&1\end{bmatrix}\begin{bmatrix}p_4\cr p_5\end{bmatrix}
=\begin{bmatrix}a_{20}\cr a_{11}\end{bmatrix};\qquad
\begin{bmatrix}p_4\cr p_5\end{bmatrix}=
\begin{bmatrix}1&0\cr\o &1\end{bmatrix}
\begin{bmatrix}a_{20}\cr a_{11}\end{bmatrix}.
$$\\
Thus
 $$
  \begin{cases}
p_1=1\\
p_2=-\o\\
p_4=a_{20}\\
p_5=a_{20} \o +a_{11}
  \end{cases}
  \eqno (2{\it Pol})
 $$\\
and choice of a root $\o$ generates different possible
factorizations of operator $A_2$.\\

Having computed $p_2, p_4, p_5$ one can plug them into two next
equations of {\it 2SysP}
$$
  \begin{cases}
  a_{10} &= \ \mathcal{L}(p_4)+p_3p_4+p_1p_6\\
  a_{01} &= \ \mathcal{L}(p_5)+p_3p_5+p_2p_6
  \end{cases}\\
 $$
 and get a {\it linear} system of
equations in two variables $p_3,p_6$ which can easily be solved
\begin{eqnarray}
  p_3 =  \frac{\o a_{10}+a_{01} -\o\mathcal{L}a_{20}- \mathcal{L}(a_{20} \o+a_{11})}
{2a_{20}\o+a_{11}},\nonumber\\
  p_6 =\frac{ (a_{20}\o+a_{11})(a_{10}-\mathcal{L}a_{20})-a_{20}(a_{01}
 -\mathcal{L}(a_{20}\o+a_{11}))}{2a_{20}\o+a_{11}}.\nonumber\\
  \end{eqnarray}

if $\mathcal{P}_2'(\o)=2a_{20}\o+a_{11}\ne 0$, i.e. $\o$ is a
simple root. At this point all coefficients $p_1, p_2, ..., p_6$
have been computed
  and  condition of factorization

\be\label{cond2} a_{00} = \mathcal{L}(p_6)+p_3p_6 \ee

takes form\\
\be\label{cond2explicit}
\begin{cases}
a_{00} = \mathcal{L} \left\{
 \frac{\o a_{10}+a_{01} - \mathcal{L}(2a_{20} \o+a_{11})}
{2a_{20}\o+a_{11}}\right\}+ \frac{\o a_{10}+a_{01} -
\mathcal{L}(2a_{20} \o+a_{11})}
{2a_{20}\o+a_{11}}\times\\
\times\frac{ a_{20}(a_{01}-\mathcal{L}(a_{20}\o+a_{11}))+
(a_{20}\o+a_{11})(a_{10}-\mathcal{L}a_{20})}{2a_{20}\o+a_{11}}.
\end{cases}
 \ee

\subsection{LPDO of order 3}

Now we consider an operator
\begin{eqnarray}\label{A3}
A_3=\sum_{j+k\le3}a_{jk}\partial_x^j\partial_y^k =a_{30}\p_x^3 +
a_{21}\p_x^2 \py + a_{12}\px \py^2 +a_{03}\p y^3\\ \nonumber +
a_{20}\p_x^2+a_{11}\px\py+a_{02}\py^2+a_{10}\px+a_{01}\py+a_{00}.
\end{eqnarray}

with smooth coefficients and look for a factorization
$$
A_3=(p_1\px+p_2\py+p_3)(p_4 \p_x^2 +p_5 \px\py  + p_6 \py^2 + p_7
\px + p_8 \py + p_9).
$$

The conditions of factorization are described
 by the following system:
 $$
  \begin{cases}
  a_{30} &= \ p_1p_4\\
  a_{21} &= \ p_2p_4+p_1p_5\\
  a_{12} &= \ p_2p_5+p_1p_6\\
  a_{03} &= \ p_2p_6\\
  a_{20} &= \ \mathcal{L}(p_4)+p_3p_4+p_1p_7\\
  a_{11} &= \ \mathcal{L}(p_5)+p_3p_5+p_2p_7+p_1p_8\\
  a_{02} &= \ \mathcal{L}(p_6)+p_3p_6+p_2p_8\\
  a_{10} &= \ \mathcal{L}(p_7)+p_3p_7+p_1p_9\\
  a_{01} &= \ \mathcal{L}(p_8)+p_3p_8+p_2p_9\\
  a_{00} &= \ \mathcal{L}(p_9)+p_3p_9
   \end{cases}
  \eqno (\it{3SysP})
 $$
with $\mathcal{L} = p_1 \px + p_2 \py $.\\

Once again we may assume without loss of generality that the
coefficient of the term of highest order in $\p_x$ does not
vanish, and that the linear factor is normalized:
$$
a_{30}\ne 0,\qquad p_1=1.
$$
The first four equations of
 {\it 3SysVar} describing the highest order terms are
equations in the variables $p_2, p_4, p_5, p_6$.  Solving these
equations requires the choice of a root $-p_2$ of a certain
polynomial of third degree.  Once this choice has been made, the
remaining top order coefficients $p_4, p_5, p_6$ are easily found.
The top order coefficients can now be plugged into the next four
equations of {\it 3SysP}.  The first three of these four equations
will now be a {\it linear} system of equations in the variables
$p_3,p_7, p_8$ which is easily solved. The next equation is now a
{\it linear} equation on variable $p_9$ which means that all
variables $p_i, i=1,...,9$ have been found. The last two equations
of {\it 3SysP} will give us then the {\it conditions  of
factorization}.\\

 Namely, at {\bf the first step} from

$$
  \begin{cases}
  a_{30} = p_4 \\
  a_{21} = p_2p_4+p_5 \\
  a_{12} = p_2p_5+p_6\\
  a_{03} = p_2p_6
  \end{cases}\\
$$
it follows that
$$
\mathcal{P}_3(-p_2):=  a_{30}(-p_2)^3 +a_{21}(- p_2)^2 +
a_{12}(-p_2)+a_{03}=0.
$$
As for the case of second order, taking $p_2=-\o$, where $\o$ is a
root of the characteristic polynomial $\mathcal{P}_3$ we get a
linear system in $p_4, p_5, p_6$ with $\o$ as parameter.  Then
again
$$
p_2=-\o, \qquad
 \mathcal{L} = \px - \o \py,
$$
which leads to

$$
  \begin{cases}
  a_{30} = p_4 \\
  a_{21} = -\o p_4 + p_5 \\
  a_{12} = -\o p_5+p_6
  \end{cases}\\
$$
i.e.

$$
\begin{bmatrix} 1&0&0\cr -\o&1&0\cr 0&-\o&1\end{bmatrix}\begin{bmatrix}p_4\cr p_5\cr p_6\end{bmatrix}
=\begin{bmatrix}a_{30}\cr a_{21}\cr a_{12}\end{bmatrix};\qquad
\begin{bmatrix}p_4\cr p_5\cr p_6\end{bmatrix}=
\begin{bmatrix}1&0&0\cr\o &1&0 \cr \o^2& \o& 1 \end{bmatrix}
\begin{bmatrix}a_{30}\cr a_{21}\cr a_{12} \end{bmatrix}.
$$

Thus

 $$
  \begin{cases}
p_1=1\\
p_2=-\o\\
p_4=a_{30}\\
p_5=a_{30} \o+a_{21}\\
p_6=a_{30}\o^2+a_{21}\o+a_{12}.
  \end{cases}
  \eqno (3{\it Pol})
 $$

 At {\bf the second step}, from

$$
  \begin{cases}
  a_{20} &= \ \mathcal{L}(p_4)+p_3p_4+p_1p_7\\
  a_{11} &= \ \mathcal{L}(p_5)+p_3p_5+p_2p_7+p_1p_8\\
  a_{02} &= \ \mathcal{L}(p_6)+p_3p_6+p_2p_8\\
  \end{cases}\\
 $$
 and (2{\it Pol}) we get

 $$
  \begin{cases}
  a_{20}-\mathcal{L} a_{30} &=\ p_3 a_{30} +p_7\\
  a_{11}-\mathcal{L}(a_{30} \o+a_{21}) &=\ p_3(a_{30}\o+a_{21})- \o p_7+p_8\\
  a_{02}-\mathcal{L}(a_{30}\o^2+a_{21}\o+a_{12})&=\ p_3
(a_{30}\o^2+a_{21}\o+a_{12})-\o p_8.
  \end{cases}
  \eqno (3{\it Lin*})\\
 $$
As a linear system for $p_3$, $p_7$, $p_8$ this has determinant
$$
3a_{30}\o^2+2a_{21}\o+a_{12}=\mathcal{P}'_3(\o),
$$
so if $\o$ is a simple root the system has unique solution
\begin{eqnarray}
p_3 = \frac{\o^2 (a_{20} -\mathcal{L} a_{30})
+\o(a_{11}-\mathcal{L}(a_{30}\o+a_{21}))+a_{02}-\mathcal{L}
(a_{30}\o^2+a_{21}\o+a_{12})}{3a_{30}\o^2+2a_{21}\o+a_{12}};\nonumber\\
 p_7= \frac{a_{20}-\mathcal{L}a_{30}}{3a_{30}\o^2+2a_{21}\o+a_{12}}
-\frac{a_{30}}{3a_{30}\o^2+2a_{21}\o+a_{12}}\cdot p_3;\nonumber\\
 p_8= \frac{\o(a_{20}-\mathcal{L}a_{30})+a_{11}-\mathcal{L}(a_{30}\o+a_{21})}
{3a_{30}\o^2+2a_{21}\o+a_{12}} -\frac{a_{30}\o
+a_{21}}{3a_{30}\o^2+2a_{21}\o+a_{12}}\cdot p_3.\nonumber
\end{eqnarray}

 In order to find the last coefficient $p_9$ we use the next
equation of ({\it3SysVar}), namely:

$$
a_{10}  =  \mathcal{L}(p_7)+p_3p_7+p_1p_9, \eqno (3{\it Lin**}).
$$

At this point all coefficients $ p_i, i=1,...9$ have been
computed,
under the assumption that $\o$ is a simple root. \\

At {\bf the third step} from\\

\be \label{cond3}
\begin{cases}
a_{01} = \ \mathcal{L}(p_8)+p_3p_8+p_2p_9\\
a_{00} = \ \mathcal{L}(p_9)+p_3p_9
\end{cases}
\ee

 all the {\it necessary conditions} for factorization can be
written out.  We do not do so here because the formulas are
tedious and do not add anything to understanding the main idea. If
the conditions are satisfied,  the explicit factorization formulae
could be written out as for the second-order operator. The
difference is that in this case the polynomial defined by the
highest order terms is of degree 3
and we have not one but two conditions of factorization.\\

\subsection{Constant coefficients}
\begin{itemize}

\item{} Obviously, in case of constant coefficients $a_{ij}$ all
formulae above can be simplified considerably and used for
classical factorization problem of a polynomial. For instance, a
bivariate second order polynomial
$$
\mathcal{W}=X^2-Y^2+a_{10}X+a_{01}Y+a_{00}
$$
can be factorized into two linear polynomials,
$$
X^2-Y^2+a_{10}X+a_{01}Y+a_{00}=(p_1X+p_2Y+p_3)(p_4X+p_5Y+p_6),
$$
{\bf iff}
$$
a_{00}=\pm\frac{a_{01}^2-a_{10}^2}{4}.
$$

In each case coefficients $p_i$ can be written out explicitly, for
instance if
$$
a_{00}=\frac{a_{01}^2-a_{10}^2}{4},
$$
then
$$
X^2-Y^2+a_{10}X+a_{01}Y+a_{00}=(X+Y+\frac{a_{01}-a_{10}}{2})(X-Y+\frac{a_{01}+a_{10}}{2}).
$$

\item{}  As in case of order two,  constant coefficients $a_{ij}$
simplify all the formulae and reduce the problem under
consideration to the classical factorization  of a polynomial. For
instance, a bivariate third order polynomial
$$
\mathcal{W}=X^2Y+XY^2+a_{20}X^2+a_{11}XY+a_{02}Y^2+a_{10}X+a_{01}Y+a_{00}
$$
can be factorized into the product of one linear and one second
order polynomials,
$$
X^2Y+XY^2+a_{20}X^2+a_{11}XY+a_{02}Y^2+a_{10}X+a_{01}Y+a_{00}=$$$$=(p_1X+p_2Y+p_3)(p_4X^2+p_5XY+p_6Y^2+p_7
X + p_8 Y + p_9)
$$
for instance,  if
$$
a_{01}=a_{10}+(a_{20}+1)(a_{11}-a_{20}-a_{02}),$$$$
a_{00}=(a_{11}-a_{20}-a_{02})[a_{10}+a_{20}(a_{11}-a_{20}-a_{02})],
$$

and the result of factorization then is (with notation $\gamma=
a_{11}-a_{20}-a_{02}$):
$$
X^2Y+XY^2+a_{20}X^2+a_{11}XY+a_{02}Y^2+a_{10}X+a_{01}Y+a_{00}=$$
$$=(X+Y+\gamma)(XY-a_{20}X + (a_{20}-a_{11}+\gamma ) Y +
a_{10}+a_{20}\gamma).
$$
\end{itemize}

\section{Laplace transformation}
\subsection{Main notions}
The most important question now is - what to do when
conditions of factorization are violated? Do we still have a way
to solve an equation $\mathcal{L}(\psi)=0$ corresponding to the
initial operator? In order to answer these questions let us
re-write results of BK-factorization for generic case of second
order hyperbolic operator as

\begin{equation}\label{dar}
\mathcal{L}: \quad \p_x \p_y + a\p_x + b\p_y + c =
\left\{\begin{array}{c}
(\p_x + b)(\p_y + a) - ab - a_x + c\\
(\p_y + a)(\p_x + b) - ab - b_y + c
\end{array}\right.
\end{equation}

and corresponding LPDE as $(\p_x \p_y + a\p_x + b\p_y + c)\psi_1
=0$ and introduce new function $\psi_2= (\p_y + a)\psi_1$. Our
main goal now is to construct some {\bf new} LPDE having $\psi_2$
as a solution and to check its factorization property. If this new
LPDE is factorizable, then its solution is written out explicitly
and due to the invertibility of a transformation $\psi_1
\rightarrow \psi_2$ the
formula for solution of initial LPDE can also be obtained immediately. \\

Let us first introduce some definitions.

\paragraph{Definition 3.1 } Two operators of order $n$
$$
\mathcal{L}_1=\sum_{j+k\le n}a_{jk}\partial_x^j\partial_y^k \quad
\mbox{and} \quad \mathcal{L}_2=\sum_{j+k\le
n}b_{jk}\partial_x^j\partial_y^k
$$
are called {\it equivalent operators} if  there exists some
function $f=f(x,y)$ such that
$$
f \mathcal{L}_1= \mathcal{L}_2 \circ f.
$$
\paragraph{Definition 3.2 } Expressions $$\hat{a}=  ab  +a_x - c \quad
\mbox{and} \quad \hat{b}=  ab  +b_y - c$$ are called {\it Laplace
invariants}.

\paragraph{Lemma 3.3 } Two hyperbolic operators $\mathcal{L}_1$ and
$\mathcal{L}_2$ of the form (\ref{dar}) are equivalent {\bf iff}
their Laplace invariants coincide
pairwise.\\

$\blacktriangleright$ Indeed,
$$
f\mathcal{L}_1(\psi)=f\p_x \p_y + fa_1\p_x + fb_1\p_y + fc_1
$$
and
$$
\mathcal{L}_2(f\psi )=\p_x \p_y(f )\psi + \p_yf \p_x \psi + \p_x
\p_y (\psi )f + \p_y \psi \p_x f+$$$$
 a_2\p_x( f) \psi + a_2\p_x (\psi) f+
b_2\p_y (f) \psi + b_2\p_y (\psi) f + f\psi c_2,
$$
i.e. $ f \mathcal{L}_1= \mathcal{L}_2 \circ f $ iff
$$
a_1=a_2+ \frac{\p_yf}{f}=a_2+ (\log f)_y;
$$
$$
b_1=b_2+ \frac{\p_xf}{f}=b_2+ (\log f)_x;
$$
$$
c_1=c_2+ \frac{\p_x\p_yf}{f}+a_2\frac{\px f}{f} + b_2\frac{\py
f}{f}=c_2+(\log f)_{xy}+ (\log f)_x (\log f)_y+$$$$+ a_2(\log
f)_x+b_2 (\log f)_y.
$$

Direct substitution of these expressions into formulae for Laplace
invariants gives (we use notation $\varphi = \log f$):

$$
\hat{a}_1=a_1b_1 + a_{1,x} - c_1=(a_2+
\varphi_y)(b_2+\varphi_x)+(a_2+ \varphi_y)_x -
$$
$$
c_2-\varphi_{xy}- \varphi_x \varphi_y- a_2\varphi_x-b_2 \varphi_y=
a_2b_2+ a_{2,x}-c_2= \hat{a}_2.
$$

Analogously one can obtain $\hat{b}_1=\hat{b}_2$ and it means that
for two equivalent hyperbolic operators their Laplace invariants
do coincide.\qed \\

$\blacktriangleright$ First of all, let us notice that  two
operators
$$
 \mathcal{L}_1= (\p_x + b_1)(\p_y + a_1)+A_1
$$
and
$$
 \mathcal{L}_2= (\p_x + b_2)(\p_y + a_2)+A_2
$$
can be transformed into some equivalent form
$$
 \tilde{\mathcal{L}}_1= (\p_x + b_1)\p_y + A_1
$$
and
$$
 \tilde{\mathcal{L}}_2= (\p_x + b_2)\p_y + A_2
$$
(perhaps) by different functions $f_1, \ f_2$ such that
$$
f_1 \mathcal{L}_1=  \tilde{\mathcal{L}}_1 \circ f_1, \ \ f_1
\mathcal{L}_2=  \tilde{\mathcal{L}}_2 \circ f_2
$$

and as was proven above, Laplace invariants of the initial
operators coincide with those of the equivalent ones. Operator of
the form
$$
\mathcal{L}= (\p_x + b)\p_y + A
$$
has following Laplace invariants: $ A$ and  $-b_y +A$, i.e. $
A_1=A_2 $ and $b_{1y}=b_{2y} $  for operators $\mathcal{L}_1$ and
$\mathcal{L}_2$ with the same Laplace invariants. It yields to
$$
b_1=b_2 + \varphi (x), \ \   \tilde{\mathcal{L}}_2= (\p_x + b_1 +
\varphi (x))\p_y + A_1
$$
with some arbitrary smooth function $\varphi (x)$. Now operator
$\tilde{\mathcal{L}}_2$ differs from $\tilde{\mathcal{L}}_1$ only
by term $\varphi (x)$ which can be "killed" by one more equivalent
transformation, namely, for some function $f_3=f_3(x)$
$$
\tilde{\tilde{\mathcal{L}}}_2= f_3^{-1} \tilde{\mathcal{L}}_2
\circ f_3=(\p_x + b_1 + \varphi (x)+ (\log f_3)_x)\p_y + A_1
$$
and choice  $\varphi (x)=f_3^{'}/f$ completes the proof.
 \qed \\

Now let us rewrite  initial operator (\ref{dar}) as
$$
\mathcal{L}_1: \quad \p_x \p_y + a_1\p_x + b_1\p_y + c_1
$$

and notice that
$$(\p_x + b_1)\psi_2=\hat{a}_1\psi_1 $$ leads to
$$
\left\{\begin{array}{c}
(\p_x + b_1)\psi_2= \hat{a}_1\psi_1 \\
  (\p_y + a_1)\psi_1=\psi_2
\end{array}\right.
\RA \left\{\begin{array}{c}
\frac{1}{\hat{a}_1}(\p_x + b_1)\psi_2= \psi_1 \\
  (\p_y + a_1)\psi_1=\psi_2
\end{array}\right.
 $$

 and standard formula for  {\bf log-derivative}:
 $$ e^{\varphi}\p_y  e^{-\varphi}= \p_y -
\varphi_y \quad \mbox{with} \quad \varphi=\log{\hat{a}},$$
 gives finally
a new operator $\mathcal{L}_2$ with corresponding LPDE
\begin{equation}\label{hatL}
\mathcal{L}_2(\psi_2)\equiv \Big[\Big(\p_y + a_1 -
(\log{\hat{a}_1})_y \Big)(\p_x +
b_1)-\hat{a}_1\Big]\psi_2=0.\end{equation}

In order to check whether these two operators $\mathcal{L}_1$ and
$\mathcal{L}_2$ are different, let us compute  Laplace
invariants of the new operator $\mathcal{L}_2$:
$$
\begin{cases}
\hat{a}_1=  a_1b_1  +a_{1,x} - c_1 \\
\hat{b}_1=  a_1b_1  +b_{1,y} - c_1
\end{cases}
\RA
\begin{cases}
\hat{a}_2= \hat{a}_1- \hat{a}_{1,y}- (\log \hat{a}_1)_{xy}+a_{1,x} \\
\hat{b}_2=  \hat{a}_1
\end{cases}
$$
Now one can see that operators   $\mathcal{L}_1$ and
$\mathcal{L}_2$ are  not equivalent and  operator $\mathcal{L}_2$
is factorizable if $\hat{a}_2 =0$ or
$\hat{b}_2=0$ (see example below).\\

\paragraph{Definition 3.4}
Transformation $\mathcal{L}_1 \RA \mathcal{L}_2$, i.e.
$$
\begin{cases}
\hat{a}_1=  a_1b_1  +a_{1,x} - c_1 \\
\hat{b}_1=  a_1b_1  +b_{1,y} - c_1
\end{cases}
\RA
\begin{cases}
\hat{a}_2= \hat{a}_1- \hat{a}_{1,y}- (\log \hat{a}_1)_{xy}+a_{1,x} \\
\hat{b}_2=  \hat{a}_1
\end{cases}
$$
is called {\it Laplace transformation.}\\

If first new operator is also not factorizable, the procedure can
be carried out for as many steps as necessary in order to get some
factorizable operator.  At the step $N$ when the first
factorizable operator is found, algorithm stops because the
division on corresponding $\hat{a}_N =0$ is not possible any
more. In fact, it is possible to write out formulae for Laplace transformation in terms of
Laplace invariants only.

\paragraph{Theorem 3.5 } Let $u_n$ is one of Laplace invariants $\hat{a}_n,
\hat{b}_n$ obtained at the step $n$. Then

\be \label{nLaplace}u_{n+1}=2u_n+(\log{u_n})_{xy}-u_{n-1}. \ee

$\blacktriangleright$ Indeed, due to Lemma 3.3 it is enough to regards sequence
of operators of the form
$$
\mathcal{L}_n: \quad \p_x \p_y + a_n\p_x + b_n\p_y + c_n \quad
\mbox{with } a_n=0
$$
because it is easy to find some function $f_{a_n}$ (for instance,
$f_{a_n}=e^{-\int a_ndy}$) such that

\be \label{A0}\mathcal{A}_{n,0}=f_{a_n}^{-1}\mathcal{L}_n \circ
f_{a_n} = \p_x \p_y + \tilde{b}_n\p_y + \tilde{c}_n. \ee

From now on
$$
\mathcal{L}_n: \quad \p_x \p_y + b_n\p_y + c_n
$$
and tilde-s are omitted for simplicity of notations. Now formulae
for Laplace transformation take form

\be \label{LaplaceTrans}\psi_{n,y}=-c_n\psi_{n+1}, \quad
\psi_n=[\px+b_n+(\log c_n)_x]\psi_{n+1} \ee

and Eq.(\ref{hatL}) can be rewritten for the function $\psi_{n+1}$
as
$$
[\px+b_n+(\log {c_n})_x]\py
\psi_{n+1}+[c_n+b_{n,y}+(\log{c_n})_{xy}]\psi_{n+1}=0,
$$
i.e.
$$
c_{n+1}-c_n=b_{n,y}+(\log{c_n})_{xy}, \quad b_{n+1}-b_n=
(\log{c_n})_{x},
$$
$$
c_{n+2}-c_{n+1}=b_{n+1,y}+(\log{c_{n+1}})_{xy}, \quad
b_{n+2}-b_{n+1}= (\log{c_{n+1}})_{x}
$$
and finally \be
\label{TodaC}c_{n+2}=2c_{n+1}+(\log{c_{n+1}})_{xy}-c_n. \ee Notice
that in this case, first Laplace invariant is
$$\hat{a}_n=  a_nb_n  +a_{n,x} - c_n =-c_n $$
and obviously satisfies to Eq.(\ref{TodaC}), i.e. for the first
invariant the statement of the theorem is proven. In order to
prove it for the second invariant $\hat{b}_n$ one has to choose
another sequence of operators with $b_n=0$ generated by some
function $f_{b_n}$ such that

\be \label{B0}\mathcal{B}_{n,0}=f_{b_n}^{-1}\mathcal{L}_n \circ
f_{b_n} = \p_x \p_y + \tilde{a}_n\p_x + \tilde{c}_n. \ee
 \qed

Notice that in order to obtain the recurrent formula for Laplace
invariants, we used separation of variables $x$ and $y$ given by
(\ref{LaplaceTrans}). Moreover, introduction of a new discrete
variable $n$ allows us to regards these equations as
difference-differential ones. In order to deal with this sort of
equations one needs a couple of definitions.

\paragraph{Definition 3.6 } An operator $T$ acting on the
infinite sequences of functions
$$(...,\psi_{-2},
\psi_{-1}, \psi_0, \psi_1, \psi_2 ,... ,\psi_n,...)$$
 as $$ T\psi_n=\psi_{n+1}, \quad T^{-1}\psi_n=\psi_{n-1}$$
is called {\bf shift operator}. For convenience of notation
sometimes infinite vector-function $ \vec{\psi}_{\infty}$ is
introduced
$$
\vec{\psi}_{\infty}=(...,\psi_{-2}, \psi_{-1}, \psi_0, \psi_1,
\psi_2 ,... ,\psi_n,...)
$$
and  matrix of operator $T$ then has the following form:

\be\label{T}   T=
 \left(
 \begin{array}{llllllll}
... & 0 &  1  & 0 & 0 & \dots  & 0 & ...\\
... & 0 &  0  & 1 & 0 &\dots  &  0 & ... \\
 ... & 0 & 0 & 0 & 1 & \dots  &  0  & ...\\
 ... & && \ddots  & \ddots & \ddots  &  & ...\\
 ... & 0 & 0 & \dots & 0 & 0 & 1 & ...\\
... & 0 &   0     & \dots & 0 & 0 &  0& ...
\end{array}
\right), \ee i.e. it is infinite matrix with all zero elements but
the elements over main diagonal - they are equal to 1.

\paragraph{Definition 3.7 } Commutator $\mathcal{C}=[A,B]$ of two operators $A$ and $B$
is defined as $$\mathcal{C}=AB-BA.$$\\

Obviously, following properties hold true:
\begin{itemize}
\item{} $[\px,\py]=0$ (cross-derivative rule), \item{}
$[\px,T]=[\py,T]=0$, \item{}$[T,a]=T\circ a-a \circ T=
(T(a)-a)\circ T$ (Leibnitz rule analog).
\end{itemize}

Let us now regard two operators corresponding
 Laplace transformations from Theorem 3.5 rewriting slightly
formulae (\ref{LaplaceTrans}) in terms of shift operator: \be
\label{LAXb}\psi_{n,y}=-c_nT\psi_n, \quad \psi_{n,x}+b_n\psi_{n}=
T^{-1}\psi_n.\ee

The use of shift operator makes it possible to present their
commutator
$$
\mathcal{C}=[\py+c_nT, \ \ \px+b_n-T^{-1}] \equiv [\py+cT,\ \
\px+b-T^{-1}]
$$
omitting low index $n$.
\paragraph{Lemma 3.8 }
Commutator  $$\mathcal{C}=[\py+cT,\ \ \px+b-T^{-1}]$$ is equal to
zero {\bf iff}

 \be \label{TodaB}
c_{x}=c(T(b)-b), \quad b_{y}=c -T^{-1}(c). \ee

$\blacktriangleright$ Indeed, by definition
$$
\mathcal{C}=b_y-c_xT+cTb-bcT+T^{-1}cT-c=-c+b_y+T^{-1}(c)-(c_x-cT(b)+bc)T.
$$
Now, if $ \mathcal{C}=0$, then $-c+b_y+T^{-1}(c)=0$ and
$c_x-cT(b)+bc=0$, i.e. (\ref{TodaB}) holds.

If (\ref{TodaB}) holds, then coefficients of $ \mathcal{C}$ are
equal to zero, i.e. $ \mathcal{C}=0$. \qed \\

 At the end of this Section let us notice that in the original Eq.(\ref{dar}) two variables $x$ and
$y$ played symmetrical role which can also be observed in
commutation relation of Lemma 3.8 after appropriate gauge
transformation:
$$
e^{q_n}(\py+c_nT)e^{-q_n}=\py-a_n+T, \ \
e^{q_n}(\px+b_n+T^{-1})e^{-q_n}=\px+c_{n-1}T^{-1}
$$
with
$$
c_n=e^{q_{n+1}-q_n},\  \ b_n=q_{nx}, \ \  a_n=q_{ny}.
$$
It can be shown that $q_n$ satisfy the following equation \be
\label{toda} q_{n,xy}=e^{q_{n+1}-q_n}-e^{q_{n}-q_{n-1}}.\ee

The equation (\ref{toda}) is usually called {\bf two-dimensional
Toda lattice} and plays fundamental role in the theory of Laplace
transformations.

\subsection{Truncation condition}

\paragraph{Definition 3.9 } Truncation condition for
Eq.(\ref{nLaplace}), namely
$$(\log{u_n})_{xy}=u_{n+1}-2u_n+u_{n-1}, \quad n=1,...,N+1$$
is defined by Dirichlet boundary conditions, i.e. $u_0=
u_{N+1}=0.$\\

For example, case $\boxed{N=1}$ gives us {\bf Liouville
equation}{\footnote{See Ex.1}} $$(\log u_1)_{xy}+2u_1=0$$ while
case $\boxed{N=2}$ yields to the system
$$
\begin{cases}
(\log{u_1})_{xy}=u_{2}-2u_1+u_{0}\\
(\log{u_2})_{xy}=u_{3}-2u_2+u_{1}
\end{cases} \RA
\begin{cases}
(\log{u_1})_{xy}  =  -2u_1+u_{2}\\
(\log{u_2})_{xy}  = \ \ u_{1}-2u_2
\end{cases}
$$
These both cases are known to be integrable in quadratures. \\

Case of arbitrary given $N$ corresponds to the system of equations
with
 following matrix

 \be\label{AN}   A_N=
 \left(
 \begin{array}{llllll}
 -2 &  1  & 0 & 0 & \dots  & 0\\
1 &  -2  & 1 & 0 &\dots  &  0  \\
 0 & 1 & -2 & 1 & \dots  &  0  \\
 && \ddots  & \ddots & \ddots  &  \\
 0 & 0 & \dots & 1 & -2 & 1 \\
0 &   0     & \dots & 0 & 1 &  -2
\end{array}
\right) \ee

on the right hand. This matrix is some {\bf Cartan matrix} and
another choice of boundary conditions leads to  Cartan matrices of
other form. It is interesting to notice that {\bf all} Cartan
matrices can be constructed in this way. Moreover,
 it is proven that  system of equations corresponding to each Cartan matrix is
integrable in quadratures (\cite{lez}). Most well-known source of
Cartan matrices is semi-simple classification of algebras Lie -
there is exists one-to-one correspondence between these matrices
and semi-simple algebras Lie (\cite{lez}).\\

Explicit expression for any Laplace invariant $u_n$ after $n$
Laplace transformations is given by following lemma (\cite{lez})
which for simplicity is formulated for the special case Toda
chain.

\paragraph{Lemma 3.10 } Let us regard an infinite sequence of functions
$$\{d_n\}, \ n=0,1,2,...$$ such that
$$\px\py \log{d_n}=\frac{d_{n+1}d_{n-1}}{d^2_{n}}, \ \ \forall n$$
and
$$d_0=1, \ \ d_1=w(x,y)$$
for some smooth function $w(x,y)$ of two variables $x,y$. Then
$$
d_n=\det (\px^i\py^j w), \ i,j=1,...,n.
$$
\paragraph{Corollary 3.11 }
Sequence of functions
$$ u_n=\frac{d_{n+1}d_{n-1}}{d^2_{n}}$$
is solution of  Eq.(\ref{nLaplace}) while sequence of functions
$$e^{q_n}=\frac{d_n}{d_{n-1}}$$
is solution of Eq. (\ref{toda}).\\

Theorem 3.5  describes an infinite chain of equations
corresponding to Laplace transformations and to start with this
chain, we need nothing more then two invariants. On the other
hand, many applications of this theorem are connected with some
special problems in which different sort of finite chains are
considered. In the next Section we will  discuss  two most usable
ways to construct some finite chain of invariants and we close
this Section with an example of 2-steps chain.

\paragraph{Example 3.12 }

Let us regard operator
$$
\mathcal{L}_1= \p_x \p_y + x\p_x  + 2,
$$
then its Laplace transformation gives
$$
\begin{cases}
\hat{a}_1=  -1 \\
\hat{b}_1=  -2
\end{cases}
\RA
\begin{cases}
\hat{a}_2= 0 \\
\hat{b}_2=  -1
\end{cases}
$$
and operator $\mathcal{L}_2$ is factorizable:
$$
\mathcal{L}_2=\px\p_y+x\px+1=\px(\py+x).
$$
It is a simple task to write out explicitly solution $\psi_2$ of
LPDE \be \label{step1} \mathcal{L}_2(\psi_2):=\px(\py+x)(\psi_2)=0
\ee and afterwards solution $\psi_1$ of
$$ \mathcal{L}_1(\psi_1):= (\p_x \p_y + x\p_x  + 2)(\psi_1)=0 $$
can be computed by formula \be \label{end1}\psi_1= -\p_x
\psi_2.\ee

Indeed, introducing in Eq.(\ref{step1}) notation
$\varphi=(\py+x)\psi_2$ we find that $\varphi=Y(y)$ is arbitrary
function of one variable (...)
$$
\psi_2= X(x)e^{-xy}+ \int e^{x(y'-y)}Y(y')dy'.
$$

\vspace{5mm}

\subsection{Periodic closure}

\paragraph{Definition 3.13 } {\bf Classical periodic closure} for
equation on Laplace invariants
$$(\log{u_n})_{xy}=u_{n+1}-2u_n+u_{n-1}, \quad n=1,...,N$$
is defined by periodic boundary conditions, i.e. $u_{n+N}=
u_{n}.$\\

In this case Cartan matrix $A_N$ is replaced by matrix
$\tilde{A}_N$ and for $N \ge 3$ its form is

 \be\label{tildeAN}\tilde{A}_N=
 \left(
 \begin{array}{llllll}
 -2 &  1  & 0 & 0 & \dots  & 1\\
1 &  -2  & 1 & 0 &\dots  &  0  \\
 0 & 1 & -2 & 1 & \dots  &  0  \\
 && \ddots  & \ddots & \ddots  &  \\
 0 & 0 & \dots & 1 & -2 & 1 \\
1 &   0     & \dots & 0 & 1 &  -2
\end{array}
\right) \ee

Notice{\footnote{Ex.2}} that matrix $\tilde{A}_N$ {\bf is
degenerated}. It will be shown below that for $N=1$ initial equation can be solved explicitly,
while for $N=2,3$ initial system of equations allows some reduction to {\bf one} scalar equation. \\

Let us regard first case $\boxed{N=1}$ , it yields to
$(\log{u_1})_{xy}=u_{2}-2u_1+u_{0}=0$ and obviously $u_1=
g_1(x)g_2(y)$ with  arbitrary smooth functions $g_1(x),g_2(y)$.\\

Case $\boxed{N=2}$ is more interesting due to the huge amount of
applications (surfaces with constant curvature, relativity theory,
etc. ) and gives rise to  the system of equations

\be \label{z2} \begin{cases}(\log{u_1})_{xy}=2(u_{1}-u_2),\\
(\log{u_2})_{xy}=2(u_{2}-u_{1})\end{cases}\ee

with many important properties: it has conservation laws,
symmetries, soliton-type particular solutions, etc. In particular,
the reduction of this system can easily be constructed to the one
scalar equation:
$$
(\log{u_1})_{xy}+(\log{u_2})_{xy}=0 \RA (\log{u_1 u_2})_{xy}=0 \RA
u_1 u_2 = X(x) Y(y)
$$
with two smooth arbitrary functions $X(x), Y(y)$. Suppose now $u_1
u_2=1$, then setting $u_1=e^\theta, \ \ u_2=e^{-\theta}$ we find
solutions of Sys.(\ref{z2}) from solutions of equation
 \be \label{sinh} \theta_{xy}+\sinh \theta=0. \ee

This equation is called {\bf sinh-Gordon equation} \cite{bianchi}.\\

Case $\boxed{N=3}$ corresponds to the system
$$
\begin{cases}
(\log{u_1})_{xy}=u_2+u_3-2u_1\\
(\log{u_2})_{xy}=u_1+u_3-2u_2\\
(\log{u_3})_{xy}=u_1+u_2-2u_3
\end{cases}
$$
which allows reduction $ u_1=u_3$
$$
\begin{cases}
(\log{u_1})_{xy}=u_2+u_1-2u_1\\
(\log{u_2})_{xy}=u_1+u_1-2u_2\\
(\log{u_1})_{xy}=u_1+u_2-2u_1
\end{cases}
\RA
\begin{cases}
(\log{u_1})_{xy}=u_2-u_1\\
(\log{u_2})_{xy}=2(u_1-u_2)
\end{cases}
$$
and it can be treated analogously with the case above:
$$u_1u_2u_3=1, \ \ u_2=e^{-2\theta}, \ \ u_1=e^{\theta}, \ \ u_3=
e^{\theta} \RA$$
 \be \label{z3}
\theta_{xy}=e^{-2\theta}-e^{\theta}. \ee

Eq.(\ref{z3}) is called {\bf Tzitzeica equation} and  its
solutions give solution of initial system of equations. Tzitzeica
equation is also
very important for various applications \cite{ziz}.\\

For both Eq.(\ref{sinh}) and Eq.(\ref{z3}) their general solutions
{\bf are not available} as well as for the case of general $N$.
Attempts to solve appearing systems of equations directly demand
some tedious technique of inverse scattering
 and produce partial solutions with singularities.
 Method to reduce the initial system to one equation
simplifies drastically construction of smooth solutions. \\

At the end of this section let us notice that the same three
equations which were obtained while studying  truncated and
periodical cases - namely, Liouville, sinh-Gordon and Tzizeica
equations - do appear together in some other context \cite{sha1}:

\paragraph{Theorem.} Nonlinear PDE of the form $u_{xy}=f(u)$ has higher symmetries
{\bf iff} one of three cases take place: \be \label{3func}f=e^u,\
\ f=e^u+e^{-u}, \ \ f=e^u+e^{-2u}.
\ee\\

As we know already (see Chapter 1) that integrability of a
differential equation is intrinsically related to its symmetry
properties. Of course, possession  of a symmetry {\bf does not
mean} that equation is integrable in some sense but this fact
gives us a good hint on what equations {\bf might be integrable}.
Moreover, for some classes of differential  equations it is proven
\cite{sha2} that integrability is equivalent to some well-defined
symmetry properties.\\

 From this point of view the theorem above justifies
  hypothesis that two of these three equations (integrability in quadratures
  of Liouville equation we have demonstrated already) have good integrability properties.
  As we will show in the next section,  integrability of these PDEs
can be reduced to the integrability of well-know ODEs.
\\

 Thus, in
contrast to the truncation condition which leads to integrability
in quadratures for arbitrary order $N$, periodical closure
generates a more complicated situation. Here we are not able to
get the answer directly in terms of invariants but we need first
to study the properties of solutions $\psi_n$ at the all $N$ steps
simultaneously.

\subsection{Separation of variables}

Obviously, in case of periodical closure with period $N$ functions
$\psi_1$ and $\psi_{N+1}$ satisfy the same equation which was
presented in Theorem 3.6:

\be \label{psina} \psi_{n,xy}+b_{n}\psi_{n,y}+c_{n}\psi_{n}=0, \ \
\forall \ \ n=0, \pm1, \pm2,... \ee

or, in matrix form,
$$
(\px\py+\{b\}\py+\{c\})\vec{\psi}_{\infty}=0
$$
with diagonal  matrices $\{b\}$ and $\{c\}$. This fact allows us
to regard {\bf finite} vector-function $\vec{\psi}$,
$$
\vec{\psi}=(\psi_1,...\psi_{N}),
$$
which defines completely all the properties of the initial
infinite system of equations. Notice that the fact of coincidence
for coefficients of two LPDEs {\bf does not mean} that their
solutions $\psi_{n+N}$ and $\psi_{n}$ also do coincide - they
might differ, for instance, by a constant multiplier. Therefore we
need now some notion of periodic closure for solutions.

\paragraph{Definition 3.14 } {\bf Bloch periodic closure} is
defined for the components of function $ \vec{\psi}$ as follows:
$$
\psi_{n+N}=k^N\psi_{n}.
$$

Notice once more that truncation and classical periodical closure
are defined for Laplace invariants, i.e. for {\bf coefficients} of
Eqs.(\ref{psina}), while Bloch closure deals with {\bf solutions}
of the same equations.

\paragraph{Corollary 3.15 }  Shift
matrix $T_N$ in case of Bloch periodical closure has form

\be\label{T_N}   T_N=
 \left(
 \begin{array}{llllll}
 0 &  1  & 0 & 0 & \dots  & 0\\
0 &   0  & 1 & 0 &\dots  &  0  \\
 0 & 0 &  0 & 1 & \dots  &  0  \\
 && \ddots  & \ddots & \ddots  &  \\
 0 & 0 & \dots & 0 &  0 & 1 \\
k^N &   0     & \dots & 0 & 0 &   0
\end{array}
\right) \ee where $k \in \Complex$ is a free parameter.

\paragraph{Corollary 3.16 }
 Basic Lemma 3.8 holds true also for the
case of periodic closure, i.e. commutator of Laplace
transformations $$\mathcal{C}=[\py+cT_N,\ \ \px+b-T_N^{-1}] = 0$$
{\bf iff}
$$
c_{n,x}=c_n(b_{n+1}-b_{n}), \quad b_{n,y}=c_n -c_{n-1}, \quad n
\in \Integer_N,$$ which can be checked directly.

\paragraph{Example 3.17 }

As it was shown above, in case of $N=2$ which corresponds to the
classical periodic closure of $u_n$, chain of invariants
degenerates into sinh-Gordon equation  and this closure has the
form (\ref{z2}). Corollary 3.14 allows us to construct a
connection between closure
for invariants and closure for solutions of(\ref{LaplaceTrans}) :\\

$$
\begin{bmatrix}\psi_1\cr \psi_2\end{bmatrix}_x=
\begin{bmatrix}-b_1 & k^{-2} \cr 1 & -b_2\end{bmatrix}
\begin{bmatrix}\psi_1\cr \psi_2 \end{bmatrix}, \quad
\begin{bmatrix}\psi_1\cr \psi_2\end{bmatrix}_y=
\begin{bmatrix}0 & -c_1 \cr -k^2c_2 & 0\end{bmatrix}
\begin{bmatrix}\psi_1\cr \psi_2 \end{bmatrix}.
$$
Then
$$
\psi_{1x}+b_1\psi_1=k^{-2}\psi_2, \ \
\psi_{1}=b_2\psi_2+\psi_{2x},
$$
and excluding  $\psi_1$ or $\psi_2 $ we get an equation of the
second order on {\bf one} scalar function, for instance
$$
\psi_{2,xx}+(b_1+b_2)\psi_{2x}+(b_{2x}+b_1b_2-k^{-2})\psi_2=0.
$$
This equation is obviously equivalent to linear Shrödinger
equation
$$
\psi_{xx}=(\lambda + u)\psi, \quad \lambda= k^{-2},
$$
i.e. Sinh-Gordon equation is S-integrable. This important fact
plays role, for instance, while constructing surfaces of constant
curvatures (see very exhaustive review \cite{bob}).\\

\section{General invariants and semi-invariants}

Before discussing the notion of general invariant, let us notice
that  {\bf arbitrary} LPDO of second order, $A_2$,  can be
represented in the form  of factorization with reminder
$$
 A_2
=(p_1\px+p_2\py+p_3)(p_4\px+p_5\py+p_6)-l_2
$$
where reminder $l_2$ is defined by (\ref{cond2}):

\be\label{l2} l_2=  a_{00} - \mathcal{L} \left\{
 \frac{\o a_{10}+a_{01} - \mathcal{L}(2a_{20} \o+a_{11})}
{2a_{20}\o+a_{11}}\right\}- \frac{\o a_{10}+a_{01} -
\mathcal{L}(2a_{20} \o+a_{11})} {2a_{20}\o+a_{11}}\times$$$$
\times\frac{ a_{20}(a_{01}-\mathcal{L}(a_{20}\o+a_{11}))+
(a_{20}\o+a_{11})(a_{10}-\mathcal{L}a_{20})}{2a_{20}\o+a_{11}}.
 \ee

Similar to the case of order two,
 arbitrary LPDO of third order $A_3$ can be represented
 in the following form
$$ A_{3}=(p_1\px+p_2\py+p_3)(p_4 \p_x^2 +p_5 \px\py  + p_6 \py^2 +
p_7 \px + p_8 \py + p_9)-l_3\py-l_{31}. $$ In contrast to the
second order LPDO, in this case factorization with reminder gives
us not a function but a linear first order operator and it is
convenient for our further investigations to regard in this case
two "reminders" $l_3$ and $l_{31}$ which are defined by
Sys.(\ref{cond3}):
$$l_3= a_{01}- (p_1\px+p_2\py+p_3)p_8-p_2p_9
$$
$$ l_{31}=a_{00}-(p_1\px+p_2\py+p_3)p_9.$$

In the next Section it will be shown that  "reminders" $l_2, \
l_3, \ l_{31}$ are invariants of the corresponding operators uner
the equivalence transformations.

\subsection{Construction of invariants}
Let us first recollect  definition of two equivalent operators.

\paragraph{Definition 4.1 } Two operators of order $n$
$$
\mathcal{L}_1=\sum_{j+k\le n}a_{jk}\partial_x^j\partial_y^k \quad
\mbox{and} \quad \mathcal{L}_2=\sum_{j+k\le
n}b_{jk}\partial_x^j\partial_y^k
$$
are called {\it equivalent operators} if  there exists some
function $f=f(x,y)$ such that
$$
f \mathcal{L}_1= \mathcal{L}_2 \circ f.
$$

The definition is given for an operator  $ \mathcal{L}$ of
arbitrary order $n$ and obviously {\bf any} factorization
$$
\mathcal{L}=\mathcal{L}_1\mathcal{L}_2
$$
can be written out in equivalent form
$$
f^{-1} \mathcal{L}\circ f=f^{-1}\mathcal{L}_1\mathcal{L}_2\circ
f=(f^{-1}\mathcal{L}_1f)(f^{-1}\mathcal{L}_2\circ f),
$$
as well as sum of operators
$\mathcal{L}=\mathcal{L}_1+\mathcal{L}_2$:
$$
f^{-1} \mathcal{L}\circ f=f^{-1}(\mathcal{L}_1+\mathcal{L}_2)\circ
f=(f^{-1}\mathcal{L}_1\circ f)+(f^{-1}\mathcal{L}_2\circ f).
$$
Below we will take function $f$  in a form $f=e^{\varphi}$ for
convenience. In order to formulate theorem on invariants we need
following notations:

$$ A_{2a}=
a_{20}\p_x^2+a_{11}\px\py+a_{02}\py^2+a_{10}\px+a_{01}\py+a_{00},
$$

$$ A_{2p}=(p_1\px+p_2\py+p_3)(p_4 \p_x +p_5 \py+
p_6)-l_2, $$

$$ A_{3a}=a_{30}\p_x^3 + a_{21}\p_x^2 \py + a_{12}\px
\py^2 + a_{03}\py^3 +
a_{20}\p_x^2+$$$$+a_{11}\px\py+a_{02}\py^2+a_{10}\px+a_{01}\py+a_{00},
$$

$$ A_{3p}=(p_1\px+p_2\py+p_3)(p_4 \p_x^2 +p_5 \px\py
+ p_6 \py^2 + $$$$+p_7 \px + p_8 \py + p_9)-l_3\py-l_{31}, $$

$$\tilde{p}_{i}=f^{-1}p_{i}\circ f, \ \
\tilde{a}_{i,j}=f^{-1}a_{i,j}\circ f,$$

$$\tilde{A}_{i}=f^{-1}A_{i}\circ f, \ \
i=2a,2p,3a,3p.$$

Above $A_{2p}=A_{2a}$, i.e. $A_{2p}$ and $A_{2a}$ are different
forms of the same operator - its initial form and its form after
the factorization with reminder. The same keeps true for $A_{3p}$
and $A_{3a}$, i.e. $A_{3p}=A_{3a}$.

\paragraph{Theorem 4.2 }
 For an operator of order 2, its reminder $l_2$ is
its {\bf invariant} under
 the equivalence transformation, i.e.
$$ \tilde{l}_{2}=l_{2}. $$ For an operator of order 3, its reminder
${l}_{3}$ is its {\bf invariant}, i.e. $$\tilde{l}_{3}=l_{3},$$
while reminder ${l}_{31}$ changes its form as follows:
 $$\tilde{l}_{31}=l_{31}+l_{3}\varphi_y. $$

$\blacktriangleright$ Indeed, for operator of order 2
$$
A_{2a}=A_{2p}-l_2,
$$
i.e.
$$
\tilde{A}_{2a}=f^{-1}A_{2a}\circ f= f^{-1} (A_{2p}-l_2)\circ
f=$$$$=\tilde{A}_{2p}-f^{-1}(l_2)\circ f=\tilde{A}_{2p}-l_2.
$$
For operator of order 3
$$
A_{3a}=A_{3p}-l_{3}\py-l_{31},
$$
i.e.
$$
\tilde{A}_{3a}=f^{-1}A_{3a}\circ f= f^{-1}
(A_{3p}-l_{3}\py-l_{31})\circ
f=$$$$=\tilde{A}_{3p}-f^{-1}(l_{31}+l_{3}\py)\circ
f=\tilde{A}_{3p}-l_{3}\py-l_{3}\varphi_y-l_{31}.\qed
$$

\paragraph{\bf Corollary 4.3: }
If $l_{3}=0$, then $l_{31}$ becomes invariant.\\

That is the reason why we call $l_{31}$ further {\bf
semi-invariant}.

\paragraph{\bf Corollary 4.4: }
If $l_{3} \neq 0$, it is always possible to choose some function
$f: \quad \tilde{l}_{31}= l_{3}\varphi_y+l_{31}=0$.\\

Notice that for second order operator, if its invariant
 $l_2=0$ then operator is factorizable while for third order
 operator two its invariants have to be equal to zero,
 $l_{3}=l_{31}=0$. On the other hand, if operator of third order
 is not factorizable we can always regard it as an operator with
only one non-zero invariant. Of course, all this is true for {\bf
each} distinct root of characteristic polynomial, so that one
expression, say, for $l_{3}$ will generate three invariants in
case of three distinct roots of corresponding polynomial.
Expressions for invariants $l_2$ and $l_{3}$ and also for
semi-invariant $l_{31}$ can be easily written out explicitly using
formulae given by BK-factorization (Section 2.1 and 2.2).\\

As it was show already, for an important particular case -
hyperbolic operator of second order in the form \be
\label{Lap}\p_x \p_y + a\p_x + b\p_y + c \ee - there exist two
Laplace invariants which coincide pairwise for equivalent
operators (Lemma 3.3). After rewriting hyperbolic operator in the
form \be \label{BK}\p_x^2 - \p_y^2 + a\p_x + b\p_y + c \ee by
appropriate change of variables, we can construct Laplace
invariants as a simple particular case from the formulae for
general invariants (see next
Section).\\

\paragraph{\bf Corollary 4.5: } Two hyperbolic second order
operators having the same normal form, (\ref{Lap}) or (\ref{BK}),
are equivalent {\bf iff} their general invariants coincide.

\subsection{Hierarchy of invariants}

As it was shown above, every general invariant is a function of a
distinct root $\o$ of the characteristic polynomial and each
distinct root provides one invariant. It means that for operator
of order $n$ we can get no more than $n$ different invariants.
Recollecting that BK-factorization in this case gives us one first
order operator and one operator of order $n-1$, let us put now
following question: are general invariants of operator of order
$n-1$ also invariants of
corresponding operator of order $n$?\\

Let regard, for instance, operator of order 3:
$$
{A}_{3a}={A}_{3p}={A}_{1}{A}_{2a}-l_{3}\py-l_{31}=
{A}_{1}({A}_{2p}-l_2)-l_{3}\py-l_{31}={A}_{1}{A}_{2p}-l_2{A}_{1}-l_{3}\py-l_{31}$$
and obviously
$$
\tilde{A}_{3a}={A}_{1}{A}_{2p}-l_2{A}_{1}-l_{3}\py-l_{31}=
\tilde{A}_{1}\tilde{A}_{2p}-l_2\tilde{A}_{1}-l_{3}\py-\tilde{l}_{31},
$$
i.e. $l_2$ is also invariant of operator ${A}_{3a}$. Let us notice
that general invariant $l_3=l_3(\o^{(3)})$ is a function of a
distinct root $\o^{(3)}$ of the polynomial
$$
\mathcal{P}_3(z)= a_{30}z^3+a_{21}z^2+a_{12}z+a_{03}
$$
while general invariant $l_2=l_2(\o^{(2)})$ is a function of a
distinct root $\o^{(2)}$ of the polynomial
$$
\mathcal{R}_2(z)=p_4 z^2 +p_5 z  + p_6
$$
with $p_4, \ p_5, \ p_6$ given by ({\it 3Pol}) for $\o=\o^{(3)}$.
In case of all distinct roots of both polynomials
$\mathcal{P}_3(z)$ and $\mathcal{R}_2(z)$, one will get maximal
number of invariants, namely 6 general invariants. Repeating the
procedure for an operator of order $n$, we get maximally $n!$
general invariants. In this way for operator of arbitrary order
$n$ we can construct the hierarchy of its general invariants
$$
l_n, l_{n-1}, ..., l_2
$$
and their explicit form is given by BK-factorization.\\

For instance, let us regard a third order hyperbolic operator in
the form

\be \label{ex4} C=a_{30}\p_x^3 + a_{21}\p_x^2 \py + a_{12}\px
\py^2 +a_{03}\p y^3+\mbox{terms of lower order} \ee

 with constant high order coefficients, i.e.
$a_{ij}=\const \ \forall \ i+j=3$ and all roots of characteristic
polynomial
$$
a_{30}\o^3+a_{21}\o^2+a_{12}\o+a_{03}=\mathcal{P}_3(\o)
$$
are distinct and real.  Then we can construct three simple
independent general invariants in following way. Notice first that
in this case high terms of (\ref{ex4}) can be written in the form
$$
(\a_1 \px + \b_1 \py)(\a_2 \px + \b_2 \py)(\a_3 \px + \b_3 \py)
$$
 for all non-proportional $\a_j, \b_i$ and after appropriate change
 of variables this expression can easily be reduced to

 $$
\px \py (\px+ \py).
 $$

Let us introduce notations

$$\pa_1=\pa_x,\ \
\pa_2=\pa_y,\ \ \pa_3=\pa_1+\pa_2=\p_t,$$

then all terms of the third and second order can be written out as

$$C_{ijk}=(\pa_i+a_i)(\pa_j+a_j)(\pa_k+a_k)=\pa_i\pa_j\pa_k+
a_k\pa_i\pa_j+a_j\pa_i\pa_k+a_i\pa_j\pa_k+$$
$$+
(\pa_j+a_j)(a_k)\pa_i+(\pa_i+a_i)(a_k)\pa_j+(\pa_i+a_i)(a_j)\pa_k+
(\pa_i+a_i)(\pa_j+a_j)(a_k)$$

 with $$a_{20}=a_2, \
a_{02}=a_1, \ a_{11}=a_1+a_2+a_3$$ and $c_{ijk}=C-C_{ijk}$ is an
operator of the first order which can be written out explicitly.
As it was shown above, coefficients of $c_{ijk}$ in front of first
derivatives {\bf are invariants} and therefore, any linear
combination of invariants is an invariant itself. These invariants
have the form:\\

for $c_{123}$ we have
$$
a_2a_3+a_1a_2+\py(a_3)+\px(a_2) -a_{10}, \ \
a_1a_3+a_1a_2+\px(a_3)+\px(a_2) -a_{01};
$$

for $c_{312}$ we have
$$
a_2a_3+a_1a_2+\p_t(a_2)+\px(a_2) -a_{10}, \ \
a_1a_3+a_1a_2+\p_t(a_1)+\px(a_2) -a_{01};
$$

for $c_{231}$ we have
$$
a_2a_3+a_1a_2+\py(a_3)+\py(a_1) -a_{10}, \ \
a_1a_3+a_1a_2+\p_t(a_1)+\py(a_1) -a_{01}.
$$

Direct calculation gives us three simplest general invariants of
the initial operator $C$:
$$l_{21}=a_{2,x}-a_{1,y}, \ l_{32}=a_{3,y}-a_{2,t}, \ l_{31}=a_{3,x}-a_{1,t}.$$

\paragraph{Proposition 4.6} General invariants $l_{21}, \ l_{32}, \ l_{31}$
are all equal to zero {\bf iff} operator $C$ is equivalent to an
operator
 \be \label{Prop}L=\pa_1\pa_2\pa_3+b_1\pa_1+b_2\pa_2+c,\ee
i.e. $\exists \ \mbox{function } \ f: \ \   f^{-1}C \circ f =L.$\\

 $\blacktriangleright$ Obviously
 $$
f^{-1}(\px \py \p_t )\circ f= (\px +(\log f)_x)(\py +(\log
f)_y)(\p_t +(\log f)_t)
 $$
for any smooth function $f$. Notice that it is the form of an
operator $C_{ijk}$ and introduce a function $f$ such that

$$
a_1=(\log f)_x, \ \ a_2= (\log f)_y, \ \ a_3= (\log f)_t.
$$

This system of equations on $f$ is over-determined
 and it has  solution $f_0$ {\bf iff}
 $$a_{2,x}-a_{1,y}=0, \ a_{3,y}-a_{2,t}=0,
\ a_{3,x}-a_{1,t}=0,$$ i.e. $l_{21}= l_{32}= l_{31}=0.$ \qed

$\blacktriangleleft$ Indeed, if $C$ is equivalent to (\ref{Prop}),
then $a_{20}=a_{02}=a_{11}=0$ and obviously
$l_{21}=l_{32}=l_{31}=0.$ \qed \\

At the end of this section let us notice that the reasoning above
can be carried out for the hyperbolic operator of order $n$ with
constant leading coefficients and analog of  Proposition 4.6 keeps
true with $n$ order terms as
$$
\p_1  \p_2 \p_3.... \p_n $$ with $$ \p_1=\px, \ \p_2=\py, \
\p_3=\px+\py, \ \p_4=\a_4 \px+ \b_4 \py, \ ... \ \p_n=\a_n \px+
\b_n \py
$$
and terms of order $n$ and $n-1$ can be written as
$$
C_{i_1,i_2,...,i_n}=(\p_1+a_1)(\p_2+a_2)...(\p_n+a_n)
$$

while corresponding linear general invariants will take form

$$
l_{ij}=\p_i(a_j)-\p_j(a_i).
$$

\subsection{Examples}
\subsubsection{Operator of order 2}
Let us regard first for simplicity  LPDO of the second order with
constant leading coefficients, i.e.
$$
a_{ij}=\const \quad \forall (i+j)=2
$$
and all roots of characteristic polynomial are distinct. Then
obviously any root $\o$ also does not depend on $x,y$ and
expressions for $p_{ij}$ can be simplified substantially. Let us
introduce notations
$$
a_{00}\o^0=\mathcal{P}_0(\o),
$$
$$
a_{10}\o^1+a_{01}=\mathcal{P}_1(\o),
$$
$$
a_{20}\o^2+a_{11}\o+a_{02}=\mathcal{P}_2(\o)
$$
and notice that now $\o$ and $ \mathcal{P}' _2(\o) \neq 0$ are
constants.\\

 Using formulae from Section 2.1 we get

\begin{eqnarray}\label{invar2}
p_1=1\nonumber\\
p_2=-\o \nonumber\\
p_3 =\frac{\mathcal{P}_1(\o)} {\mathcal{P}' _2(\o)}\nonumber\\
 p_4=a_{20}\nonumber\\
 p_5= a_{20} \o +a_{11}\nonumber\\
 p_6 =\frac{ (a_{20}\o+a_{11})a_{10}-a_{20}a_{01}
}{\mathcal{P}' _2(\o)}\nonumber
\end{eqnarray}

which yields to
$$
\mathcal{L}(p_6)=\frac{a_{20}\o+a_{11}}{\mathcal{P}' _2(\o)}
\mathcal{L}(a_{10})-\frac{a_{20}}{\mathcal{P}'
_2(\o)}\mathcal{L}(a_{01})
$$
and invariant $l_2$ takes form

\be \label{p7} l_2=
-\mathcal{L}(p_6)-p_3p_6+a_{00}=\frac{a_{20}\o+a_{11}}{\mathcal{P}'
_2(\o)} \mathcal{L}(a_{10})-\frac{a_{20}}{\mathcal{P}'
_2(\o)}\mathcal{L}(a_{01})+$$$$+\frac{\mathcal{P}_1(\o)}
{\mathcal{P}' _2(\o)}\frac{ (a_{20}\o+a_{11})a_{10}-a_{20}a_{01}
}{\mathcal{P}' _2(\o)}-a_{00}. \ee \\

\begin{itemize}
\item{} Let us regard hyperbolic operator in the form
\begin{equation}\label{ex1}
\partial _{xx} - \partial_{y y} + a_{10} \partial_{ x} +
a_{01} \partial_{ y} + a_{00},
\end{equation}
 i.e. $a_{20}=1, a_{11}=0, a_{02}=-1$ and $\o=\pm 1, \ \mathcal{L}=\px-\o \py.$
Then $l_2$ takes form
$$
l_2=a_{00}-\mathcal{L}(\frac{\o a_{10}-a_{01}}{2\o})-\frac{\o
a_{10}-a_{01}}{2\o} \frac{\o a_{10}+a_{01}}{2\o}
$$
which yields, for instance for the root $\o=1$, to
$$
l_2=a_{00}-\mathcal{L}(\frac{a_{10}-a_{01}}{2})-\frac{
a_{10}^2-a_{01}^2}{4}=a_{00}-(\px-
\py)(\frac{a_{10}-a_{01}}{2})-\frac{ a_{10}^2-a_{01}^2}{4}
$$
and after obvious change of variables in (\ref{ex1}) we get
finally first Laplace invariant $\hat{a}$
$$
l_2= c-\p_{\tilde{x}}a -ab = \hat{a},
$$
where
$$
a=\frac{a_{10}-a_{01}}{2}, \ \ b=\frac{a_{10}+a_{01}}{2}, \ \
c=a_{00}.
$$
Choice of the second root, $\o=-1$, gives us the second Laplace
invariant $\hat{b}$, i.e. Laplace invariants are particular cases
of the general invariant so that each Laplace invariant
corresponds to a special choice of $\o$.

\item{} Let us proceed analogously with an elliptic operator
\begin{equation}\label{ex2}
\partial _{xx} + \partial_{y y} + a_{10} \partial_{ x} +
a_{01} \partial_{ y} + a_{00},
\end{equation}
then $\o=\pm i, \ \mathcal{L}= \px-\o \py$ and
$$
l_2=a_{00}+(\px \mp i \py)(\frac{\pm a_{10}+a_{01}i}{2})+i\frac{
a_{10}^2+a_{01}^2}{4}
$$
where choice of upper signs corresponds to the choice of the root
$\o=i$ and choice of lower signs corresponds to $\o=-i$.
\end{itemize}

\subsubsection{Operator of order 3}

 Now let us regard  LPDO of the third
order with constant leading coefficients, i.e.
$$
a_{ij}=\const \quad \forall (i+j)=3
$$
with at least one  root distinct of characteristic polynomial
$$
a_{30}\o^3+a_{21}\o^2+a_{12}\o+a_{03}=\mathcal{P}_3(\o)
$$
and notice that now $\o$ and $ \mathcal{P}' _3(\o) \neq 0$ are
constants.\\

 Using formulae from Section 2.2 we get
\begin{eqnarray}
p_1=1\nonumber\\
p_2=-\o \nonumber\\
p_3 =
\frac{\mathcal{P}_2(\o)}{\mathcal{P}'_3(\o)}\nonumber\\
p_4=a_{30}\nonumber\\
p_5=a_{30} \o+a_{21}\nonumber\\
p_6=a_{30}\o^2+a_{21}\o+a_{12}\nonumber\\
 p_7=
\frac{a_{20}}{\mathcal{P}'_3(\o)}-\frac{a_{30}}
{\mathcal{P}'_3(\o)}\frac{\mathcal{P}_2(\o)}{\mathcal{P}'_3(\o)}\nonumber\\
 p_8= \frac{\o
a_{20}+a_{11}}{\mathcal{P}'_3(\o)}-\frac{\o a_{30}
+a_{21}}{\mathcal{P}'_3(\o)}\frac{\mathcal{P}_2(\o)}{\mathcal{P}'_3(\o)}\nonumber\\
p_9=a_{10}
-\frac{\mathcal{L}(a_{20})}{\mathcal{P}'_3(\o)}+\frac{a_{30}}
{\mathcal{P}'_3(\o)}\frac{\mathcal{L}(\mathcal{P}_2(\o))}{\mathcal{P}'_3(\o)}-
\frac{\mathcal{P}_2(\o)}{\mathcal{P}'_3(\o)}(
\frac{a_{20}}{\mathcal{P}'_3(\o)}-\frac{a_{30}}
{\mathcal{P}'_3(\o)}\frac{\mathcal{P}_2(\o)}{\mathcal{P}'_3(\o)})\nonumber
\end{eqnarray}
and $l_3, l_{31}$ are
$$-(p_1\px+p_2\py+p_3)p_8-p_2p_9+ a_{01} =l_3,$$
$$-(p_1\px+p_2\py+p_3)p_9+ a_{00}=l_{31}.$$
The formulae are still complicated and in order to show the use of
them let us regard here one simple example of an operator

\be\label{ex3} B=\p_x^2 \py + \px \py^2 +a_{11}\px\py
+a_{10}\px+a_{01}\py+a_{00}, \ee

with $a_{30}=a_{03}=a_{20}=a_{02}=0,\ \ a_{21}=a_{12}=1$. Then its
invariant
$$l_{3}=\px a_{11}-a_{01}$$
and semi-invariant $$l_{31}=\px a_{10}-a_{00}$$ have very simple
forms and gives us immediately a lot of information about the
properties of operators of the form (\ref{ex3}), for instance,
these operators are factorizable, i.e. has zero invariants
$l_3=l_{31}=0$, {\bf iff}
$$a_{11}=\int a_{01}dx + f_1(y), \ \ a_{10}= \int a_{00}dx +
f_2(y)
$$ with two arbitrary functions on $y$, $f_1(y)$ and $f_2(y)$.
Another interesting fact is that if coefficient $a_{11}=a_{11}(y)$
is function of one variable $y$, then $a_{00}$ is general
invariant and there definitely should be some nice geometrical
interpretation here, etc.

\section{Summary}

At the end of this Chapter we would like to notice following very
interesting fact - beginning with operator of order 4, maximal
number of general invariants is bigger then number of coefficients
of a given operator, $$\frac{(n+1)(n+2)}{2} < n! \ \ \forall
n>3.$$ It means that general invariants are dependent on each
other and it will be a challenging task to extract the subset of
independent general invariants,
i.e. basis in the finite space of general invariants.\\

 As to semi-invariants,
notice that an operator of arbitrary order $n$ can always be
rewritten in the form of factorization with reminder of the form
$$
l_n\px^k+l_{n,1}\px^{k-1}+...+l_{n,k-1}, \ \ k<n
$$
and exact expressions for all $l_i$ are provided by
BK-factorization procedure. The same reasoning as above will show
immediately that $l_n$ is always  general invariant, and each
$l_{n,k-i_0}$ is $i_0$-th semi-invariant, i.e. it becomes
invariant in case if $l_{n,k-i}=0, \ \forall i<i_0$.\\

In this paper, explicit formulae for $l_2, \ l_3, \ l_{31}$ are
given. Formulae for higher order operators can be obtained by pure
algebraic procedure described in \cite{bk2005} but they are too
tedious to be derived by hand, i.e. programm package for
symbolical computations is needed.\\

Already in the case of three variables, the factorization problem
of a corresponding operator and also constructing of its
invariants becomes more complicated, even for constant
coefficients. The reason of it is that in bivariate case we needed
just to factorize leading term polynomial which is always possible
over $\Complex$. It is not the case for more then 2 independent
variables where a counter-example is easily to find (see Ex.5),
i.e. there exist some non-trivial conditions to be found for
factorization of polynomials in more then two variables.

\section{Exercises for Chapter 6}

\paragraph{1.} Using this particular solution $u=-\frac{1}{(x+y)^2}$ of
the Liouville equation
$$(\log u)_{xy}+2u=0$$ find general solution of this equation.

{\bf Hint:} Liouville equation is invariant under the following
change of variables: $$\hat{x}=X(x),\quad \hat{y}=Y(y), \quad
\hat{u}(\hat{x},\hat{y})=\frac{u(x,y)}{X'(x)Y'(y)}.$$

\paragraph{2.} Prove that matrix (\ref{tildeAN}) is degenerate.

\paragraph{3.} Let Laplace invariants are equal, i.e.
$\hat{a}=\hat{b}$. Prove that initial operator $$ \p_x \p_y +
a\p_x + b\p_y + c$$ is equivalent to the operator $$ \p_x \p_y  +
c.$$

\paragraph{4.} Let in Lemma 3.8 function $w(x,y)$ is chosen as
$$w(x,y)= X_1(x)Y_1(y)+ X_2(x)Y_2(y).$$ Prove that $d_3=0.$

\paragraph{5.} Check that
$$
x^3+y^3+z^3-3xyz=(x+y+z)(x^2+y^2+z^2-xy-xz-zy)
$$
and prove that $ x^3+y^3+z^3$ is not divisible by a linear
polynomial $$\a x+ \b y+ \g z + \d$$ for any complex coefficients
$\a,\ \b,\ \g, \ \d$.

\chapter{Commutativity of linear operators}

\section{Introduction}

 In this Chapter, notion of commutativity
for two linear operators will be studied which leads us to
construction of LAX pairs consisting of a given linear operator
and a new one commuting with a given operator.
 This technique is
used afterwards for construction of a commuting operator for a
{\bf linearized}  operator  with coefficients {\bf depending on
the solution of nonlinear equation}. Existence of LAX pair for
nonlinear differential equation is closely connected with
symmetries of this equation and can be regarded as a definition of
its integrability. Connections of this new definition of
integrability with definitions given in Chapter 1 will be
demonstrated as well as some examples.

\section{Operators of one variable}
\subsection{Commutativity of formal series}

Commutativity of differential polynomials is a classical problem,
highly non-trivial because polynomials constitute a ring. On the
other hand, formal series constitute a field and their
commutativity problem can be solved easily and necessary and
sufficient conditions can be written out explicitly. Therefore,
our first step here is to study commutativity properties for
Lorant series which demands to generalize for the case $n \in
\Integer$ formulae (introduced at the beginning of Chapter 1) for
differential operator

\be  \label{Da}D^n \cdot a=\sum_{k} \left( \ba{c}n\\k \ea \right)
D^k(a)D^{n-k} \ee

and binomial coefficients

\be \label{binom1}\left( \ba{c}n\\k \ea
\right)=\frac{n(n-1)...(n-k+1)}{1\cdot 2 \cdots  k}
\quad \mbox{with} \quad \left( \ba{c}n\\0 \ea \right)=1 .\ee \\

A pseudo-differential operator (PSDO) is introduced \cite{shub} as
a following formal series  \be\label{dser}
 A=\sum_{k=-\infty}^n a_kD^k=
 a_nD^n+ \dots +a_0+a_{-1}D^{-1}+a_{-2}D^{-2}+\dots\ .
\ee while the  product of PSDOs  defined as consequent monomial
multiplication using (\ref{Da}) and  standard formula
  $D^jD^k=D^{j+k}$  where $j$ and $k$ are not necessary positive
  integers, with
 the same formula (\ref{binom1}) for the superposition with
 $n\in\Integer.$ \\

 For example, using for simplicity notation $$D(a)=a_x, \ D^2(a)=a_{xx}, \ ....,$$
we can compute

$$ D^{-1}\circ a=\sum_{k=0}^\infty
 \left(\ba{c} -1\\ k\ea\right)D^k(a)D^{-1-k}= a D^{-1}-a_x D^{-2}+a_{xx}
 D^{-3}+\dots \, ;$$

 $$ D^{-2}\circ a=\sum_{k=0}^\infty
 \left(\ba{c} -2\\ k\ea\right)D^k(a)D^{-2-k}= a D^{-2}-2a_x D^{-3}+3a_{xx}
 D^{-4}+\dots \, .$$\\

It is important to understand clearly that notation $D^k(a)$ means
action of operator $D^k$ on $a$ in case $k \ge 0$ and it is just a
formal description in case $k < 0.$\\

It follows from definition that a product of two PSDOs of order
$m$ and $n$ has order $m+n$ and its leading term is product of two
leading terms correspondingly. Associativity of so defined
multiplication
 can be derived directly using the properties of binomial
 coefficients.\\

\paragraph{Definition 4.1} An  element $A$ given by (\ref{dser}) of associative algebra $\cal{F}{D}$,
$A\in\cal{F}{D}$, is called {\bf pseudo differential operator}
(PSDO) of order $n$ with leading term $a_nD^n$.

 Notation  $A_+$ is used
for Taylor part of $A$, \be\label{a+} A_+=\sum_{k=0}^n a_kD^k=
 a_nD^n+ \dots +a_0
\ee and notation $A_-$ is used for Lorant part of $A$,
\be\label{a-} A_-=\sum_{k=-\infty}^{-1} a_kD^k=
a_{-1}D^{-1}+a_{-2}D^{-2}+\dots. \ee

That produces presentation of algebra $\cal{F}{D}$ as  direct sum
of subalgebras \be\label{dirsum}
\cal{F}{D}=\cal{F}_+(D)+\cal{F}_-{D} \ee

Obviously,  in case when a PSDO under consideration coincide with
its Taylor part $A_+$, i.e. a formal series is in fact a finite
sum,  multiplication defined for PSDOs gives the same result as
multiplication of usual differential operators.\\

\paragraph{Lemma 4.2} Let $A$ be a PSDO
(\ref{dser}) of order $n$ with 1 as leading coefficient. Then
following keeps true:\\

1. for arbitrary $n \in \Integer$ there exists  unique formal
series $L$ such
that $AL=LA=1$ ;\\

2.   for positive $n>0$ there exists
 unique  formal series $M$ of the first order, with leading coefficient 1
 and such that $M^n=A$ and $M^nA=AM^n$.\\

$\blacktriangleleft$ Starting with
 \be \label{series}
  B=D+b_0+b_{-1}\circ D^{-1}+b_{-2}\circ D^{-2}+\dots
 \ee
one finds by induction that
\[ B^n=D^n+nb_0D^{n-1}+b_{1,n}D^{n-2}+b_{2,n}D^{n-3}+\dots   \]
where the new coefficients
     \[ b_{j,n}= nb_{-j}+f[b_0,\, b_{-1},\dots,b_{-j+1}]. \]
Hereafter $f[\dots]$ means that the function $f$ is a differential
polynomial in its arguments. For example
\[ b_{1,n}=nb_{-1}+f[b_0]=nb_{-1} + \frac{n(n-1)}{2}(b_{0,x}+b_0^2).   \]
The system of equations
\[ a_{n-1}=nb_0,\quad a^{-1}=b_{1,n},\quad  a_{n-3}=b_{2,n}, \dots \]
has the triangular form and therefore uniquely solvable.\qed

\paragraph{Definition 4.3} Two PSDOs $L$ and $M$ obtained in Lemma
4.2 are
 called {\bf inverse}  and {\bf n-th root} PSDO correspondingly.
 They are denoted as $A^{-1}$ and $A^{1/n}$.

\paragraph{Example 4.4} As an illustration let us compute powers $A^2$ and $A^3$ of
 the first order PSDO of the general form
 \be\label{10}
 A=D+a_0+D^{-1}a_1+D^{-2}a_2+D^{-3}a_3+\dots \ee
We find
$$ A^2=D^2+2a_0D+2a_1+a_0^{2}+ a_{0,x}+a_{21}D^{-1}+a_{22}D^{-2}+\dots,
$$
$$ A^3=D^3+3a_0D^2+(3a_1+3a_0^{2}+3a_{0,x})D+a_{30}+a_{31}D^{-1}+
+\dots $$ where
$$     a_{21}=2a_2+a_{1,x}+2a_0a_1, $$
$$     a_{30}=3a_2+3a_{1,x}+ 6a_0a_1+a_0^3+3a_0a_{0,x}+a_{0,xxx}.$$

Formulae become longer with each step but in principle the
procedure goes smoothly and can be continued up to any power.
Representing $n$-th power of original PSDO $A$ as $A^n=A^n_+
+A^n_-,$ we can, using  formulae above, compute Taylor´s part for
$n=1, 2$:

$$ A^1_+=D+a_0,\quad A^2_+=D^2+2a_0D+2a_1+a_0^{2}+ a_{0,x}+2a_{1}.
$$\\

For instance,   the first three coefficients of the series
$A=\sqrt{L}$ for Schrödinger operator $L=D^2-u$ (introduced in
Chapter 2) have form

$$2a_0=0, \quad 2a_1=-u, \quad a_{21}=2a_2-\frac12 u_x=0.$$

\paragraph{Proposition 4.5} Let $A$ be a PSDO of the of the form
(\ref{dser})  with 1 as leading coefficient. Then arbitrary PSDO
$B$ commutes with a given $A$ {\bf iff} $B$ can be represented as
formal series

\be \label{ser2} B=\sum \b_k A^{k/n},  \ee

where all $\b_k$ are constants, $\b_k \in \Complex$.\\

\paragraph{Definition 4.6} Pseudo-differential operator   $C=AB-BA$ constructed
from  two PSDOs $A$ and $B$ is called their {\bf commutator} and
it is denoted as $C=[A,B].$

\paragraph{Proposition 4.7} Let $A$ is a PSDO of order $n$ and
$B$ is a PSDO of order $m$, then their commutator has order $k <
m+n$,
$$
[A,B]=\sum_{k<m+n} c_kD^k,
$$
and its leading term has form
$$
c_{n+m-1}=na_nb_{m,x}-mb_ma_{n,x}.
$$

In particular, by $a_n=1,\ n\neq0$ and $[A,B]=0$ this formula
implies that the series $B$ has constant leading coefficient
$b_{m,x}=0.$

 \subsection{Commutativity of differential operators}

Ordinary differential operators

$$ A=a_nD^n+ a_{n-1}D^{n-1}+\dots+a_1D +a_0 $$

can obviously be regarded as a particular case of PSDOs and the
operation of
 multiplication defined in previous section,
can be understood in usual sense, as composition of ordinary
differential operators. Also result of Prop. 4.7 keep true, i.e.
 if $A$ and $B$ are differential operators of orders $m$ and $n$
respectively, then their commutator $C$ is a differential operator
of order $\le m+n-1$.\\

On the other hand, in comparison with PSDO, differential operators
have some specific properties which we are going to present below,
beginning with following definition.

\paragraph{Definition 4.8}  {\bf Centralizer} $C(A)$ of an ordinary  differential operator
is a set
 differential operators $\{B\}$ which commute  with $A$, i.e.
 $[A,B]=AB-BA=0$.
Centralizer $C(A)$ is called {\bf trivial} if it contains only
polynomials in $A$ with constant coefficients.\\

In order to construct classes of equivalence on the set
differential operators, let us introduce two transformations of
differential operators as follows:

\begin{itemize}
\item{} change of independent variable $x\to \hat{x}$

\be\label{xx} a\frac{d}{dx}=\frac{d}{d\hat{x}} \ \LRA \
\frac{d\hat{x}}{dx}=a(x) \ee

\item{} similarity transformation

\be\label{sim} A\mapsto \hat{A}=f^{-1}\circ A \circ f\ee

where $a=a(x)$ and $f=f(x)$ are appropriate chosen smooth
functions.
\end{itemize}

It is easy to see that transformations (\ref{xx}) and (\ref{sim})
allows bring the differential operator $A$ of order $n>0$ to the
{\it canonical form} \be\label{tido} \ti{A}=\hat{D}^n+
\ti{a}_{n-2}\hat{D}^{n-2}+\dots +\ti{a}_0 \ee in which the first
two coefficients are fixed by $1$ and $0$, respectively.
Particularly, the change of variable (\ref{xx}) with $a=a_n^{1/n}$
leads to the operator with unitary leading coefficient. An example
of the differential operator in the canonical form is given by
Schrödinger operator $L=D^2-u$. Any differential operator of
second order is equivalent Schrödinger operator $L=D^2-u$  up to
transformations (\ref{xx}) and (\ref{sim}) as has been indicated
above.

\paragraph{Example 4.9} Let us describe centralizer for first
order differential operator  $A=a(x)D_x+b(x)$. After change of a
variable (\ref{xx}) it takes form
$$
\hat{A}=\hat{D}+\hat{b}, \  \hat{D}=\frac{d}{d\hat{x}} $$ and the
second transformation  another change of variables
$$\ti A=\alpha^{-1}A\alpha, \quad \mbox{with} \quad
D\log\alpha=\ti b$$  gives finally
$$
A=D
$$
where all "tildes" and "hats" over $A$ are omitted for simplicity
of notations. It can be shown that both transformation (\ref{xx})
and (\ref{sim}) does not change structure of centralizer and
Proposition 4.5 gives its description as
$$
\{D^0, \ D, D^2, D^3,...\ \}.
$$
Thus, centralizer constructed in this example, is trivial.\\

\paragraph{ Theorem 4.10 } Let   $C(A)$ is centralizer of
a differential operator $A$ of order $n>0$, then any two elements
of the centralizer, $B_1, B_2 \in C(A)$ commute with each other:
$$[B_1,B_2]=0.$$

{\it Proof.}  As in examples above, an appropriate change of
independent variable produces constant leading coefficient of
 $A$ and without loss of generality let us put 1 as
 leading coefficient. Then statement of the theorem is direct
 corollary of Proposition 4.5. \qed \\

It follows from Theorem 4.10 that
$$ [A_1, A_2]=0, \ [A_2, A_3]=0\RA A_1, A_3 \in C(A_2)\RA [A_1, A_3]=0. $$
Therefore the binary relation $[A,B]=0$ is an equivalence and one
can replace the operator $A$ in $C(A)$ by any nontrivial (of non
zero order) element lying in centralizer. The minimal order $n>0$
of nontrivial elements of $C(A)$ gives some indication what is a
structure of the centralizer. Thus, for $n=1$ a centralizer is
always trivial as it was demonstrated in Example 4.9. But in the
case with $n=2$ it is not true. In order to construct full
description of nontrivial centralizers in the case $n=2$ we need
 to study first a  problem about commutative pairs of operators
of the second and the third orders. In this case we need just
compute the commutator $C=[A,B]$ and write down the condition when
it vanish.

\paragraph{Example 4.11 }Let $A=D^2-u$ and $B=D^3+aD+b.$  Then
$$  [A,\, B]=a_2D^2+a_1D+a_0,$$ $$
a_2=2a_x+3u_x,\ \ a_1=a_{xx}+3u_{xx}+2b_x,\ \
a_0=u_{xxx}+au_x+b_{xx}.$$ Excluding coefficients of the third
order operator one gets equation for the potential $u$
 \be\label{kort}
u_{xxx}-6uu_x=\eps u_x\ee where $\eps\in \Complex$ is  a constant
of integration. Any solution of Eq. (\ref{kort}) gives rise a
commuting pair of operators. Particularly, in the case
$$
A=D^2+x^{-2}\eps,\quad B=D^3+x^{-2}\alpha_1 D+x^{-3}\alpha_2
$$
the condition of commutativity $[A,B]=0$ reduces to linear algebra
as follows
$$3\eps=2\alpha_1,\quad\alpha_1+\alpha_2=0,\quad
12\eps+\alpha_1\eps+6\alpha=0\RA \eps=-2,\, \alpha_1=-3,\,
\alpha_2=3.
$$\\

Thus, we have now the operator $A=D^2-2x^{-2}$ of the second order
such that $C(A)$ contains the third order operator $B.$ Since all
polynomials in $A$ gives operators of {\bf even} orders,  the
third order operator $B$ can not be written as polynomial in $A$
and thus we have at last example of nontrivial centralizer $C(A)$.
 One can verify directly that $A^3=B^2$ in this case (Cf. Prop.
 4.5).\\

This way it was shown that any nontrivial centralizer $C(A)$ of
Schrödinger operator $A=D^2-u$ contains some odd order operator
$B.$ In the opposite case, when all orders of operators in
centralizer are even one can easily prove that any operator $B\in
C(A)$ of order $2n$ could be represented in the polynomial form
$$ B=c_nA^n+c_{n-1}A^{n-1}+\dots+c_0, \ \ c_k\in\Complex.$$
Indeed, the leading coefficient should be constant due to
Proposition 4.7. Denote this coefficient by $c_n$ and consider the
operator $\ti B=B-c_nA^n\in  C(A). $ By condition its order is
even $2\ti n,\  \ti n<n$ again and we can proceed further in order
to diminish it.

Analogous reasoning in the case of nontrivial centralizer $C(A)$
of Schrödinger operator leads to following algebraic formula for
any operator $B_2\in C(A):$ \be
\label{bba}B_2^2=P_2(A)B_1+Q_2(A)\ee where $B_1\in C(A)$ has {\bf
minimal} odd order and $P_2(A),$ $Q_2(A)$ are polynomials in $A$
with constant coefficients. Particularly, apply this formula to
$B_1$ we get
$$ B_1^2=P_1(A)B_1+Q_1(A),\ B=B_1-\frac12 P_1(A)\RA B^2=Q(A).$$
It is easy to see that new operator $B\in C(A)$ has the same
minimal odd order $m =2n+1$ as $B_1$ and can be used as generator
as well as $B_1$ in the general formula (\ref{bba}) and, hence
degree of polynomial $Q(A)$ in the last equation is the same $m
=2n+1.$ In the next Section we describe how to reconstruct
centralizer using this polynomial $Q(A)$  which defines in certain
sense multiplication in the commutative ring $C(A)$ with two
generators $A$ and $B$ related by the equation \be\label{ba}
 B^2=Q(A). \ee

For this construction we need to introduce a very important notion
of common eigenfunctions for two commuting differential operators
$L$ and $M$ of orders $n$ and $m$, respectively. Let us notice
first that if

$$
 L\psi=\la\psi, \ \ \mbox{and} \ \ \quad [L,M]=0,
$$

then for some function $\hat\psi$ such that $\hat\psi=M\psi$ we
have

$$L\hat\psi=\la\hat\psi.$$

It means that for any fixed $\la$ the $n-$dimensional subspace
$\ker
 (L-\la)$ constituted by solutions of equation $L\psi=\la\psi$ for eigenfunctions of the operator
 $L,$ is invariant under the action of operator $M.$

 \paragraph{Definition 4.12.} Let $L$ and $M$ are commuting differential operators and
  $\ti M$ is the restriction of the operator $M$ on an invariant subspace $\ker (L-\la).$
  The eigenvector $\ti{M} \psi=\mu\psi$ is called {\bf common eigenfunction} of two differential
  operators $L$ and $M$.
  \vspace{3mm}

  It follows from Definition 4.12 that this common eigenfunction $\psi$ solves two
  equations:
 \be\label{mula} L\psi=\la\psi,\quad
M\psi=\mu(\la)\psi.\ee

 In  particular, for Schrödinger operator $L=A=D^2-u$ and $M=B$ given by (\ref{ba}),
the last formula above implies $\mu(\la)=\sqrt{Q(\la)}$ and
therefore Eq. (\ref{mula}) takes the form
$$
\psi_{xx} = (\lambda+u)\psi,\quad B\psi=\sqrt{Q(\la)}\psi.
$$
It could be shown that this over determined system of two equation
for one function $\psi$ (common eigenfunction) can be reduced to
the first order linear ODE (due the fact that greatest common
divisor of orders of operators $L$ and $M$ is equal 1.)

\paragraph{Example 4.13} The differential operators from Ex. 4.11 are related by the equation
 $B^2=A^3$ and equations (\ref{mula}) for common eigenfunction give
$$
\psi_{xx}=(k^2+\frac{2}{x^2})\psi,\quad
\psi_{xxx}=\frac{3}{x^2}\psi_x+(k^3-\frac{3}{x^3})\psi,\quad\la=k^2.
$$
After differentiation with respect to $x$ the first equation and
comparison it with the second one, we obtain the first order ODE:
$$x\frac{\psi_x}{\psi}=\frac{x^3k^3+1}{x^2k^2-1}=kx+\frac{1}{kx-1}.
$$

\paragraph{Remark.}

In many cases it is convenient to rewrite the scalar differential
equation $L\psi=\la\psi$ for eigenfunctions as the systems of
first order equations for the vector function $\vec\psi:$
\be\label{can1} \vec\psi_x=V(x,\la)\vec\psi  \ee
 For the general differential
operator $L$ of order $n$ it is $n-$th order system with
$\vec\psi=(\psi^1,\dots,\psi^n)^\tau.$ For instance, for
Schrödinger operator we have
$$ \psi^1_x=\psi^2,\quad \psi^2_x=(\la+u)\psi^1 $$
and therefore the $2\times 2$ matrix $V$ for Schrödinger operator
is
\be\label{sch1}V=\left(\ba{cc} 0& 1\\
                  \la+u& 0\ea\right).\ee

It is possible to proceed with the same reasoning as above using
matrix representation of operators and generalize the results
obtained for Schrödinger operator on the operators of higher order
but this task lies beyond the scope of this text.

\section{Operators of two variables}
Above we regarded operators of one variable that implies a
differentiation $D_x$ with respect to single independent variable
$x$. Now we try to insert additional differentiations $D_{t_k},$
compatible with $D_x$, and to introduce a corresponding set of
independent variables $t_k, \ k=1,\ 2,\ 3,\ \dots\ .$ First,
 we regard this problem algebraically and consider action of
these additional
 differentiations $D_{t_k},$ upon elements of the ring of formal series
 in powers of $D$. It immediately produces a nonlinear system of PDEs for
  coefficients of the formal series. In particular these PDEs include in itself famous
 Kadomtzev-Petviashvili equation (KP):
\be\label{kp}
 u_{tx}=(u_{xxx}-6uu_{x})_x+ 3 u_{yy}.\ee

Further we consider a broad class of evolutionary equations
\be\label{Evol} u_t=f(x, u, u_x, u_{xx},u_{xxx}\dots)\ee with one
space variable $x.$ We  show how  to build up constructively a Lax
pair for equations (\ref{Evol}) possessing infinite number of
higher symmetries. It is remarkable that in this case basic
theorem is formulated again in terms of formal series in powers of
$D$ and highlights notions of algebraic theory of the first
sections from a new point of view.

\subsection{KP-model}

In order to introduce an action additional differentiations
$D_{t_k}$ acting upon elements of the set of PSDO (\ref{dser}) it
is sufficient to define it action on the coefficients of formal
series $A=\sum a_kD^k.$ We postulate the formula as follows
\be\label{dtk} D_{t_k}(A)= a_{n,t_k}D^n+ \dots
+a_{0,t_k}+a_{-1,t_k}D^{-1}+\dots \defeq [B_k, A] \ee where the
formal series $B_k, \ k=1,\ 2,\ 3,\ \dots\ $ will be defined
exactly below and
 $n\in\Integer$ denotes an order of $A.$

 Firstly, we remark that the basic differentiation $D$ in the set of formal series
  could be written down in this form and corresponds to the choice $B_1\defeq D.$
Namely, use the definition of multiplication of the formal series
$A=\sum a_kD^k$ and $D$ we have \be\label{d1} [D, A]=DA-AD=\sum
D\circ a_kD^k- a_kD^{k+1}=\sum  a_{k,x}D^k \ee Thus, we verified
Eq.(\ref{dtk}) for $k=1$ with $B_1=D$ and $ D_{t_1}\equiv D_x.$

 Secondly, we remind that the Leibnitz rule (general feature of any differentiation)
 is satisfied for the operation of commutation with a formal series $B$ since
 $$ [B, A_1A_2]=BA_1A_2-A_1BA_2+A_1BA_2-A_1A_2B=[B, A_1]A_2+A_1[B,A_2].$$
 Thus, left and hand sides of Eq.(\ref{dtk}) obey Leibnitz rule and we wish now to compare
 orders of the two formal series in the formula under discussion. Due Proposition 3.7 the order of commutator
 $[B,A]$ is equal $n+m-1$ if the formal series $B$ of order $m$ is chosen independently
 from the series $A$ of order $n.$ On the other hand the order of $D_{t_k}(A)$ has to be less or equal
of the order $n$ of $A.$ In the case $m=1$ all agreed since
$n+m-1=n$ and that opportunity used in
 Eq.(\ref{d1}). For $m>1$ in order to equalize the orders we must choose at least leading coefficients
 of $B_k$ in the strong accordance with leading coefficients of series $A.$ It is the source of
 {\bf nonlinearity} imprinted in Eq.(\ref{dtk}). Take in account Proposition 3.5 we formulate below very
  nice algebraic solution of this problem of equalizing orders in left and right sides of Eq.(\ref{dtk}).
 Some times next proposition is considered as a model of modern integrability theory.

\paragraph{Lemma 4.1. (Sato ??)} Let formal series $B$ of the first order is normalized as
follows \be\label{10} B=D+b_1D^{-1}+b_2D^{-2}+b_3D^{-3}+\dots \ee
 and the differentiations (\ref{dtk}) of the coefficients $b_k$ of this series defined by formulae
 \be\label{kpk}
D_k\defeq  D_{t_k}(B)= b_{1,t_k}D^{-1}+ b_{2,t_k}D^{-2}+\dots=
[B^k_+, B]
 \ee

Then the differentiation $D_1=D_x$ and $D_nD_m(B)=D_mD_n(B)$ for
any $n, \ m\ge 1,$ and moreover \be\label{dmn}
 D_n(B^m_+)-D_m(B^n_+)=[B^n_+,B^m_+], \quad  D_n(B^m_-)-D_m(B^m_-)=[B^m_+,B^n_+].  \ee

 $\blacktriangleleft$ Obviously $B_+=D$ and, therefore, $D_1(B)=B_x.$ In order to prove (\ref{dmn}) we
 start with the identity
 $$
 0=[B^{n+1},B^{m+1}]=[B^n_+,B^m_+]+[B^n_+,B^m_-]-[B^m_+,B^n_-]+[B^n_-,B^m_-].
 $$
 On the other hand
 $$[ D_m(B^n_+)=D_m(B^{n+1}-B^n_-)=[B^m_+,B^n_-]+[B^m_+,B^n_+]-D_m(B^n_-),$$
 $$ D_n(B^m_+)=D_n(B^{m+1}-B^m_-)=[B^n_+,B^m_-]+[B^n_+,B^m_+]-D_n(B^m_-).$$
and (\ref{dmn}) follows from it. It is easy to see now the
commutativity conditions $D_nD_m(B)=D_mD_n(B)$ are corollary of
Eq. (\ref{dmn}). \qed

\paragraph{Remark 4.2} In this chapter we do not consider additional differentiations $D_{t_k}$
for formal series of general form. Nevertheless, the equations
Eq.(\ref{kpk}) and Leibnitz rule imply that for powers $B^n$ of
the normalized series (\ref{10}) we have
 \be\label{kpkn}
 D_{t_k}(B^n)=[B^k_+, B^n].
 \ee
This generalization Eq.(\ref{kpkn}) and Proposition 3.4 allows to
include in domain of
 differentiations $D_{t_k}$ formal series (\ref{dser}) of any order $n$ but yet normalized
 (i.e. leading and next one coefficients equal $1$ and $0,$ respectively). In order to
 get off these restrictions one should to develop algebraic variant of transformations (\ref{xx} and
 (\ref{sim} used in previous Section yet it is not easy at all. We are going to discuss this
  normalization problem in next section from quite different point of view.

 The {\bf nonlinear} action on elements of $\cal{F}{D}$ of
 additional differentiations $D_{t_k},$ compatible with $D_x$ defined by Lemma 4.1
is not unique. In particular, one can start with the formal series
\be\label{11} A=D+a_{0}+a_{1}D^{-1}+a_{2}D^{-2}+a_{3}D^{-3}+\dots
\ee and use another way (Cf. (\ref{a+}), (\ref{a-})) to separate
polynomial and Lorant parts of $A^n:$
$$
 A^n=\ti{A^n_+}+\ti{A^n_-},\quad \mbox{where} \ \ti{A^n_+}\defeq D^n+a_{n,1}D^{n-1}+a_{n,2}D^{n-2}+\dots
+a_{n,n-1}D$$ and $ n=1, \ 2,\, 3,\dots\,  .$ The proof of next
proposition repeats the proof of Lemma 4.1 and will be omitted.

\paragraph{Lemma 4.3} Let formal series $A$ of the first order defined by  (\ref{11})
 and the differentiations (\ref{dtk}) of the coefficients $a_k$ of this series defined by formulae
 \be\label{mkpk}
\ti{D_k}\defeq  D_{t_k}(A)=a_{0,t_k}+ a_{1,t_k}D^{-1}+
a_{2,t_k}D^{-2}+\dots= [\ti{A^k_+}, A].
 \ee
Then $\ti{D_1}=D_x$ and $\ti{D_n}\ti{D_m}(A)=\ti{D_m}\ti{D_n}(A)$
for any $n, \ m\ge 1.$ \vspace{5mm}

One could notice that as well as Lemma 4.1 corresponds to
representation $\cal{F}{D}$ as the direct sum of the polynomial
and not polynomial parts (see Definition 3.1)  Lemma 4.3
corresponds to another way do define this direct sum (Cf.
(\ref{dirsum})).

We are going now to consider the action of differentiations
$D_{t_k},$ on elements $\cal{F}{D}$ in more details and write down
an explicitly some nonlinear equations for coefficients of the
formal series (\ref{11}) and (\ref{10}). In order to stress that
we take next:

\paragraph{Definition 4.3 } We call by {\bf mKP-hierarchy} and {\bf KP-hierarchy} the infinite
 system of PDEs for the coefficients $a_k$ and $b_k$ of the formal series(\ref{11}) and (\ref{10})
 defined by Eq.(\ref{mkpk}) and Eq.(\ref{kpk}), respectively.\\

 Below we provide basic, in certain sense, examples of equations of mKP and KP hierarchies.
It will be equations for $a_0$ and $b_1$ , respectively, with
derivatives on the first three independent
 variables $t_1, t_2, t_3.$ In the case of mKP-hierarchy it called modified Kadomtzev-Petviashvili equation
 (mKP  shortly) and is similar to KP-equation (\ref{kp}) mentioned in the introduction.

\paragraph{Example 4.5 } Historically, for KP-equation (\ref{kp}):
$$ u_{xt}=(u_{xxx}-6uu_x)_x+3u_{yy} $$
the auxiliary linear problem arised in the form as follows
 \be\label{kppsi} \psi_y=\psi_{xx}-u\psi,\quad
\eps\psi_t=\psi_{xxx}+a\psi_x+b\psi\ee In order to find the
compatibility conditions which gives KP-equation we have to equate
cross derivatives $(\psi_y)_t=(\psi_t)_y.$ Denote $\psi_k=D^k\psi$
we find after $x-$differentiation Eqs.(\ref{kppsi})
$$\eps(\psi_y)_t=\psi_5+(a\psi_1+b\psi)_{xx}-\eps u_t\psi-u(\psi_3+a\psi_1+b\psi),$$
$$\eps(\psi_t)_y=\psi_5-(u\psi)_{xxx}+a_y\psi_1+a(\psi_3-u\psi_1-u_x\psi)+b_y\psi+b(\psi_2-u\psi).$$
Equate now coefficients by $\psi_k$ in above formulae one gets for
$k=2,\ 1,\ 0$, respectively
$$ 2a_x+3u_x=0,\quad 2b_x=a_y-3u_{xx}-a_{xx}=a_y+a_{xx},$$
$$ b_y+\eps u_t=b_{xx}+u_{xxx}+au_x.$$
Rewrite the last equation in terms of $u$ one obtains KP.

In terms of Definition 4.3 it corresponds to differentiations
(\ref{dtk}) with $k\le3.$ We have for $D_{t_2}, D_{t_3}$:
$$\begin{cases}
\mbox{mKP}: \ \ \  \ti{A^2_+}=D^2+2a_{0}D,\quad \ti{A^3_+}=D^3+3a_{0}D^2+3(a_{1}+a_{0,x}+a_{0}^2)D. \\
\mbox{KP}: \ \ \  \ B^2_+=D^2+2b_1,\quad
B^3_+=D^3+3b_1D+3(b_{1,x}+b_2)
\end{cases}
$$
Therefore it coincides in KP-case with (\ref{kppsi}) up to
notations. Moreover, Lemma 4.1 equations
$$ D_2(B)=[B^2_+,B],\quad D_3(B)=[B^3_+,B]$$
yield
$$ b_{1,y}=b_{1,xx}+2b_{2,y},\quad  b_{2,y}=b_{2,xx}+2b_{3,x}+2b_{1,x}b_1.$$
$$ b_{1,t}=b_{1,xxx}+3b_{2,xx}+3b_{3,x}+b_1b_{1,x}$$
and, thus, exclude $b_2$ and $b_3$ we find:
$$ [b_{1,t}-\frac14b_{1,xxx}-3b_1b_{1,x}]_x=\frac34 b_{1,yy}$$
Compare these two view points on KP one could say that these are
equivalent and give the same nonlinear PDE.

Analogously, using Lemma 4.2 equations one gets mKP \be\label{mkp}
4q_{tx}=D_x(q_{xxx}-2q_x^3)+3q_{yy}+6q_{xx}q_y,\quad a_{0}=q_x.
\ee \vspace{5mm}

In algebraic approach to integrability the notion of commutativity
appears preferable
 in comparison with the notion of symmetry. Thus it seems awkward to reformulate Lemmas 4.1-2 in terms of
 symmetries of the infinite system of PDEs constituting hierarchies. Yet consider particular
 equations of hierarchies like  (\ref{kp}) and (\ref{mkp}) the notion of symmetry becomes more
 important and useful. In the next example we demonstrate reductions Eq.(\ref{kp}) related with stationary
 points of very simple symmetries transformations.

\paragraph{Example 4.6 } Since the equations contain independent variables only inexplicitly
they are keep its form invariant under shift In the case $b_1$ not
depends in $y$ KP equation reduce and gives rise KdV:
$$ b_{1,y}=0\RA  $$
Bussinesque??

\subsection{Conservation laws}
 Interplay of symmetries and conservation laws for ODEs has been discussed in Introduction.
This interconnection for PDEs is quite different but very
interesting as well and we present here a terse introduction to
pure algebraic aspects of the theory oriented on applications in
next chapters. Hereafter in this section we restrict ourself by
the case of KP-hierarchy and will denote by $\cal{B}$ the set of
polynomial functions on coefficients $b_j$ of the formal series
(\ref{10})  and its derivatives with respect to $t_i, \ i=1, 2,
\dots$.

\paragraph{Definition 4.7} Corollary of equations of KP-hierarchy written in the form of divergence i.e.
\be\label{divkp} \sum_k D_{t_k}(\beta_k)=0\ee is called {\bf
conservation law} of KP-hierarchy if sum in (\ref{divkp}) is
finite and the
 the variables $\beta_k\in\cal{B}.$

In order to build up conservation laws of KP-hierarchy one can use
lemma below which based on next
 definition.

\paragraph{Definition 4.8.} For the formal series $A\in \cal{F}{D}$ in powers of $D$
$$ A=\sum_{k=-\infty}^n a_kD^k=
 a_nD^n+ \dots +a_0+a_{-1}D^{-1}+a_{-2}D^{-2}+\dots\ .  $$
the coefficient $a_{-1}$ is called {\bf residue} and is denoted as $\res(A)$.\\

\paragraph{Lemma 4.9} For all $ m,\, n\in \Integer$
$$ \res\left([aD^m,bD^n]\right)= D_x\a_{m,n} $$
where $\a_{m,n}$ is a polynomial function on $a$, $b$ and it's
$x-$derivatives.

{\it Proof.} The residue vanish if the powers $m,\, n$ obey the
condition $mn\ge 0.$ For instance in the case $n=0$ the commutator
$aD^mb-baD^m$ is a differential operator if $m\ge 0$ and PSDO of
order $m-1\le -2$ if $m<0$ (see Prop. 3.7). Obviously the
coefficient by $D^{-1}$ is zero in both cases. Obviously as well
that the residue vanish if $m+n<0.$

 Let now $m,\, n$ have different signs and $m+n=k\ge 0.$ Then in virtue of (\ref{Da})
$$
\res\left([aD^m,bD^n]\right)= \left( \ba{c}m\\k+1 \ea \right)(
aD^{k+1}(b)+(-1)^{k}D^{k+1}(a)b)  $$ since
$$ m+n=k\RA m(m-1)\cdots(m-k)=\pm n(n-1)\cdots(n-k).  $$
Standard "integration by parts" proves now the lemma. Particularly
for $k=0, \ 1$ we have, respectively
$$ aD(b)+D(a)b=D(ab)\quad aD^2(b)-D^2(a)b=D(aD(b)-D(a)b).$$
\qed\\

Use (\ref{kpkn}) we find
$$ D_{t_k}(\res B^n)=\res D_{t_k} B^n= \res [B^k_+, B^n] $$
and Lemma 4.9 yields

\paragraph{Corollary 4.10.} KP-hierarchy possess infinite sequence of
conservation laws of the special form as follows \be\label{rhon}
 D_{t_k}(\rho_n)=D_x(\sigma_{k,n}), \quad \rho_n\defeq\res(B^n), \ \sigma_{k,n}\in \cal{B}\ee
where $\cal{B}$ denotes the set of polynomial functions on
coefficients $b_j$ of the formal series (\ref{10}).
\paragraph{Example 4.11.}
$$
\rho_1=b_1,\quad \rho_2=2b_2-b_{1,x},\quad
\rho_2=2b_2-b_{1,x},\quad \rho_3=3b_3+3b_1^2-b_{2,x}-b_{1,xx}
$$
$$D_{t_k}(\rho_1)=D_{x}(\rho_{n+1})$$
\paragraph{Theorem 4.12}(\cite{Wils}). Let in the case (\ref{10})
 $$ A=D+\sum_{i=1}^\infty b_iA^{-i},\quad A^m=(A^m)_+
 +\sum_{k=1}^\infty  b_{km} A^{-k}. $$
 Then for all $i,\, n=1,\, 2,\dots$
 \be\label{dbb}  D_n b_i=D b_{in} ,\qquad (D\equiv D_1).      \ee

 {\em Proof.} Lemma 4.1 yields
  $$
 D_n(A_1)=D(A_n)+[A_1,A_n],\quad  A_1=\sum_{i=1}^\infty b_iA^{-i},\quad
 A_m=\sum_{k=1}^\infty  b_{km} A^{-k}.  $$
 We have now
  $$
 D_n(A_1)=
 \sum_{i=1}^\infty D_n(b_i)A^{-i}+ \sum_{i=1}^\infty b_i [A^{-i}, A_n]
 \quad
 D(A_n)=
 \sum_{k=1}^\infty D(b_{kn})A^{-k}+ \sum_{k=1}^\infty b_{kn} [A^{-k}, A_1]
 $$
 and
 $$ [A_1,A_n]+\sum_{k=1}^\infty b_{kn} [A^{-k}, A_1]-
  \sum_{i=1}^\infty b_i [A^{-i}, A_n]=0.$$
  \qed
\vspace{3mm}

 In order to highlight the notion of conservation laws
let us consider $T-$periodic in $x$ solutions of KP-hierarchy.
Then
$$ D_{t_k}(\rho)=D_x(\sigma_{k}) \RA  D_{t_k}\int_{x_0}^{x_0+T}\rho dx=\int_{x_0}^{x_0+T}\sigma_{k,x}
dx=0.
$$
 Therefore, the integral $R(\vec{t})\defeq \int_{x_0}^{x_0+T}\rho
dx$ conserves it's value as times $t_k$ change and the integral
$R=\int_{x_0}^{x_0+T}\rho dx$ rather than density $\rho$ has
"physical sense." That is the reason to call two conservation laws
in the differential form $ D_{t_k}(\rho_j)=D_x(\sigma_{j,k}),\
j=1,2$ with distinct densities {\bf equivalent}
$\rho_1\equiv\rho_2$ if \be\label{equiv}\int_{x_0}^{x_0+T}\rho_1
dx=\int_{x_0}^{x_0+T}\rho_2 dx. \ee

In general not necessary periodic case, we take
 \paragraph{Definition 4.11}
 Two conservation laws of KP-hierarchy $
D_{t_k}(\rho_j)=D_x(\sigma_{j,k}),\ j=1,2$ to be said  {\bf
equivalent} iff $\rho_1-\rho_2\in \cal{B}_0$ where latter is
subspace of $\cal{B}$ constituted by $\im D.$
 \vspace{3mm}

Obviously, equivalence relation introduced in Def.4.11 implies the
equality (\ref{equiv}). Moreover, as we will see in the next
section this relation $\rho\equiv 0$ could be verified in fully
algorithmic way.

\section{Integrability problem}
In this section we discuss a general definition of {\it
integrability} for evolutionary PDEs (\ref{Evol}) with one space
variable $x$ using the theory of formal series in powers of
$D=D_x$ (sections 2 and 6). Follow traditions of differential
algebra \cite{ritt} we introduce the notation
$$ u\mapsto u_1 \mapsto u_2\mapsto \dots\mapsto u_k\mapsto\dots  $$
for the infinite sequence of derivatives with respect to $x$ of
the basic variable $u.$ \footnote{In fore cited book it called \it
differential indeterminate.} In these notations the general
evolutionary PDE is rewritten as follows \be\label{fgen} u_t=f(x,
u, u_1, u_2,\dots ,u_n). \ee The highest derivatives $u_n$ defines
the {\it order} $n$ of Eq. (\ref{fgen}) and we consider usually
equations of order $n\ge 2.$ The dependence on $x$ of the right
hand side in Eq.(\ref{fgen}) became very important in the next
Chapter in relation with godograph type transformations.

\subsection{ Basic definitions.}
Consider the set $\cal{U}$ of smooth functions of variables $x, u,
u_1, u_2,\dots $ we will use the equation Eq. (\ref{fgen}) in
order do define action of the "additional" differentiation $D_t$
on $\cal{U}$ (the first one is $D=D_x$). Namely we take as
definitions of these differentiations the formulae \bea\label{dxg}
D_x: \ \ \ & g(x, u, u_1, u_2,\dots ,u_m)\mapsto & g_x+g_*(u_1),\quad u_1=D(u), \\
\label{dtg} D_t: \ \ \ & g(x, u, u_1, u_2,\dots ,u_m)\mapsto &
g_*(f),\quad f=D_t(u) \eea where $g_x=\pa_x(g)$ is partial
derivative of the function $g$ and $g_*$ denotes differential
operator as follows
$$ g_*\defeq \sum_k \frac{\pa g}{\pa u_k} D^k=
\frac{\pa g}{\pa u}+\frac{\pa g}{\pa u_1}D+\dots+\frac{\pa g}{\pa
u_m}D^m.
$$
It is easy to see that operators (\ref{dxg}),(\ref{dtg}) act on
the "generators" of the set $\cal{U}$ in a natural way, as partial
differentiations should do:
 $$ D_x(x)=1, \ \ D_x(u)=u_1,\quad D_t(x)=0, \ \ D_t(u)=f(x, u, u_1, u_2,\dots ,u_n) $$
 and
 $$ D_xD_t(u)=D_x(f), \ \ D_tD_x(u)=D_t(u_1)=D_x(f)\RA D_xD_t(u)=D_tD_x(u).$$

It is important to notice that action of $m-$th order differential
operator $g_*$ on functions $v$ can be defined also as
\be\label{genlin}
  g_*(v)={dg[u+\eps v]\over d\eps}\vert_{\eps=0}.
\ee In particular as corollary of Eq. (\ref{genlin}) we find that
$$ (fg)_*=fg_*+gf_*. $$
 One can prove now that $D_x,$ $D_t$ and $\Delta=D_xD_t-D_tD_x$ as well, obey
 Leibnitz rule and that $\Delta=0$ on the set $\cal{U}.$

\paragraph{Definition 4.13.} The operators (\ref{dxg}) and (\ref{dtg}) on the set $\cal{U}$ are
called {\bf total differentiation with respect to $x$} and {\bf
evolutionary differentiation related with $f\in \cal{U}$},
respectively. The Gato derivative (\ref{genlin}) of a function
 $g\in \cal{U}$ we call {\bf linearization} of the function $g.$

Summing up we have

\paragraph{Proposition 4.14.} For any smooth function $f\in \cal{U}$ corresponding evolutionary
differentiation (\ref{dtg}) commutate with (\ref{dxg}).
\vspace{5mm}

This statement looks trivial but if, as in previous Section, we
try to find other evolutionary differentiations \be\label{dtfk}
  D_{t_k}(u)=f_k(x, u, u_1, u_2,\dots ,u_{n_k})
 \ee
which commutate each other we shall come to nontrivial obstacles.
Situation here in certain sense corresponds to correlation of the
Proposition 3.5 which is very general and abstract and more
interesting problem about commutating differential operators
considered in Section 4. The role of Proposition 3.5 plays now
Lemma 4.1 and commutativity conditions $[D_{t_j}, D_{t_k}]=0$ have
to play the role of commuting differential operators. This link
became more apparent and instructive if we bring up commutativity
conditions to the level of a tangent space of $\cal{U}$ using
(\ref{genlin}). Namely instead of writing down the cross
derivatives equality $D_{t_1}D_{t_2}(u)=D_{t_2}D_{t_1}(u)$ for the
two evolutionary equations \be\label{f12} D_{t_1}(u)=f_1(x, u,
u_1, u_2,\dots ,u_{n_1}),\quad D_{t_2}(u)=f_2(x, u, u_1, u_2,\dots
,u_{n_2})\ee we will consider analogous condition
$D_{t_1}D_{t_2}(v)=D_{t_2}D_{t_1}(v)$ for linearized ones
 $D_{t_k}(v)=(f_{k})_*(v), \ k=1,2.$ Translate this condition into condition of commutativity of
 corresponding operators
\be\label{d12} [D_{t_1}-(f_1)_*, D_{t_2}-(f_2)_*]=0 \ee
 we come to next
 \footnote{It can be proved that cross derivatives equation $D_{t_1}D_{t_2}(u)=D_{t_2}D_{t_1}(u)$
 implies (\ref{d12}).}

\paragraph{Definition 4.15.} Let us say that the two evolutionary equations (\ref{fi2})
 satisfy the {\bf consistency condition} and, respectively, {\bf compatibility condition}
 if the cross derivatives equation $D_{t_1}D_{t_2}(u)=D_{t_2}D_{t_1}(u)$ is fulfilled or, and in
 the second case, if differential operators equality (\ref{d12}) is valid i.e.
\be\label{crossd}
  D_{t_1}((f_2)_*)-D_{t_2}((f_1)_*)=[(f_1)_*, \ (f_2)_*]
 \ee
We remind (Cf. (\ref{dtk})) that the action of any evolutionary
differentiations $D_{t_k}$ on polynomials $A=\sum a_jD^j, \ a_j\in
\cal{U}$ is defined by $D_{t_k}(A)=\sum D_{t_k}(a_j)D^j. $
\vspace{5mm}

In order to compare two almost equivalent Definitions 4.15 we give
simple example.

\paragraph{Example 4.16.} Let us consider the pair evolutionary equations as follows
$$
D_{t_1}(u)=f(x, u, u_1, u_2,\dots ,u_{n}),\quad D_{t_2}(u)= u_1.
$$
Then, apply (\ref{dxg}) and (\ref{dtg}) and Proposition 4.14 we
get
$$ D_{t_1}D_{t_2}(u)= D_{t_1}(u_1)=D_x(f)= f_x+f_*(u_1), \ \
D_{t_2}D_{t_1}(u)= D_{t_2}(f)=f_*(u_1). $$ Thus the consistency
condition is satisfied iff  $f_x=0$ and $f=f(u, u_1, u_2,\dots
,u_n).$ On the other hand the compatibility condition
(\ref{crossd}) yields
$$ f_{u,x} =f_{u_1,x}=\dots=f_{u_n,x}=0\LRA f=f(u, u_1, u_2,\dots ,u_n)+X(x) $$
where $X(x)$ is arbitrary function of the independent variable
$x.$

\subsection{Main theorem}

We wish now to address more practical question. Let the
evolutionary equation (\ref{fgen}) of the order $n$ is given i.e.
$D_t(u)=f.$ What conditions on its right hand side $f\in \cal{U}$
or, more exactly, on the corresponding differential operator
$f_*:$ \be\label{f*} f_*= \frac{\pa f}{\pa u}+\frac{\pa f}{\pa
u_1}D+\dots+\frac{\pa f}{\pa u_n}D^n. \ee are needed in order to
build up a hierarchy of evolutionary differentiations
$D_{t_k}(u)=f_k$ compatible with fore given. Take in account that
in the compatibility condition \be\label{ffk}
D_{t}((f_k)_*)-D_{t_k}((f)_*)=[f_*, \ (f_k)_*] \ee the order $n_k$
of the differential operator $(f_k)_*$ could be arbitrary large
whilst the order of $D_{t_k}((f)_*)$ is fixed by $n$ we see that
leading terms in $D_{t}((f_k)_*)$ and in the commutator in the
right hand side should coincides. It gives rise the definition
which reproduces the main formula (\ref{dtk}) of previous section
in a new environment

\paragraph{Definition 4.16} We say that evolutionary equation
$$ u_t=f(x, u, u_1, u_2,\dots ,u_n)  $$
of order $n$ posses {\bf formal symmetry} if there is the formal
series of order $m\ne 0$
$$
 A=a_mD^m+a_{m-1}D^{m-1}+\dots+a_0+a_{-1}D^{-1}+a_{-2}D^{-2}+\dots
$$
with coefficients $a_j\in \cal{U}$ such that \be\label{afa}
D_t(A)=[f_*,A].\ee
 \vspace{3mm}

 Apply this definition to Burgers and Korteweg de Vries equations which we often use as
illustrative examples, we find
\paragraph{Example 4.17} For Burgers equation
$$ u_t=u_2+2uu_1=f,\quad f_*=D^2+2uD+2u_1$$
the formal symmetry is PSDO as follows $ A=D+u+u_1D^{-1}.$ Indeed
$$ [f_*,A]=f+D(f)D^{-1}\quad\mbox{and}\quad A=D+u+u_1D^{-1}\RA D_t(A)=u_t+u_{1,t}D^{-1} $$
and thus $D_t(A)=[f_*,A]$.

Similarly, in the case of Korteweg de Vries equation (KdV)
$$  u_t=6uu_1-u_3=f,\qquad  f_*=-D^3+6uD+6u_1$$
the second order PSDO $A=D^2-4u-2u_1D^{-1}$ is the formal symmetry
since
$$  [f_*, A]=-4f-2D(f)D^{-1}, \ \ \mbox{and}\quad A=D^2-4u-2u_1D^{-1}\RA D_t(A)=-4u_t-2u_{1,t}D^{-1} $$
and thus $D_t(A)=[f_*,A]$. \vspace{3mm}

Next statement can be considered as milestone of integrability
theory of evolutionary equations with one space variable.

\paragraph{Theorem 4.18} If there are series of evolutionary differentiations $D_{t_k}(u)=f_k$ of orders
$n_k\to \infty$ by $k\to \infty$ satisfying compatibility
conditions (\ref{ffk}) for a given evolutionary
 equation $D_t(u)=f$ of order $n\ge 2.$  Then this equation $D_t(u)=f$ possess the
formal symmetry $A$ of the first order: \be\label{fsym}
A=a_1D+a_0+a_{-1}D^{-1}+\dots    \ee \vspace{5mm}

 Construction of the solution $A$ of the basic equation $D_t(A)=[f_*,A]$ needs some technicalities
 related with the limiting procedure in Eq.(\ref{ffk}) when $k\to \infty.$ We skip that formal part of
 proof of Theorem 4.18 (see Appendix??) and go instead directly to applications.

Firstly, we explain how to get formal symmetry of the first order
starting with any solution of Eq.(\ref{afa}) and describe it's
general solution. Obviously,
$$ \a_1 A_1 +\a_2 A_2, \ \ ,\a_j\in\Complex \quad\mbox{and}\quad A_1A_2      $$
satisfy Eq.(\ref{afa}) if $A_1$ and $A_2$ do. Apply results of
Section 3 we get.

\paragraph{Lemma 4.19} Let formal series in $D$ of order $m$ i-e- $B=\sum b_jD^j, \ b_j\in \cal{U}$  is
solution of Eq.(\ref{afa}) and let $m\ne 0,$ $n\ge2.$ Then there
exist the first order formal symmetry $A$ (see (\ref{fsym})) with
the leading coefficient \be\label{a1} a_1=(f_n)^{\frac{1}{n}},
\quad f_n\defeq \frac{\pa f}{\pa u_n}   \ee and any solution
$\ti{B}$  of Eq.(\ref{afa}) of order $\ti m$ can be represented as
formal series in powers of $A$ with constant coefficients
$\a_l\in\Complex$:
$$
 \ti{B}=\sum_{k\le\ti m } \a_k A^k $$
{\it Proof.} Slight generalization of Lemma 3.2 yields the formal
series $B_1$ with coefficients from
 $\cal{U}$ such that $B_1^m=B.$ Let us consider the formal series $R=D_t(B_1)-[f_*,B_1].$ Since
$$ D_t(B)= D_t(B_1)B_1^{m-1}+B_1D_t(B_1)B_1^{m-2}+\dots+B_1^{m-1}D_t(B_1),$$
and
\[  [f_*,B]=[f_*,A]A^{m-1}+A[f_*,A]A^{m-2}+\dots+A^{m-1}[f_*,A]  \]
we have
\[   B_t-[f_*,B]=RA^{m-1}+ARA^{m-2}+\dots+A^{m-1}R=0. \]
In this sum the all $m$ series in $D^{-1}$ have the same leading
term and, hence, this leading term is zero. It is possible only if
$R=0$ and therefore we proved that $B_1$ is a solution.

In order to find the formula for leading coefficients of solutions
Eq.(\ref{afa}) one should find the leading term in $[f_*,B]-B_t.$
Since $n\ge 2$ it is contained in
$$ [f_*,B]=c_{n+m-1}D^{n+m-1}+\dots,\quad c_{n+m-1}=nf_nD(b_m)-mb_mD(f_n) $$
where $f_n$ and $b_m$ are leading coefficients in $f_*$ and $B$
(Cf  Proposition 3.7 ). Vanishing leading coefficient gives the
equation
\[ nD\log b_m=mD\log f_n \LRA n\log b_m-m\log f_n=\const  \]
which implies for $m=1$ the formula (\ref{a1}) for the leading
coefficient of the first order formal symmetry.\qed

Most known applications of Theorem 4.18 related with next
proposition quite similar to Corollary 4.10:
\paragraph{Corollary 4.20.} In conditions of Theorem 4.18 the evolutionary equation (\ref{fgen}) possess
infinite series of conservation laws as follows: \be\label{ares}
D_{t}(\res A^n)=\res D_t(A^n)= \res [f_*, A^n] \in\im D, \ \
n=-1,1,2,\dots . \ee

Hereafter $\im D$ denotes the subset of functions $g\in\cal{U}$
which can be rewritten as $g=D_x(h), h\in\cal{U}.$ In other words
Corollary 4.20 provides series of conservation laws
$$
D_t(\rho_n)= D_x(\sigma_n),\quad \rho_n, \sigma_n\in\cal{U}, \ \
\rho_n\defeq\res A^n. $$ of the equation (\ref{fgen})) under
cosideration.
\paragraph{Remark 4.20'.} Series conservation laws (\ref{ares})) should be completed
by the conservation law with the dencity
$\rho_0=\frac{a_{m-1}}{a_m}.$

Namely, rewrite equation (\ref{afa}) for $m-$th order formal
symmetry
$$
 A=a_mD^m+a_{m-1}D^{m-1}+\dots+a_0+a_{-1}D^{-1}+a_{-2}D^{-2}+\dots
$$
as follows
$$
D_t(A)A^{-1}=f_*-Af_*A^{-1}=[A^{-1}, Af_*]
$$
we find  $\res (A_tA^{-1}) \in\im D$ due Lemma 4.9 and that
$$
\res(A_tA^{-1})=\\
\res\left[(a_{m,t}D^m+a_{m-1,t}D^{m-1})(a_m^{-1}D^{-m}+\alpha
D^{-m-1})\right]=$$
$$
(a_{m,t}(\alpha-m\frac{a_{m,x}}{a_m^2})+\frac{a_{m-1,t}}{a_m}.
$$
It remains to notice that
$$
\alpha=m\frac{a_{m,x}}{a_m^2}-\frac{a_{m-1}}{a_m}.
$$

\paragraph{Example 4.21.} For $n=-1$ we have due (\ref{a1})
$$
\res
A^{-1}=\res\left(\frac{1}{a_1}D^{-1}+\dots\right)=(f_n)^{-\frac{1}{n}}.$$
Thus $\rho_{-1}=\left(\frac{\pa f}{\pa u_n}\right)^{-\frac{1}{n}}$
and by Corollary 4.20 the evolutionary equation (\ref{fgen})
should always satisfy the condition \be\label{a1res}
D_{t}(\rho_{-1})=D_t\left(\frac{\pa f}{\pa
u_n}\right)^{-\frac{1}{n}}\in\im D \ee if it is compatible with
hierarchy of evolutionary differentiations.

\paragraph{ Definition 4.22.} The conservation law (i.e. differential corollary of (\ref{fgen}))
 \be\label{rhosigma} D_t(\rho)= D_x(\sigma),\quad  \rho, \sigma\in\cal{U}            \ee
 is called trivial if $\rho\in \im D$ and there is $ h\in\cal{U}$ such that
 $$ \rho=D_x(h) \RA D_t (\rho)=D_t(D_x(h))=D_x(D_t(h))\in\im D   $$
 for any evolutionary differentiation $D_t.$ The functions $\rho$ and $\sigma$ in Eq.(\ref{rhosigma})
are called density and flux, respectively, and two conservation
laws with the densities
 $\rho_1$ and $\rho_2$ are considered equivalent if $\rho_1-\rho_2\in\im D.$
 \vspace{3mm}

Come back to Example 4.21 we see now there is an alternative or
$(\rho_{-1})\in\im D$ or $D_{t}(\rho_{-1})\in\im D.$ In both cases
this condition is non trivial one and can be verified using
classical statement by (Helmgolz??) proven in P.Olver book
\cite{Olver}.

\paragraph{Theorem 4.23.} The function $ h\in\cal{U}$ belongs to the subspace  $\im D$
iff $ \hat E(h)=0$ where {\bf Euler operator} $\hat E$ is defined
as follows \be\label{euler} \hat{E}\defeq {\pa\over \pa
u}-D{\pa\over \pa  u_1}+ D^2{\pa\over \pa u_2}-D^3{\pa\over \pa
u_3}+\dots \ .\ee \vspace{3mm}

Equations similar to Eq.(\ref{afa}) play major role in modern
integrability theory and are called commonly by {\bf Lax pair}.
The first test which have to justify that that is not trivial one
\footnote{There is known a few cases of faked Lax pairs.} could be
formulated similarly to the definition $S-$integrability below. In
other words proper Lax pair must produce series of nontrivial
conservation laws.

\paragraph{ Definition 4.24.} The evolutionary equation (\ref{fgen})) with formal symmetry (\ref{afa}))
is called $S-$integrable if the series (\ref{ares})) contains
nontrivial conservation laws of arbitrary higher order and
$C-$integrable in the opposite case. \vspace{3mm}

The famous  Korteweg de Vries (see (\ref{kdv}) in Section?) gives
us an example of $S-$integrable equation. As matter of fact the
formal symmetry for KdV indicated in Example 4.17 related very
closely with Lemma 4.6 ( see Eq.(\ref{***}) and concluding remark
in Section?). KdV hierarchy will be discussed in next section.

Although Definition 4.24 looks too complicated to be really useful
it is not the case. In the next chapter we apply this definition
to the second order evolutionary equations (\ref{fgen})) and
verify that first three of conditions (\ref{ares}) fully describe
{\bf all} second order evolutionary equations possessing hierarchy
of higher order symmetries. Burgers equation from Example 4.17 is,
in certain sense, typical representative in this class and a class
of $C-$integrable equations in general.
\subsection*{Symmetries and cosymmetries}
Concluding this chapter we discuss duality between symmetries and
conservation laws starting with "zero" dimensional evolutionary
equations i.e. dynamical systems
  \be\label{dynn}
  \frac{d\vec u}{dt}=\vec{f}(\vec{u}),\quad
  \vec{u}=(u^1,u^2,\dots,u^n)^T
  \ee
  where $()^T$ denotes hereafter matrix transposition. In this
  case the conservation law with the density $\rho=\rho(\vec{u})$ is just a
  constant of motion
  $$
\frac{d\rho}{dt}=\frac{\pa\rho}{\pa
u^1}f^1+\dots+\frac{\pa\rho}{\pa u^n}f^n =0.
$$
Denote by $\hat\rho$ the column-vector constituted by the partial
derivatives $\frac{\pa\rho}{\pa u^k}$ we notice that
$$
\hat{\rho}_t+f_*^T\hat\rho=0.
$$
\section{Summary}
The primary goal of this chapter is to establish a workable
criterion that can be readily checked to determine whether a given
evolutionary equation possess integrability features like
symmetries, local conservation laws and Lax pair.
 \vspace{5mm}

\vspace{5mm}

\section{Exercises for Chapter 4}

\paragraph{1.} Prove the set of DO of the form
$L=D^n+\sum\eps_kx^{-k}D^{n-k}$ closed under multiplication,
Define the transformation which kill the coefficient $\eps_1$ in
$L.$

\paragraph{2.} Find the fist three coefficients in the series $A=(D-f)^{-1}.$
Prove that all coefficients are "bona defined" i.e. are
differential polynomials in f.
 {\it Hint:} Use the formula $D-f=\ph D\ph^{-1},\, f=D\log\ph.$
\paragraph{3.} Verify that
\[ L=D^2-u\RA  \left(L^{\frac32}\right)_+=D^3-\frac34(uD+Du).  \]
\paragraph{4.} Prove that the change of independent variables
yields
\[
dx=g\, dy\LRA D_y=gD_x \quad\mbox{and}\quad\hat
U=g^2U+\{D_y,g(y)\} \] where a " Schwarzian derivative" $\{D,a\}$
is defined as follows \be\label{34} \{D_x,a(x)\}\defeq
\frac34\frac{a_x^2}{a^2}-\frac12\frac{a_{xx}}{a}. \ee \vspace{2mm}

One can verify readily that Liouville transformation is invertible
and
$$
\hat U=g^2 (U-\{D_x,h(x)\})\LRA U=h^2(\hat U-\{D_y,g(y)\}),\quad
g(y)h(x)=1.
$$

{\textbf{5.}} For evolution equation
\[ u_t=u_n+F(u,u_1,\dots,u_k),\quad k<n,\quad (n\ge2)  \]
possessing a formal symmetry prove that $F_k=\pa_{u_k}F$ is a
density of a local conservation law.
\paragraph{6.} Check out that
\[L=D^2-u,\quad   L_t=[(L^{\frac32})_+,L]\LRA 4u_t+u_{xxx}=6uu_x.         \]
\paragraph{7.} Find non-trivial densities $\rho=\rho(u,u_1)$ for Burgers
equation from Example 2.2. The picture would be incomplete without
next

\paragraph{8.} Prove that condition $ f_*(g)=g_*(f)$ coincides with
cross derivatives equation $D_{t_1}D_{t_2}(u)=D_{t_2}D_{t_1}(u)$
if $D_{t_1}u=f$ and $D_{t_2}u=g.$

\paragraph{9.}
 Prove that the bracket $\{f^1,f^2\}\defeq f_*^1(f^2)-f_*^2(f^1)$
 (Cf Definition 2.8) satisfies the Jacobi identity.
\paragraph{10.}
Find the constant $\eps$ such that $D_{t_2}u=\eps tu_x+1$ is the
symmetry of equations $D_{t_1}u=f$ from Examples 2.2, 2.3 in the
sense that $D_{t_1} D_{t_2}u=D_{t_2}D_{t_1}u.$

\chapter{Integrability of non-linear PDEs}

\section{Introduction}
 In this Chapter we are going to introduce some
definition of integrability for nonlinear differential equations
in partial derivatives.(Cf. Intro. in previous Ch.3) It links
together the notion of higher symmetries of PDEs and the theory of
commutativity from previous chapter. While in Chapter 3 we have
deal with a differentiation $D_x$ with respect to single
independent variable $x$, now we try to insert additional
differentiations $D_{t_k},$ compatible with $D_x$ and to introduce
a corresponding set of independent variables $t_k, \ k=1,\ 2,\ 3,\
\dots\ .$ In the first section we address this problem
algebraically and consider action these additional
 differentiations $D_{t_k},$ upon elements of the ring of formal series
 in powers of $D$ studied in Section 3.2. It immediately produces a nonlinear system of PDEs for
  coefficients of the formal series under consideration. In particular these PDEs include in itself famous
 Kadomtzev-Petviashvili equation (KP):
\be\label{kp}
 u_{tx}=(u_{xxx}-6uu_{x})_x+ 3 u_{yy}.\ee In the next, oriented on PDEs theory,
section we consider a broad class of evolutionary equations
\be\label{Evol} u_t=f(x, u, u_x, u_{xx},u_{xxx}\dots)\ee with one
space variable $x.$ We will show how constructively build up a Lax
pair for equations (\ref{Evol}) possessing infinite number of
higher symmetries. It is remarkable that in this case basic
theorem is formulated again in terms of formal series in powers of
$D$ and highlights notions of algebraic theory of the first
section from the new point of view.

\section{KP-model} In order to introduce an action additional
differentiations $D_{t_k}$ acting upon elements of the set of PSDO
(\ref{dser}) it is sufficient to define it action on the
coefficients of formal series $A=\sum a_kD^k.$ We postulate the
formula as follows \be\label{dtk} D_{t_k}(A)= a_{n,t_k}D^n+ \dots
+a_{0,t_k}+a_{-1,t_k}D^{-1}+\dots \defeq [B_k, A] \ee where the
formal series $B_k, \ k=1,\ 2,\ 3,\ \dots\ $ will be defined
exactly below and
 $n\in\Integer$ denotes an order of $A.$

 Firstly, we remark that the basic differentiation $D$ in the set of formal series
  could be written down in this form and corresponds to the choice $B_1\defeq D.$
Namely, use the definition of multiplication of the formal series
$A=\sum a_kD^k$ and $D$ we have \be\label{d1} [D, A]=DA-AD=\sum
D\circ a_kD^k- a_kD^{k+1}=\sum  a_{k,x}D^k \ee Thus, we verified
Eq.(\ref{dtk}) for $k=1$ with $B_1=D$ and $ D_{t_1}\equiv D_x.$

 Secondly, we remind that the Leibnitz rule (general feature of any differentiation)
 is satisfied for the operation of commutation with a formal series $B$ since
 $$ [B, A_1A_2]=BA_1A_2-A_1BA_2+A_1BA_2-A_1A_2B=[B, A_1]A_2+A_1[B,A_2].$$
 Thus, left and hand sides of Eq.(\ref{dtk}) obey Leibnitz rule and we wish now to compare
 orders of the two formal series in the formula under discussion. Due Proposition 3.7 the order of commutator
 $[B,A]$ is equal $n+m-1$ if the formal series $B$ of order $m$ is chosen independently
 from the series $A$ of order $n.$ On the other hand the order of $D_{t_k}(A)$ has to be less or equal
of the order $n$ of $A.$ In the case $m=1$ all agreed since
$n+m-1=n$ and that opportunity used in
 Eq.(\ref{d1}). For $m>1$ in order to equalize the orders we must choose at least leading coefficients
 of $B_k$ in the strong accordance with leading coefficients of series $A.$ It is the source of
 {\bf nonlinearity} imprinted in Eq.(\ref{dtk}). Take in account Proposition 3.5 we formulate below very
  nice algebraic solution of this problem of equalizing orders in left and right sides of Eq.(\ref{dtk}).
 Some times next proposition is considered as a model of modern integrability theory.

\paragraph{Lemma 4.1. (Sato ??)} Let formal series $B$ of the first order is normalized as
follows \be\label{10} B=D+b_1D^{-1}+b_2D^{-2}+b_3D^{-3}+\dots \ee
 and the differentiations (\ref{dtk}) of the coefficients $b_k$ of this series defined by formulae
 \be\label{kpk}
D_k\defeq  D_{t_k}(B)= b_{1,t_k}D^{-1}+ b_{2,t_k}D^{-2}+\dots=
[B^k_+, B]
 \ee

Then the differentiation $D_1=D_x$ and $D_nD_m(B)=D_mD_n(B)$ for
any $n, \ m\ge 1,$ and moreover \be\label{dmn}
 D_n(B^m_+)-D_m(B^n_+)=[B^n_+,B^m_+], \quad  D_n(B^m_-)-D_m(B^m_-)=[B^m_+,B^n_+].  \ee

 $\blacktriangleleft$ Obviously $B_+=D$ and, therefore, $D_1(B)=B_x.$ In order to prove (\ref{dmn}) we
 start with the identity
 $$
 0=[B^{n+1},B^{m+1}]=[B^n_+,B^m_+]+[B^n_+,B^m_-]-[B^m_+,B^n_-]+[B^n_-,B^m_-].
 $$
 On the other hand
 $$[ D_m(B^n_+)=D_m(B^{n+1}-B^n_-)=[B^m_+,B^n_-]+[B^m_+,B^n_+]-D_m(B^n_-),$$
 $$ D_n(B^m_+)=D_n(B^{m+1}-B^m_-)=[B^n_+,B^m_-]+[B^n_+,B^m_+]-D_n(B^m_-).$$
and (\ref{dmn}) follows from it. It is easy to see now the
commutativity conditions $D_nD_m(B)=D_mD_n(B)$ are corollary of
Eq. (\ref{dmn}). \qed

\paragraph{Remark 4.2} In this chapter we do not consider additional differentiations $D_{t_k}$
for formal series of general form. Nevertheless, the equations
Eq.(\ref{kpk}) and Leibnitz rule imply that for powers $B^n$ of
the normalized series (\ref{10}) we have
 \be\label{kpkn}
 D_{t_k}(B^n)=[B^k_+, B^n].
 \ee
This generalization Eq.(\ref{kpkn}) and Proposition 3.4 allows to
include in domain of
 differentiations $D_{t_k}$ formal series (\ref{dser}) of any order $n$ but yet normalized
 (i.e. leading and next one coefficients equal $1$ and $0,$ respectively). In order to
 get off these restrictions one should to develop algebraic variant of transformations (\ref{xx} and
 (\ref{sim} used in previous Section yet it is not easy at all. We are going to discuss this
  normalization problem in next section from quite different point of view.

 The {\bf nonlinear} action on elements of $\cal{F}{D}$ of
 additional differentiations $D_{t_k},$ compatible with $D_x$ defined by Lemma 4.1
is not unique. In particular, one can start with the formal series
\be\label{11} A=D+a_{0}+a_{1}D^{-1}+a_{2}D^{-2}+a_{3}D^{-3}+\dots
\ee and use another way (Cf. (\ref{a+}), (\ref{a-})) to separate
polynomial and Lorant parts of $A^n:$
$$
 A^n=\ti{A^n_+}+\ti{A^n_-},\quad \mbox{where} \ \ti{A^n_+}\defeq D^n+a_{n,1}D^{n-1}+a_{n,2}D^{n-2}+\dots
+a_{n,n-1}D$$ and $ n=1, \ 2,\, 3,\dots\,  .$ The proof of next
proposition repeats the proof of Lemma 4.1 and will be omitted.

\paragraph{Lemma 4.3} Let formal series $A$ of the first order defined by  (\ref{11})
 and the differentiations (\ref{dtk}) of the coefficients $a_k$ of this series defined by formulae
 \be\label{mkpk}
\ti{D_k}\defeq  D_{t_k}(A)=a_{0,t_k}+ a_{1,t_k}D^{-1}+
a_{2,t_k}D^{-2}+\dots= [\ti{A^k_+}, A].
 \ee
Then $\ti{D_1}=D_x$ and $\ti{D_n}\ti{D_m}(A)=\ti{D_m}\ti{D_n}(A)$
for any $n, \ m\ge 1.$ \vspace{5mm}

One could notice that as well as Lemma 4.1 corresponds to
representation $\cal{F}{D}$ as the direct sum of the polynomial
and not polynomial parts (see Definition 3.1)  Lemma 4.3
corresponds to another way do define this direct sum (Cf.
(\ref{dirsum})).

We are going now to consider the action of differentiations
$D_{t_k},$ on elements $\cal{F}{D}$ in more details and write down
an explicitly some nonlinear equations for coefficients of the
formal series (\ref{11}) and (\ref{10}). In order to stress that
we take next:

\paragraph{Definition 4.3 } We call by {\bf mKP-hierarchy} and {\bf KP-hierarchy} the infinite
 system of PDEs for the coefficients $a_k$ and $b_k$ of the formal series(\ref{11}) and (\ref{10})
 defined by Eq.(\ref{mkpk}) and Eq.(\ref{kpk}), respectively.\\

 Below we provide basic, in certain sense, examples of equations of mKP and KP hierarchies.
It will be equations for $a_0$ and $b_1$ , respectively, with
derivatives on the first three independent
 variables $t_1, t_2, t_3.$ In the case of mKP-hierarchy it called modified Kadomtzev-Petviashvili equation
 (mKP  shortly) and is similar to KP-equation (\ref{kp}) mentioned in the introduction.

\paragraph{Example 4.5 } Historically, for KP-equation (\ref{kp}):
$$ u_{xt}=(u_{xxx}-6uu_x)_x+3u_{yy} $$
the auxiliary linear problem arised in the form as follows
 \be\label{kppsi} \psi_y=\psi_{xx}-u\psi,\quad
\eps\psi_t=\psi_{xxx}+a\psi_x+b\psi\ee In order to find the
compatibility conditions which gives KP-equation we have to equate
cross derivatives $(\psi_y)_t=(\psi_t)_y.$ Denote $\psi_k=D^k\psi$
we find after $x-$differentiation Eqs.(\ref{kppsi})
$$\eps(\psi_y)_t=\psi_5+(a\psi_1+b\psi)_{xx}-\eps u_t\psi-u(\psi_3+a\psi_1+b\psi),$$
$$\eps(\psi_t)_y=\psi_5-(u\psi)_{xxx}+a_y\psi_1+a(\psi_3-u\psi_1-u_x\psi)+b_y\psi+b(\psi_2-u\psi).$$
Equate now coefficients by $\psi_k$ in above formulae one gets for
$k=2,\ 1,\ 0$, respectively
$$ 2a_x+3u_x=0,\quad 2b_x=a_y-3u_{xx}-a_{xx}=a_y+a_{xx},$$
$$ b_y+\eps u_t=b_{xx}+u_{xxx}+au_x.$$
Rewrite the last equation in terms of $u$ one obtains KP.

In terms of Definition 4.3 it corresponds to differentiations
(\ref{dtk}) with $k\le3.$ We have for $D_{t_2}, D_{t_3}$:
$$\begin{cases}
\mbox{mKP}: \ \ \  \ti{A^2_+}=D^2+2a_{0}D,\quad \ti{A^3_+}=D^3+3a_{0}D^2+3(a_{1}+a_{0,x}+a_{0}^2)D. \\
\mbox{KP}: \ \ \  \ B^2_+=D^2+2b_1,\quad
B^3_+=D^3+3b_1D+3(b_{1,x}+b_2)
\end{cases}
$$
Therefore it coincides in KP-case with (\ref{kppsi}) up to
notations. Moreover, Lemma 4.1 equations
$$ D_2(B)=[B^2_+,B],\quad D_3(B)=[B^3_+,B]$$
yield
$$ b_{1,y}=b_{1,xx}+2b_{2,y},\quad  b_{2,y}=b_{2,xx}+2b_{3,x}+2b_{1,x}b_1.$$
$$ b_{1,t}=b_{1,xxx}+3b_{2,xx}+3b_{3,x}+b_1b_{1,x}$$
and, thus, exclude $b_2$ and $b_3$ we find:
$$ [b_{1,t}-\frac14b_{1,xxx}-3b_1b_{1,x}]_x=\frac34 b_{1,yy}$$
Compare these two view points on KP one could say that these are
equivalent and give the same nonlinear PDE.

Analogously, using Lemma 4.2 equations one gets mKP \be\label{mkp}
4q_{tx}=D_x(q_{xxx}-2q_x^3)+3q_{yy}+6q_{xx}q_y,\quad a_{0}=q_x.
\ee \vspace{5mm}

In algebraic approach to integrability the notion of commutativity
appears preferable
 in comparison with the notion of symmetry. Thus it seems awkward to reformulate Lemmas 4.1-2 in terms of
 symmetries of the infinite system of PDEs constituting hierarchies. Yet consider particular
 equations of hierarchies like  (\ref{kp}) and (\ref{mkp}) the notion of symmetry becomes more
 important and useful. In the next example we demonstrate reductions Eq.(\ref{kp}) related with stationary
 points of very simple symmetries transformations.

\paragraph{Example 4.6 } Since the equations contain independent variables only inexplicitly
they are keep its form invariant under shift In the case $b_1$ not
depends in $y$ KP equation reduce and gives rise KdV:
$$ b_{1,y}=0\RA  $$
Bussinesque??

\subsection{Conservation laws}
 Interplay of symmetries and conservation laws for ODEs has been discussed in Introduction.
This interconnection for PDEs is quite different but very
interesting as well and we present here a terse introduction to
pure algebraic aspects of the theory oriented on applications in
next chapters. Hereafter in this section we restrict ourself by
the case of KP-hierarchy and will denote by $\cal{B}$ the set of
polynomial functions on coefficients $b_j$ of the formal series
(\ref{10})  and its derivatives with respect to $t_i, \ i=1, 2,
\dots$.

\paragraph{Definition 4.7} Corollary of equations of KP-hierarchy written in the form of divergence i.e.
\be\label{divkp} \sum_k D_{t_k}(\beta_k)=0\ee is called {\bf
conservation law} of KP-hierarchy if sum in (\ref{divkp}) is
finite and the
 the variables $\beta_k\in\cal{B}.$

In order to build up conservation laws of KP-hierarchy one can use
lemma below which based on next
 definition.

\paragraph{Definition 4.8.} For the formal series $A\in \cal{F}{D}$ in powers of $D$
$$ A=\sum_{k=-\infty}^n a_kD^k=
 a_nD^n+ \dots +a_0+a_{-1}D^{-1}+a_{-2}D^{-2}+\dots\ .  $$
the coefficient $a_{-1}$ is called {\bf residue} and is denoted as $\res(A)$.\\

\paragraph{Lemma 4.9} For all $ m,\, n\in \Integer$
$$ \res\left([aD^m,bD^n]\right)= D_x\a_{m,n} $$
where $\a_{m,n}$ is a polynomial function on $a$, $b$ and it's
$x-$derivatives.

{\it Proof.} The residue vanish if the powers $m,\, n$ obey the
condition $mn\ge 0.$ For instance in the case $n=0$ the commutator
$aD^mb-baD^m$ is a differential operator if $m\ge 0$ and PSDO of
order $m-1\le -2$ if $m<0$ (see Prop. 3.7). Obviously the
coefficient by $D^{-1}$ is zero in both cases. Obviously as well
that the residue vanish if $m+n<0.$

 Let now $m,\, n$ have different signs and $m+n=k\ge 0.$ Then in virtue of (\ref{Da})
$$
\res\left([aD^m,bD^n]\right)= \left( \ba{c}m\\k+1 \ea \right)(
aD^{k+1}(b)+(-1)^{k}D^{k+1}(a)b)  $$ since
$$ m+n=k\RA m(m-1)\cdots(m-k)=\pm n(n-1)\cdots(n-k).  $$
Standard "integration by parts" proves now the lemma. Particularly
for $k=0, \ 1$ we have, respectively
$$ aD(b)+D(a)b=D(ab)\quad aD^2(b)-D^2(a)b=D(aD(b)-D(a)b).$$
\qed\\

Use (\ref{kpkn}) we find
$$ D_{t_k}(\res B^n)=\res D_{t_k} B^n= \res [B^k_+, B^n] $$
and Lemma 4.9 yields

\paragraph{Corollary 4.10.} KP-hierarchy possess infinite sequence of
conservation laws of the special form as follows \be\label{rhon}
 D_{t_k}(\rho_n)=D_x(\sigma_{k,n}), \quad \rho_n\defeq\res(B^n), \ \sigma_{k,n}\in \cal{B}\ee
where $\cal{B}$ denotes the set of polynomial functions on
coefficients $b_j$ of the formal series (\ref{10}).
\paragraph{Example 4.11.}
$$
\rho_1=b_1,\quad \rho_2=2b_2-b_{1,x},\quad
\rho_2=2b_2-b_{1,x},\quad \rho_3=3b_3+3b_1^2-b_{2,x}-b_{1,xx}
$$
$$D_{t_k}(\rho_1)=D_{x}(\rho_{n+1})$$
\paragraph{Theorem 4.12}(\cite{Wils}). Let in the case (\ref{10})
 $$ A=D+\sum_{i=1}^\infty b_iA^{-i},\quad A^m=(A^m)_+
 +\sum_{k=1}^\infty  b_{km} A^{-k}. $$
 Then for all $i,\, n=1,\, 2,\dots$
 \be\label{dbb}  D_n b_i=D b_{in} ,\qquad (D\equiv D_1).      \ee

 {\em Proof.} Lemma 4.1 yields
  $$
 D_n(A_1)=D(A_n)+[A_1,A_n],\quad  A_1=\sum_{i=1}^\infty b_iA^{-i},\quad
 A_m=\sum_{k=1}^\infty  b_{km} A^{-k}.  $$
 We have now
  $$
 D_n(A_1)=
 \sum_{i=1}^\infty D_n(b_i)A^{-i}+ \sum_{i=1}^\infty b_i [A^{-i}, A_n]
 \quad
 D(A_n)=
 \sum_{k=1}^\infty D(b_{kn})A^{-k}+ \sum_{k=1}^\infty b_{kn} [A^{-k}, A_1]
 $$
 and
 $$ [A_1,A_n]+\sum_{k=1}^\infty b_{kn} [A^{-k}, A_1]-
  \sum_{i=1}^\infty b_i [A^{-i}, A_n]=0.$$
  \qed
\vspace{3mm}

 In order to highlight the notion of conservation laws
let us consider $L-$periodic in $x$ solutions of KP-hierarchy.
Then
$$ D_{t_k}(\rho)=D_x(\sigma_{k}) \RA  D_{t_k}\int_{x_0}^{x_0+L}\rho dx=\int_{x_0}^{x_0+L}\sigma_{k,x}
dx=0.
$$
 Therefore, the integral $R(\vec{t})\defeq \int_{x_0}^{x_0+L}\rho
dx$ conserves it's value as times $t_k$ change and the integral
$R=\int_{x_0}^{x_0+L}\rho dx$ rather than density $\rho$ has
"physical sense." That is the reason to call two conservation laws
in the differential form $ D_{t_k}(\rho_j)=D_x(\sigma_{j,k}),\
j=1,2$ with distinct densities {\bf equivalent}
$\rho_1\equiv\rho_2$ if \be\label{equiv}\int_{x_0}^{x_0+L}\rho_1
dx=\int_{x_0}^{x_0+L}\rho_2 dx. \ee

In general not necessary periodic case, we take
 \paragraph{Definition 4.11}
 Two conservation laws of KP-hierarchy $
D_{t_k}(\rho_j)=D_x(\sigma_{j,k}),\ j=1,2$ to be said  {\bf
equivalent} iff $\rho_1-\rho_2\in \cal{B}_0$ where latter is
subspace of $\cal{B}$ constituted by $\im D.$
 \vspace{3mm}

Obviously, equivalence relation introduced in Def.4.11 implies the
equality (\ref{equiv}). Moreover, as we will see in the next
section this relation $\rho\equiv 0$ could be verified in fully
algorithmic way.

\newpage

\section{Integrability problem}
In this section we discuss a general definition of {\it
integrability} for evolutionary PDEs (\ref{Evol}) with one space
variable $x$ using the theory of formal series in powers of
$D=D_x$ (sections 2 and 6). Follow traditions of differential
algebra \cite{ritt} we introduce the notation
$$ u\mapsto u_1 \mapsto u_2\mapsto \dots\mapsto u_k\mapsto\dots  $$
for the infinite sequence of derivatives with respect to $x$ of
the basic variable $u.$ \footnote{In fore cited book it called \it
differential indeterminate.} In these notations the general
evolutionary PDE is rewritten as follows

 \be\label{fgen} u_t=f(x,u, u_1, u_2,\dots ,u_n). \ee

The highest derivatives $u_n$ defines the {\it order} $n$ of Eq.
(\ref{fgen}) and we consider usually equations of order $n\ge 2.$
The dependence on $x$ of the right hand side in Eq.(\ref{fgen})
became very important in the next Chapter in relation with
godograph type transformations.

\subsection{ Basic definitions.}
Consider the set $\cal{U}$ of smooth functions of variables $x, u,
u_1, u_2,\dots $ we will use the equation Eq. (\ref{fgen}) in
order do define action of the "additional" differentiation $D_t$
on $\cal{U}$ (the first one is $D=D_x$). Namely we take as
definitions of these differentiations the formulae \bea\label{dxg}
D_x: \ \ \ & g(x, u, u_1, u_2,\dots ,u_m)\mapsto & g_x+g_*(u_1),\quad u_1=D(u), \\
\label{dtg} D_t: \ \ \ & g(x, u, u_1, u_2,\dots ,u_m)\mapsto &
g_*(f),\quad f=D_t(u) \eea where $g_x=\pa_x(g)$ is partial
derivative of the function $g$ and $g_*$ denotes differential
operator as follows
$$ g_*\defeq \sum_k \frac{\pa g}{\pa u_k} D^k=
\frac{\pa g}{\pa u}+\frac{\pa g}{\pa u_1}D+\dots+\frac{\pa g}{\pa
u_m}D^m.
$$
It is easy to see that operators (\ref{dxg}),(\ref{dtg}) act on
the "generators" of the set $\cal{U}$ in a natural way, as partial
differentiations should do:
 $$ D_x(x)=1, \ \ D_x(u)=u_1,\quad D_t(x)=0, \ \ D_t(u)=f(x, u, u_1, u_2,\dots ,u_n) $$
 and
 $$ D_xD_t(u)=D_x(f), \ \ D_tD_x(u)=D_t(u_1)=D_x(f)\RA D_xD_t(u)=D_tD_x(u).$$

It is important to notice that action of $m-$th order differential
operator $g_*$ on functions $v$ can be defined also as
\be\label{genlin}
  g_*(v)={dg[u+\eps v]\over d\eps}\vert_{\eps=0}.
\ee In particular as corollary of Eq. (\ref{genlin}) we find that
$$ (fg)_*=fg_*+gf_*. $$
 One can prove now that $D_x,$ $D_t$ and $\Delta=D_xD_t-D_tD_x$ as well, obey
 Leibnitz rule and that $\Delta=0$ on the set $\cal{U}.$

\paragraph{Definition 4.13.} The operators (\ref{dxg}) and (\ref{dtg}) on the set $\cal{U}$ are
called {\bf total differentiation with respect to $x$} and {\bf
evolutionary differentiation related with $f\in \cal{U}$},
respectively. The Gato derivative (\ref{genlin}) of a function
 $g\in \cal{U}$ we call {\bf linearization} of the function $g.$

Summing up we have

\paragraph{Proposition 4.14.} For any smooth function $f\in \cal{U}$ corresponding evolutionary
differentiation (\ref{dtg}) commutate with (\ref{dxg}).
\vspace{5mm}

This statement looks trivial but if, as in previous Section, we
try to find other evolutionary differentiations \be\label{dtfk}
  D_{t_k}(u)=f_k(x, u, u_1, u_2,\dots ,u_{n_k})
 \ee
which commutate each other we shall come to nontrivial obstacles.
Situation here in certain sense corresponds to correlation of the
Proposition 3.5 which is very general and abstract and more
interesting problem about commutating differential operators
considered in Section 4. The role of Proposition 3.5 plays now
Lemma 4.1 and commutativity conditions $[D_{t_j}, D_{t_k}]=0$ have
to play the role of commuting differential operators. This link
became more apparent and instructive if we bring up commutativity
conditions to the level of a tangent space of $\cal{U}$ using
(\ref{genlin}). Namely instead of writing down the cross
derivatives equality $D_{t_1}D_{t_2}(u)=D_{t_2}D_{t_1}(u)$ for the
two evolutionary equations \be\label{f12} D_{t_1}(u)=f_1(x, u,
u_1, u_2,\dots ,u_{n_1}),\quad D_{t_2}(u)=f_2(x, u, u_1, u_2,\dots
,u_{n_2})\ee we will consider analogous condition
$D_{t_1}D_{t_2}(v)=D_{t_2}D_{t_1}(v)$ for linearized ones
 $D_{t_k}(v)=(f_{k})_*(v), \ k=1,2.$ Translate this condition into condition of commutativity of
 corresponding operators
\be\label{d12} [D_{t_1}-(f_1)_*, D_{t_2}-(f_2)_*]=0 \ee
 we come to next
 \footnote{It can be proved that cross derivatives equation $D_{t_1}D_{t_2}(u)=D_{t_2}D_{t_1}(u)$
 implies (\ref{d12}).}

\paragraph{Definition 4.15.} Let us say that the two evolutionary equations (\ref{fi2})
 satisfy the {\bf consistency condition} and, respectively, {\bf compatibility condition}
 if the cross derivatives equation $D_{t_1}D_{t_2}(u)=D_{t_2}D_{t_1}(u)$ is fulfilled or, and in
 the second case, if differential operators equality (\ref{d12}) is valid i.e.
\be\label{crossd}
  D_{t_1}((f_2)_*)-D_{t_2}((f_1)_*)=[(f_1)_*, \ (f_2)_*]
 \ee
We remind (Cf. (\ref{dtk})) that the action of any evolutionary
differentiations $D_{t_k}$ on polynomials $A=\sum a_jD^j, \ a_j\in
\cal{U}$ is defined by $D_{t_k}(A)=\sum D_{t_k}(a_j)D^j. $
\vspace{5mm}

In order to compare two almost equivalent Definitions 4.15 we give
simple example.

\paragraph{Example 4.16.} Let us consider the pair evolutionary equations as follows
$$
D_{t_1}(u)=f(x, u, u_1, u_2,\dots ,u_{n}),\quad D_{t_2}(u)= u_1.
$$
Then, apply (\ref{dxg}) and (\ref{dtg}) and Proposition 4.14 we
get
$$ D_{t_1}D_{t_2}(u)= D_{t_1}(u_1)=D_x(f)= f_x+f_*(u_1), \ \
D_{t_2}D_{t_1}(u)= D_{t_2}(f)=f_*(u_1). $$ Thus the consistency
condition is satisfied iff  $f_x=0$ and $f=f(u, u_1, u_2,\dots
,u_n).$ On the other hand the compatibility condition
(\ref{crossd}) yields
$$ f_{u,x} =f_{u_1,x}=\dots=f_{u_n,x}=0\LRA f=f(u, u_1, u_2,\dots ,u_n)+X(x) $$
where $X(x)$ is arbitrary function of the independent variable
$x.$

\subsection{Main theorem}

We wish now to address more practical question. Let the
evolutionary equation (\ref{fgen}) of the order $n$ is given i.e.
$D_t(u)=f.$ What conditions on its right hand side $f\in \cal{U}$
or, more exactly, on the corresponding differential operator
$f_*:$ \be\label{f*} f_*= \frac{\pa f}{\pa u}+\frac{\pa f}{\pa
u_1}D+\dots+\frac{\pa f}{\pa u_n}D^n. \ee are needed in order to
build up a hierarchy of evolutionary differentiations
$D_{t_k}(u)=f_k$ compatible with fore given. Take in account that
in the compatibility condition \be\label{ffk}
D_{t}((f_k)_*)-D_{t_k}((f)_*)=[f_*, \ (f_k)_*] \ee the order $n_k$
of the differential operator $(f_k)_*$ could be arbitrary large
whilst the order of $D_{t_k}((f)_*)$ is fixed by $n$ we see that
leading terms in $D_{t}((f_k)_*)$ and in the commutator in the
right hand side should coincides. It gives rise the definition
which reproduces the main formula (\ref{dtk}) of previous section
in a new environment

\paragraph{Definition 4.16} We say that evolutionary equation
$$ u_t=f(x, u, u_1, u_2,\dots ,u_n)  $$
of order $n$ posses {\bf canonical Lax pair} if there is the
formal series of order $m\ne 0$
$$
 A=a_mD^m+a_{m-1}D^{m-1}+\dots+a_0+a_{-1}D^{-1}+a_{-2}D^{-2}+\dots
$$
with coefficients $a_j\in \cal{U}$ such that
\be\label{afa}
D_t(A)=[f_*,A].                                            \ee
Some times solutions $A$ of non-zero order of this equation are
called {\bf formal symmetries}.
 \vspace{3mm}

 Apply this definition to Burgers and Korteweg de Vries equations which we often use as
illustrative examples, we find
\paragraph{Example 4.17} For Burgers equation
$$ u_t=u_2+2uu_1=f,\quad f_*=D^2+2uD+2u_1$$
the formal symmetry is PSDO as follows $ A=D+u+u_1D^{-1}.$ Indeed
$$ [f_*,A]=f+D(f)D^{-1}\quad\mbox{and}\quad A=D+u+u_1D^{-1}\RA D_t(A)=u_t+u_{1,t}D^{-1} $$
and thus $D_t(A)=[f_*,A]$.

Similarly, in the case of Korteweg de Vries equation (KdV)
$$  u_t=6uu_1-u_3=f,\qquad  f_*=-D^3+6uD+6u_1$$
the second order PSDO $A=D^2-4u-2u_1D^{-1}$ is the formal symmetry
since
$$  [f_*, A]=-4f-2D(f)D^{-1}, \ \ \mbox{and}\quad A=D^2-4u-2u_1D^{-1}\RA D_t(A)=-4u_t-2u_{1,t}D^{-1} $$
and thus $D_t(A)=[f_*,A]$. \vspace{3mm}

Next statement can be considered as milestone of integrability
theory of evolutionary equations with one space variable.

\paragraph{Theorem 4.18} If there are series of evolutionary differentiations $D_{t_k}(u)=f_k$ of orders
$n_k\to \infty$ by $k\to \infty$ satisfying compatibility
conditions (\ref{ffk}) for a given evolutionary
 equation $D_t(u)=f$ of order $n\ge 2.$  Then this equation $D_t(u)=f$ possess the
formal symmetry $A$ of the first order: \be\label{fsym}
A=a_1D+a_0+a_{-1}D^{-1}+\dots    \ee \vspace{5mm}

 Construction of the solution $A$ of the basic equation $D_t(A)=[f_*,A]$ needs some technicalities
 related with the limiting procedure in Eq.(\ref{ffk}) when $k\to \infty.$ We skip that formal part of
 proof of Theorem 4.18 (see Appendix??) and go instead directly to applications.

Firstly, we explain how to get formal symmetry of the first order
starting with any solution of Eq.(\ref{afa}) and describe it's
general solution. Obviously,
$$ \a_1 A_1 +\a_2 A_2, \ \ ,\a_j\in\Complex \quad\mbox{and}\quad A_1A_2      $$
satisfy Eq.(\ref{afa}) if $A_1$ and $A_2$ do. Apply results of
Section 3 we get.

\paragraph{Lemma 4.19} Let formal series in $D$ of order $m$ i-e- $B=\sum b_jD^j, \ b_j\in \cal{U}$  is
solution of Eq.(\ref{afa}) and let $m\ne 0,$ $n\ge2.$ Then there
exist the first order formal symmetry $A$ (see (\ref{fsym})) with
the leading coefficient \be\label{a1} a_1=(f_n)^{\frac{1}{n}},
\quad f_n\defeq \frac{\pa f}{\pa u_n}   \ee and any solution
$\ti{B}$  of Eq.(\ref{afa}) of order $\ti m$ can be represented as
formal series in powers of $A$ with constant coefficients
$\a_l\in\Complex$:
$$
 \ti{B}=\sum_{k\le\ti m } \a_k A^k $$
{\it Proof.} Slight generalization of Lemma 3.2 yields the formal
series $B_1$ with coefficients from
 $\cal{U}$ such that $B_1^m=B.$ Let us consider the formal series $R=D_t(B_1)-[f_*,B_1].$ Since
$$ D_t(B)= D_t(B_1)B_1^{m-1}+B_1D_t(B_1)B_1^{m-2}+\dots+B_1^{m-1}D_t(B_1),$$
and
\[  [f_*,B]=[f_*,A]A^{m-1}+A[f_*,A]A^{m-2}+\dots+A^{m-1}[f_*,A]  \]
we have
\[   B_t-[f_*,B]=RA^{m-1}+ARA^{m-2}+\dots+A^{m-1}R=0. \]
In this sum the all $m$ series in $D^{-1}$ have the same leading
term and, hence, this leading term is zero. It is possible only if
$R=0$ and therefore we proved that $B_1$ is a solution.

In order to find the formula for leading coefficients of solutions
Eq.(\ref{afa}) one should find the leading term in $[f_*,B]-B_t.$
Since $n\ge 2$ it is contained in
$$ [f_*,B]=c_{n+m-1}D^{n+m-1}+\dots,\quad c_{n+m-1}=nf_nD(b_m)-mb_mD(f_n) $$
where $f_n$ and $b_m$ are leading coefficients in $f_*$ and $B$
(Cf  Proposition 3.7 ). Vanishing leading coefficient gives the
equation
\[ nD\log b_m=mD\log f_n \LRA n\log b_m-m\log f_n=\const  \]
which implies for $m=1$ the formula (\ref{a1}) for the leading
coefficient of the first order formal symmetry.\qed

Most known applications of Theorem 4.18 related with next
proposition quite similar to Corollary 4.10:
\paragraph{Corollary 4.20.} In conditions of Theorem 4.18 the evolutionary equation (\ref{fgen}) possess
infinite series of conservation laws as follows
$$
 D_{t}(\res A^n)=\res D_t(A^n)= \res [f_*, A^n] \in\im D, \ \ n=-1,1,2,\dots .
$$

Hereafter $\im D$ denotes the subset of functions $g\in\cal{U}$
which can be rewritten as $g=D_x(h), h\in\cal{U}.$ In other words
Corollary 4.20 provides series of conservation laws :
\be\label{ares} D_t(\rho_n)= D_x(\sigma_n),\quad \rho_n,
\sigma_n\in\mathcal{U}, \quad  \rho_n \defeq \res(A^n). \ee of the
equation (\ref{fgen})) under consideration.
 This series should be completed by the conservation law with the density \be\label{logres}
\rho_0=\frac{a_{m-1}}{a_m}.\ee

Indeed, rewrite equation (\ref{afa}) for $m-$th order formal
symmetry
$$
 A=a_mD^m+a_{m-1}D^{m-1}+\dots+a_0+a_{-1}D^{-1}+a_{-2}D^{-2}+\dots
$$
as follows
$$
D_t(A)A^{-1}=f_*-Af_*A^{-1}=[A^{-1}, Af_*]
$$
we find  $\res (A_tA^{-1}) \in\im D$ due Lemma 4.9 and that
$$
\res(A_tA^{-1})=\\
\res\left[(a_{m,t}D^m+a_{m-1,t}D^{m-1})(a_m^{-1}D^{-m}+\alpha
D^{-m-1})\right]=$$
$$
a_{m,t}(\alpha-m\frac{a_{m,x}}{a_m^2})+\frac{a_{m-1,t}}{a_m}.
$$
It remains to notice that
$$
\alpha=m\frac{a_{m,x}}{a_m^2}-\frac{a_{m-1}}{a_m}.
$$

\paragraph{Example 4.21.} For $n=-1$ we have due (\ref{a1})
$$
\res
A^{-1}=\res\left(\frac{1}{a_1}D^{-1}+\dots\right)=(f_n)^{-\frac{1}{n}}.$$
Thus $\rho_{-1}=\left(\frac{\pa f}{\pa u_n}\right)^{-\frac{1}{n}}$
and by Corollary 4.20 the evolutionary equation (\ref{fgen})
should always satisfy the condition \be\label{a1res}
D_{t}(\rho_{-1})=D_t\left(\frac{\pa f}{\pa
u_n}\right)^{-\frac{1}{n}}\in\im D \ee if it is compatible with
hierarchy of evolutionary differentiations.
\paragraph{Definition 4.22.} The series of conservation laws
(\ref{ares})), (\ref{logres})) we call {\bf integrability
conditions.}

\paragraph{ Definition 4.23.} The conservation law is differential corollary of
(\ref{fgen})) written in a form of divergency
 \be\label{rhosigma} D_t(\rho)= D_x(\sigma),\quad  \rho, \sigma\in\cal{U}            \ee
 where functions $\rho$ and $\sigma$ are called density and flux, respectively.
 It is called trivial if $\rho\in \im D$ and there is $ h\in\cal{U}$
such that $\rho=D_x(h).$ In this case
$$ D_t(\rho)=D_t(D_x(h))=D_x(D_t(h))\in\im D   $$
 for any evolutionary differentiation $D_t.$ Two conservation
laws Eq.(\ref{rhosigma})with the densities
 $\rho_1$ and $\rho_2$ are considered equivalent if $\rho_1-\rho_2\in\im D.$
 \vspace{3mm}

Come back to Example 4.21 we see now there is an alternative or
$(\rho_{-1})\in\im D$ or $D_{t}(\rho_{-1})\in\im D.$ In both cases
this condition is non trivial one and can be verified using
classical statement by (Helmgolz??) proven in P.Olver book
\cite{Olver}.

\paragraph{Theorem 4.24.} The function $ h\in\cal{U}$ belongs to the subspace  $\im D$
iff $ \hat E(h)=0$ where {\bf Euler operator} $\hat E$ is defined
as follows \be\label{euler} \hat{E}\defeq {\pa\over \pa
u}-D{\pa\over \pa  u_1}+ D^2{\pa\over \pa u_2}-D^3{\pa\over \pa
u_3}+\dots \ .\ee \vspace{3mm}

Equations similar to Eq.(\ref{afa}) play major role in modern
integrability theory and are called commonly by {\bf Lax pair}.
The first test which have to justify that that is not trivial one
\footnote{There is known a few cases of faked Lax pairs.} could be
formulated similarly to the definition $S-$integrability below. In
other words proper Lax pair must produce series of nontrivial
conservation laws.

\paragraph{ Definition 4.25.} The evolutionary equation (\ref{fgen})) with formal symmetry (\ref{afa}))
is called $S-$integrable if the series (\ref{ares})) contains
nontrivial conservation laws of arbitrary higher order and
$C-$integrable in the opposite case. \vspace{3mm}

The famous  Korteweg de Vries (see (\ref{kdv}) in Section?) gives
us an example of $S-$integrable equation. As matter of fact the
formal symmetry for KdV indicated in Example 4.17 related very
closely with Lemma 4.6 ( see Eq.(\ref{***}) and concluding remark
in Section?). KdV hierarchy will be discussed in next section.

Although Definition 4.25 looks too complicated to be useful it is
not the case. In the next chapter we apply this definition to the
second order evolutionary equations (\ref{fgen})) and verify that
first three of conditions (\ref{ares}) fully describe {\bf all}
second order evolutionary equations possessing hierarchy of higher
order symmetries. Burgers equation from Example 4.17 is, in
certain sense, typical representative in this class and a class of
$C-$integrable equations in general.

\subsection*{Symmetries and cosymmetries}
Conclude this chapter we discuss duality between symmetries and
conservation laws starting with "zero" dimensional evolutionary
equations i.e. dynamical systems
  \be\label{dynn}
  \frac{d\vec u}{dt}=\vec{f}(\vec{u}),\quad
  \vec{u}=(u^1,u^2,\dots,u^n)^T
  \ee
  where $()^T$ denotes hereafter matrix conjugation. In this
  case the conservation law with the density $\rho=\rho(\vec{u})$ is just a
  constant of motion
  $$
\frac{d\rho}{dt}=\frac{\pa\rho}{\pa
u^1}f^1+\dots+\frac{\pa\rho}{\pa u^n}f^n =0.
$$
Infinitesimal version of conservation law is obtained by
replacement $\vec u$ by $\vec u+\eps\vec v$ in above equation (Cf
Definition 4.13):
\paragraph{Proposition 4.25.} Let $\rho=\rho(\vec{u})$ is the density of conservation law
for Eq.(\ref{dynn})) and $\hat\rho$ denotes column-vector
constituted by the partial derivatives $\frac{\pa\rho}{\pa u^k}.$
Then \be\label{hatrho} \hat{\rho}_t+f_*^T\hat\rho=0,\quad
(f_*)_{ij}=\frac{\pa f^i}{\pa u^j} \ee

{\it Proof.} The dynamical system Eq.(\ref{dynn})) yields
 \be\label{dynlin} \frac{d\vec v}{dt}=f_*\vec{v}\ee
 for infinitesimal dynamical variables $\vec v.$ The same
 change $\vec u$ by $\vec u+\eps\vec v$ gives in the infinitesimal
 approximation
 $$
 \rho(\vec u+\eps\vec v)=\rho(\vec u)+\eps\hat\rho^T\vec v.
$$
Substitute this expression into conservation law with the density
$\rho$ we get \be\label{product} \frac{d}{dt}<\hat\rho,\vec v>=
<\frac{d\hat\rho}{dt},\vec v>+<\hat\rho,\frac{d\vec v}{dt}>
\ee
where bracket denote the scalar product. Take into account
Eq.(\ref{dynlin})) one gets Eq.(\ref{hatrho})). \qed

In the case of evolutionary equation (\ref{fgen}) the
linearization (\ref{genlin}) gives for infinitesimal variable $v$
PDE as follows

 \be\label{evlin} v_t=f_*v=(f_u+f_{u_1}D+\dots)v. \ee

Similarly to zero-dimensional case Definition 4.22 of conservation
laws implies
\paragraph{Proposition 4.26.} Let $\rho=\rho(x,u,u_1,u_2,\dots)$ is the density of conservation
law (\ref{rhosigma})) for evolutionary equation (\ref{fgen}) and
 \be\label{drho}\hat\rho=\rho_u-D(\rho_{u_1})+D^2(\rho_{u_2})-\dots \ .
\ee
 Then
 \be\label{claw}
 D_t\hat\rho+f_*^T\hat\rho=0
 \ee
where $f_*^T$ is the differential operator
$$
f_*^T=f_u-D\circ f_{u_1}+D^2\circ f_{u_2}+\dots
$$
formally conjugated to the operator $f_*.$

The proof of Proposition 4.27 (see \cite{Olver}) is analogous to
the zero-dimensional case above. The function $\hat\rho$ defined
by Eq.(\ref{drho}) is called {\bf variational derivative} of the
function $\rho.$ One has to notice that in the $L-$periodic case
for the variation of the functional
$$ R[u]\defeq \int_{x_0}^{x_0+L}\rho(x,u,u_1,u_2,\dots ) dx $$
we have in the infinitesimal approximation
 \be\label{varrho} R[u+\eps v]-R[u]=\eps\int_{x_0}^{x_0+L}\hat\rho_*
 v dx=
\eps\int_{x_0}^{x_0+L}\hat\rho v dx. \ee The last formula in
particular plays the role of the scalar product (\ref{product}).

Compare Propositions 4.26 and 4.27 one can notice that in both
cases $\hat\rho\in\ker{D_t+f_*^T}$ while symmetries are in
$\ker(D_t-f_*).$ \vspace{3mm}

 For future references we formulate
\paragraph{Definition 4.28.} The order of nontrivial conservation law
(\ref{rhosigma})) with the density $\rho$ is defined as order of
variational derivative (\ref{drho}).

We remind that density $\rho$ is called trivial if $\rho\in\im{D}$
and in virtue of Theorem 4.23 variational derivative $\hat\rho=0$
in this case. One can verify that the order $n$ of variational
derivative $\hat\rho$ always is even number and if $n=2m>0$ then
$$
\hat\rho=a(x,u,u_1,\dots,u_m)u_{2m}+\dots,\quad
a=(-1)^m\frac{\pa^2\rho}{\pa u_m^2}\ne 0.
$$
For $n=0$ one can assume without loss of generality that $\rho=u.$
The latter case take place iff the right hand side $f$ of
evolutionary equation (\ref{fgen}) satisfies a condition $f\in\im
D.$

\section{Summary}
The primary goal of this chapter is to establish a workable
criterion that can be readily checked to determine whether a given
evolutionary equation possess integrability features like
symmetries, local conservation laws and Lax pairs. A systematic
linearization of all equations is one of basic tools in this
chapter.
 \vspace{5mm}

\section{Exercises for Chapter 4}

\paragraph{1.} For evolution equation
\[ u_t=u_n+F(u,u_1,\dots,u_k),\quad k<n,\quad (n\ge2)  \]
possessing a formal symmetry prove that $F_k=\pa_{u_k}F$ is a
density of a local conservation law.
\paragraph{2.} Check out that
\[L=D^2-u,\quad   L_t=[(L^{\frac32})_+,L]\LRA 4u_t+u_{xxx}=6uu_x.         \]
\paragraph{3.} Find non-trivial densities $\rho=\rho(u,u_1)$ for Burgers
equation from Example 2.2. The picture would be incomplete without
next

\paragraph{6.} Prove that condition $ f_*(g)=g_*(f)$ coincides with
cross derivatives equation $D_{t_1}D_{t_2}(u)=D_{t_2}D_{t_1}(u)$
if $D_{t_1}u=f$ and $D_{t_2}u=g.$

\paragraph{7.}
 Prove that the bracket $\{f^1,f^2\}\defeq f_*^1(f^2)-f_*^2(f^1)$
 (Cf Definition 2.8) satisfies the Jacobi identity.
\paragraph{8.}
Find the constant $\eps$ such that $D_{t_2}u=\eps tu_x+1$ is the
symmetry of equations $D_{t_1}u=f$ from Examples 2.2, 2.3 in the
sense that $D_{t_1} D_{t_2}u=D_{t_2}D_{t_1}u.$

\chapter{Burgers-type equations}

\section{Introduction}

We are going to prove follow by \cite{svin} that any evolutionary
equation of the second order
 \be\label{2ord} u_t=F(x,u,u_1,u_2),\quad u_j=D^j(u) \ee
 which possess a canonical Lax pair can be reduced by
invertible transformation to one of the following three canonical
forms
 \bea\label{sv1}
 u_t&=&D(u_1+u^2+f(x))\\  \label{sv2}
 u_t&=&D(\frac{u_1}{u^2}-2x)\\ \label{sv3}
 u_t&=&D(\frac{u_1}{u^2}+\eps_1xu+\eps_2u)
 \eea
or to the linear equation $u_t=u_2+f(x)u$ with the arbitrary
function $f(x).$ How to do it constructively??? .....

 (\ref{sv1})--(\ref{sv3})

\subsection{Rough classification}
 The special feature of
evolutionary equations of the second order which simplifies a
general structure of integrability conditions is defined by

\paragraph{Lemma 1} A nontrivial conservation law
(\ref{rhosigma})) for evolutionary equations of the second order
(\ref{2ord}) has order $n=2m$ less or equal $2.$

{\it Proof.} One can use the formulae from previous Chapter
$$\hat\rho=\rho_u-D(\rho_{u_1})+D^2(\rho_{u_2})-\dots=\hat\rho=a(x,u,u_1,\dots,u_m)u_{2m}+\dots
$$
and the equation  (\ref{claw}) for variational derivatives
$$
D_t\hat\rho+F_*^T\hat\rho=0.                             $$
Separate in this equation leading terms which contain $u_{n+2}$
one finds
 \be \label{F2}
aD^n(F)+D^2(aF_2u_n)+\dots=0,\quad F_2\defeq\frac{\pa F}{\pa u_2}
\ee and if the order $n=2m>2$ that results in $2 aF_2u_{n+2}=0\RA
a=0.$ \qed

Thus the order of conservation law for evolutionary equation
(\ref{2ord}) can be equal $0$ or $2.$ If order $n=2m=2$
Eq.(\ref{F2}) in the proof of Lemma gives
 \be \label{nonlin} 2F_2+u_2F_{22}=0\LRA F=(fu_2+g)^{-1}+h           \ee
 where functions $f,\ g$ and $h$ may depend $x, u, u_1.$

 In the zero order case with $\rho=\rho(x,u),\ \hat\rho=\rho_u\ne 0$ we have
 \be \label{2lin}
 D_t\rho=\rho_u F\in\im D\RA F=f(x,u,u_1)u_2+g(x,u,u_1).\ee

 Since $\rho_{-1}=F_{2}^{-1/2}$ should be density of conservation
 law of order less or equal $2$ (see Lemma 1) this function on $x,\ u\ u_1,\ u_2$ should be linear in $u_2$ and we have
 three possibilities:
\begin{eqnarray}
 F_{2}^{-1/2} &=& D\alpha(x,u,u_1), \label{2triv}\\
 F_{2}^{-1/2} &=& D\alpha(x,u,u_1)+\beta(x,u),\quad \beta_u\ne0 \label{2zero}\\
 F_{2}^{-1/2} &=& D\alpha(x,u,u_1)+\beta(x,u,u_1),\quad \beta_{u_1u_1}\ne0. \label{2two}
\end{eqnarray}
The first case corresponds to trivial conservation law and it has
order $0$ and $2$ in the cases (\ref{2zero}), (\ref{2two}),
respectively.

Rough classification of Eqs.(\ref{2ord}) uses only the first
integrability condition $D_t(\rho_{-1})\in\im D.$

\paragraph{Svinolupov's Theorem 2.} Up to invertible change of variables
  \be \label{xuu} \hat x=\ph(x,\ u, \ u_1),\quad \hat u=\psi(x,\ u,\ u_1),\quad
 \hat u_1=\psi_1(x,\ u,\ u_1)=\frac{d\hat u}{d\hat x}  \ee
 the general nonlinear evolutionary equation (\ref{2ord}) of the second order
  can be reduced to one of two quasi-linear forms
\begin{eqnarray}
 u_t &=& u_2+f(x,u,u_1), \label{triv'}\\
 u_t &=& D(\frac{u_1}{u^2}+f(x,u)). \label{two'}
\end{eqnarray}
if it satisfies the first integrability condition.

 {\it Outline of the proof.} Since the first integrability condition (i.e. local conservation law
 with the density $F_{2}^{-1/2}$) is invariant under invertible transformations (\ref{xuu}) it is sufficient
 to find change of variables such that
 \be\label{svin}
F_{2}^{-1/2}dx= d\hat x, \quad\mbox{or}\quad
                                               F_{2}^{-1/2}dx=\hat{u}d\hat x.
   \ee
The first case corresponds to trivial conservation law with the
density (\ref{2triv}) and the second one to cases (\ref{2zero})
and (\ref{2two}).

In order to find the change of variables in the case (\ref{2zero})
it is sufficient to use point transformation:
$$\hat x=\ph(x,\ u),\quad \hat u=\psi(x,\ u)\RA \hat{u}d\hat x=\psi(\ph_x+u_1\ph_u)dx. $$
Indeed, compare formulae (\ref{2zero}) and  (\ref{2lin}) we see
that $\a(x,u,u_1)=\a(x,u)$ in  (\ref{2zero}) and we have
$$\psi(\ph_x+u_1\ph_u)=D\a+\b=u_1\a_u +\a_x+\b.                                     $$
Thus, functions $\ph$ and $\psi$ should satisfy two equations as
follows
$$ \a_x+\b=\psi\ph_x,\  \a_u=\psi\ph_u\RA
\frac{\a_x+\b}{\a_u}=\frac{\ph_x}{\ph_u}.
$$
That yields the first order PDEs for the functions $\ph$ and
$\psi:$
$$
L\ph=0,\quad L\psi=\b_u\psi,\qquad
L\defeq(\a_x+\b)\pa_u-\a_u\pa_x.
$$
In order to find exact expression in quadratures of the function
$\psi$ in terms of $\ph$ one can substitute $\psi=\Phi(\ph,u)$
into the equation $L\psi=\b_u\psi.$ That results in the ODE
equation $(\a_x+\b)\Phi_u=\beta_u\Phi$ in which $x-$variable plays
the role of parameter. Jacobian of the transformation $\hat
x=\ph,\quad \hat u=\Phi$ is $\ph_x\Phi_u\ne 0.$

In the case (\ref{2triv}) the transformation (\ref{xuu}) which
resolve the first of equations (\ref{svin}) is called contact one.
Contact transformations are fully defined (see next section) by
the first two equations in (\ref{xuu}) \be\label{phpsi}
 \hat x=\ph(x,\ u, \ u_1),\quad \hat u=\psi(x,\ u,\ u_1) \ee
which have to satisfy one and same equation \be\label{aphi}
\phi_x+u_1\phi_u=a(x,u,u_1)\phi_{u_1} \ee with arbitrary
coefficient $a=a(x,u,u_1)$ depending on three variables $x,\ u,\
 u_1.$ In our case Eq.(\ref{svin}) gives,obviously, $\ph=\a$
and Eq. (\ref{aphi}) for $\psi$ can be rewritten as follows:
$$
 D(\a)\frac{\pa\psi}{\pa u_1}=\frac{\pa\a}{\pa
u_1}D(\psi).
$$

In the case (\ref{2two}) the corresponding contact transformation
(\ref{xuu}) is defined by the second of equations (\ref{svin}):
$$\a_x+u_1\a_u+u_2\a_{u_1}+\b=\psi(x,u,u_1)[\ph_x+u_1\ph_u+u_2\ph_{u_1}].
$$
That implies
$$ \psi\ph_{u_1}=\a_{u_1},\quad \psi\ph_u=\a_u+\b_{u_1}
$$
since due Eq.(\ref{aphi}) we have
$$\ph_x\psi{u_1}-\ph_{u_1}\psi_x=u_1(\ph_{u_1}\psi_u-\ph_u\psi_{u_1}).
$$

$$
\frac{\pa^2\a}{\pa u_1^2}= \frac{\pa\ph}{\pa u}\frac{\pa\psi}{\pa
u_1}-\frac{\pa\ph}{\pa u_1}\frac{\pa\psi}{\pa u}.
$$

\subsection{Canonical densities}
\paragraph{Lemma 2}
 Canonical densities are:
\begin{eqnarray}
\rho _{-1} &=& F_{2}^{-1/2}, \label{cond2-1}\\
\rho _{0}  &=& F_{1}/F_{2}-\sigma _{-1} F_{2}^{-1/2}, \label{cond2-2}\\
\rho _{1}  &=& F_{2}^{-1/2} (\frac{1}{2}F_{0}+\frac{1}{4}\sigma
_{0}+ \frac{1}{8}\sigma _{-1}^{2}-\frac{1}{32}F_{2}^{-1}(2
F_{1}-(D (F_{2}))^{2})), \label{cond2-3}
\end{eqnarray}
and
\[ F_{2}=\frac{\partial F}{\partial u_{2}},\ \ \
F_{1}=\frac{\partial F}{\partial u_{1}},\ \ \ F_{0}=\frac{\partial
F}{\partial u}     \] and $\sigma _{-1},\sigma _{0}$ have to be
recursively found. \vspace{3mm}

\paragraph{Remark .}
The integrability conditions
 \be \label{conds}
D_t (\rho_{k})=D_x (\sigma_{k}),\quad k=-1,\, 0,\, 1. \ee

are based upon consistence conditions for
\[ u_t=F,\quad u_\tau=G,\quad\mbox{or more precisely}\quad v_t=F_*(v),\quad
v_\tau=G_*(v) \] where $F_*=F_2D^2+F_1D+F_0$ and $G_*$ defined
analogously:
\[ \frac{d}{dt}G_*+G_*F_*= \frac{d}{d\tau}F_*+F_*G_*\RA
 \frac{d}{dt} L=[F_*, G_*]
\]
It has been proved by \cite{svin} that evolutionary equations of
the second order (\ref{2ord}) which possess the symmetry of order
$m\ge3$ satisfies integrability conditions (\ref{conds}).

\paragraph{Definition }

\section{Linear case}
In the general theory one can suppose without loss of generality
the leading coefficient $u_0=1$ up to change of independent
variables introduced by

\paragraph{Definition 3.4.}
Two potentials $U(x,\la)$ and  $\hat U(y,\la)$ are said to be
related by {\it Liouville transformation} if there is function
$g=g(y)$ such that
\[
dx=g\, dy\LRA D_y=gD_x \quad\mbox{and}\quad\hat
U=g^2U+\{D_y,g(y)\} \] where a " Schwarzian derivative" $\{D,a\}$
is defined as follows \be\label{34} \{D_x,a(x)\}\defeq
\frac34\frac{a_x^2}{a^2}-\frac12\frac{a_{xx}}{a}. \ee \vspace{2mm}

One can verify readily that Liouville transformation is invertible
and
\[
\hat U=g^2 (U-\{D_x,h(x)\})\LRA U=h^2(\hat U-\{D_y,g(y)\}),\quad
g(y)h(x)=1.
\]

*****************************\\

Let us consider change of variables $x,\ u,\ u_1,\ldots$ defined
by the formula (\ref{xu}): \be\label{xu} \bar x=\ph(x,\ u,\
u_1,\ldots,u_m),\quad \bar u=\psi(x,\ u,\ u_1,\ldots,u_n) \ee and
the next condition of invariance of the vector-field $D$
\be\label{alpha} D\circ \Phi=\alpha \bar D,      \ee where $\Phi$
notates the transformation defined by (\ref{xu}) and
 \bean
 D &=& \frac{\pa}{\pa x}+u_1\frac{\pa}{\pa u}+
 \sum \limits_{i=1}^{\infty}\Bigl(u_{i+1}\frac{\pa}{\pa u_i}\Bigr), \\
 \bar D &=& \frac{\pa}{\pa \bar x}+\bar{u}_1\frac{\pa}{\pa \bar u}+
 \sum \limits_{i=1}^{\infty}\Bigl(\bar{u}_{i+1}\frac{\pa}{\pa \bar{u}_i}
 \eean
 Recall that $D=D_x$ is the total differentiation with respect to $x$
 of functions of variables $x,\ u,\ u_1=u_x,\ldots.$

Applies \ref{alpha}) to $\bar x,\ \bar u $ one finds,
respectively,
$$ D(\ph)=\alpha,\quad D(\psi)=\alpha\bar{D}(\bar u)=\alpha\bar{u}_1. $$
Therefore,
$$           \bar{u}_1= D(\psi)/D(\ph).       $$
Express $\bar{u}_1$ as a function of $x,\ u,\ u_1,\ldots$ and
apply (\ref{alpha}) to $ \bar{u}_1.$ It gives $ \bar{u}_2=
D(\bar{u}_1)/D(\ph).$ Thus (\ref{alpha}) is equivalent to formulae
$$ \bar{u}_{i+1}= D(\bar{u}_i)/D(\ph),\quad i=0,1,\ldots.$$
which together with (\ref{xu}) define the transformation totally.
\be \label{phi} \Phi:\, (x,u,u_1,u_2,\ldots)\mapsto (\bar
x,\bar{u},\bar{u}_1,\bar{u}_2,\ldots). \ee The order of the
transformation (\ref{phi}) is grater from $m$ ¨ $n$ in formulae
(\ref{xu}).

It is easy to show that the condition (\ref{alpha}) is equivalent
of invariance conditions of $\omega_i=0:$
$\omega_i=du_i-u_{i+1}dx$ contact differential forms.
Reformulation on multi-dim case is also easy.

 For zero order transformations ({\it point transformations})
 $$  \bar x=\ph(x,\ u),\quad \bar u=\psi(x,\ u), $$
 jacobian of (\ref{phi}) $\pa(\bar x,\bar u)/\pa(x,u)\ne 0.$
 Less obviously, that Invertability conditions for a general
  (\ref{xu}) demand that order of $\Phi$ does not exceed first one and
  moreover that transformation should be closed in the space of
  variables $x,\ u,\ u_1.$ Jacobian
$$      \pa(\bar x,\bar u,\bar{u}_1)/\pa(x,u,u_1)    $$
 of the transformation
 \be \label{con}
 \bar x=\ph(x,\ u,\ u_1),\quad \bar u=\psi(x,\ u,\ u_1),\quad
 \bar{u}_1=\chi(x,\ u,\ u_1),
 \ee
 where $\chi=D(\psi)/D(\ph)$ ¨ $\ph_{u_1}\chi=\psi_{u_1}$ can be rewritten
 as follows
 \be\label{cont} \psi_u-\chi \ph_u\neq 0.          \ee
 Defined by (\ref{con}), (\ref{cont}) transformations (\ref{phi}) are
 called {\it contact.} Legandre transformation. provides an example of the
 contact transformation.

\section{Contact Transformations}

The action of the differentiation $D_x$ defined by the infinite
dynamical system \be\label{dx} D_x(x)=1,\quad D_x{\bf u}={\bf
u}_1,\quad D_x{\bf u}_k={\bf u}_{k+1}. \ee The corresponding
vector field in scalar case is
\[ D_x=\frac{\pa}{\pa x}+\sum_0^{\infty}u_{k+1}\frac{\pa}{\pa u_k} \]

Contact transformation is described as invertible transformation
\be\label{hatx}
 \hat{x}=\ph(x,u,u_1,\dots),\quad \hat{u}=\psi(x,u,u_1,\dots),\quad
\hat{u_1}=\psi_1(x,u,u_1,\dots),\dots \ee which leaves the
dynamical system (\ref{dx}) invariant. This conditions means that
\[       \psi_{k+1}=D_x(\psi_k)/D_x(\ph),\quad k=0,1,2,\ldots     \]
or \be\label{alef} D_xA(x,u,u_1,\dots)= D_x(\ph)\hat{D}_{\hat
x}\hat{A}(\hat x,\hat u,\hat{u}_1,\dots). \ee We have the
classical result
\paragraph{Proposition.}
Invertible transformations (\ref{hatx}), (\ref{alef}) are\\
 point transformations $\hat{x}=\ph(x,u),\, \hat{u}=\psi(x,u)$ and contact one with
\[
\hat{x}=\ph(x,u,u_1),\quad\hat{u}=\psi(x,u,u_1),\quad\ph_{u_1}\psi_{u_1}\neq0
\]\[
\psi_{u_1}(u_1\ph_u+\ph_x)=\ph_{u_1}(u_1\psi_u+\psi_x),\quad
\ph_u\psi_{u_1}-\psi_u\ph_{u_1}\neq0.
\]

 Composition of two contact transformations is contact again.

In order to define a contact transformation one can fix the
function $a=a(x,u,u_1)$ and choose the functions $\ph,\, \psi$ as
two distinct solutions of the first order PDE

\[
\phi_x+u_1\phi_u=a(x,u,u_1)\phi_{u_1}.
\]
The usual example of the legandre transformation with $\ph=u_1,\,
\psi=xu_1-u$ corresponds to $a=0.$ On the other hand the
infinitesimal description of the one parameter group of contact
transformations is equivalent to more common theory of the first
order PDE
\[           u_t=A(x,u,u_x).                 \]

 One-parameter group of CTs (contact transf.) are generated by the
 first order PDE
 \be \label{fo} u_t=F(x,u,u_x),  \ee
 and corresponding ODEs (bi-characteristic system of eqs.):
\be \label{kh}
 {dx\over dt}=-\gamma,\quad {du\over dt}=F-\gamma u_1,\quad
{du_1\over dt}=D_x(F)-\gamma u_2=F_x+u_1F_u \ee where $\gamma=\pa
F/\pa u_1.$

 Let us consider the vector-field
$$ D_t=F\frac{\pa}{\pa u}+D_x(F)\frac{\pa}{\pa u_x}+ \ldots $$
 related with (\ref{fo}). It is easy to see that the CT
 $$
 \bar t=t,\quad \bar x=\ph(x,\ u,\ u_1),\quad \bar u=\psi(x,\ u,\ u_1),
 $$
 where $\ph(x,\ u,\ u_1)$ is any first integral of (\ref{kh}) reduces $F$ to
 the form $F=a(x,u)u_x+b(x,u).$ Related to the latter quasi-linear case the
 dynamical system (Cf. (\ref{kh})) simplifies:
 $$ {dx\over dt}=-a(x,u),\quad {du\over dt}=b(x,u).$$
 If $\ph(x,u),\  \ph_u\neq 0$ is first integral of these then the point
 transformation
 $$
 \bar t=t,\quad \bar x=\ph(x,\ u),\quad \bar u=\psi(x,\ u),
 $$
 reduces $F$ to $\bar F=b\psi_u-a\psi_x=c(\bar x,\bar u).$
 In other words, a composition above transformations (CT and point one)
 reduces PDE (\ref{fo}) to ODE $u_t=c(x,u).$

\fbox{Contact Transformations, end}

\paragraph{B\"acklund problem.} Invertability conditions for (\ref{phi})
in the case of two $\b u=(u,v)$ or more fields.

Hodograph and legandre
 $$ \left\{\ba{lll} \bar x &=& u,\\ \bar u &=& x \ea\right. \quad
\left\{\ba{rrr} \bar x &=& u_1,\\ \bar u &=& xu_1-u.\ea\right.  $$
are involutions.

 For differential substitutions (\ref{xu}) of the first order
 $\chi=D(\psi)/D(\ph)$ generally depends of $u_2$ but in the scalar case
 it cancelled for invertable transformations. In the case
  $\b u=(u,v)$ invertability does not guarantee cancellation $\b u_2$
 in $\bar{\b u}_1=D(\psi)/D(\ph)$. For instance it is the case
 for the transformation
$$
\bar x=x,\quad \bar u={v_x\over u_x},\quad \bar v=-v+u{v_x\over
u_x}
$$
which is involution since $ \bar{v}_x=u\bar{u}_x$. Notice, that
 $h(\b u)=u v_x/ u_x$ keeps invariant under this transformation.

 Generalization of criterion of inversability of DS of the first order,
 indicated above, on the case $\b u=(u,v)$ is interesting open problem.
 For {\it degenerated} DS there is invariant of the first order like
 example above. For {\it non-degenerated} DS we assume inversability and
 the following condition:

 \fbox{order of composition of DS is sum of individual ones.}

   In this case DS defines shift transformation $n\rightarrow n+1$, and
 corresponds with the chain of equations like follows
 $$  u_{nx}=a(u_n)(v_{n+1}-v_n),\quad v_{nx}=b(v_n)(u_{n-1}-u_n).$$

\section{Scalar differential substitutions}
Let us say that there exist scalar DS $w=P(x,u,u_x)$ from \be
\label{utx}
    u_t = f(x,u,u_x,u_{xx},\dots),
\ee into \be
\label{wtx}
    w_t = g(x,w,w_x,w_{xx},\dots),
\ee if $P_{u_x}\neq 0$ and \be\label{1}
(P_{u_x}D_x+P_u)(f)=g(x,P,D_x(P),D_x^2(P),\ldots ) . \ee

Differential substitutions (DSs) defined by module of point and
contact ones. We say that DS $\alpha_1$ from (\ref{utx}) into
(\ref{wtx}) is equivalent DS $\alpha_2$, if
$\alpha_1=\beta\circ\alpha_2\circ\gamma,$ where $\beta$ and
$\gamma$ inversable transformations eqs. (\ref{utx}) and
(\ref{wtx}), respectively.

Let us relate with $P(x,u,u_x)$ the hyperbolic equation
\be\label{jib} u_{xy}+b(x,u,u_x)u_y=0, \quad
b=\frac{P_u}{P_{u_x}},   \ee and its linearization \be\label{2}
 [D_xD_y + bD_y +u_y(b_u+b_{u_x}D_x)]f= (D_y+u_yb_{u_x})\circ (D_x+b)f=0,
\ee where $f=f(x,u,u_x,\ldots$ and $D_x,$ $D_y$ denote
differentiations $f$ with respect $x$ and $y$ on solutions of the
equation $u_{xy}=-u_yb.$ Recall that solutions of (\ref{2}) are
symmetries of the hyperbolic equation.

 It is easy to see that $D_y P=0$ and that (\ref{2}) follows from
 (\ref{1}). Thus if
\be \label{puu} w=P(x,u,u_x). \ee
 is DS from (\ref{utx}) then $f$ is symmetry of corresponding
 hyperbolic equation. Inverse is also true  (see \cite{sok}).

\begin{paragraph} {\it Definition.} Substitution (\ref{puu}) called
regular if the corresponding hyperbolic equation has as
$x-$invariant as well as $D_yP(x,u,u_x)=0$ $y-$invariant
$D_xQ(y,u,u_y,u_{yy},\ldots)=0.$
\end{paragraph}

\begin{paragraph} {\it Theorem} \cite{star}. DS (\ref{puu}) is regular iff
 $$ u_x=\alpha(x,w)u^2+\beta(x,w)u+\gamma(x,w). $$
\end{paragraph}

Examples:
$$ P=u_x,\quad P=u_x+e^u,\quad P=u_x-u^2,\quad P=u_x+e^u+e^{-u}.   $$

It can be proven that any DS of the form
$$ w=\psi(x,u,u_x),\quad \bar{x}=\ph(x,u,u_x) $$
is equivalent with (\ref{puu}).

\vspace{3mm}

\centerline{\bf Differential substitutions.} \vspace{3mm}

The problem of classifications {\bf all} evolutionary equations
(\ref{Evol}) of orders $2--5$ with formal symmetries considered by
many publications. Progress here depends mainly from a development
an appropriate theory of the differential substitutions.

We say that two evolutionary equations
\[ u_t=f(x,u,u_1,u_2,\dots),\quad v_t=g(x,v,v_1,v_2,\dots)    \]
are related by the differential substitution
\[   u=h(x,v,v_1,\dots\, .  \]
if in result of {\bf total linearization}
\[ \ph_t=f_*(\ph),\quad \ph=h_*(\psi),\quad \psi_t=g_*(\psi).   \]
one arrives to the equations \be\label{fgh}    D_t-g_*=h_*^{-1}
(D_t-f_*)h_*.\LRA (h_*)_t+h_*g_*=f_*h_*. \ee which could be
rewritten in $(x,u)$ as well in $(x,v)$ variables. That implies,
in particular, that the two equaations connected by differential
substitution have the one and same order.

In the simplest case
\[ u_t=Dg(u,u_1,\dots),\quad v_t=g(v_1,v_2,\dots),\quad u=D(v)  \]
having in mind the general formula (\ref{fgh}) one proves
\paragraph{2.9 Proposition} Let $A$ be a formal symmetry of the evolutionary
equation $u_t=Dg.$ Then the similarity transformation
\[  \hat A=DAD^{-1} \LRA A=D^{-1}\hat{A}D \]
defines the formal symmetry for the equation $v_t=g,\, v=u_1.$
Vice versa if $\hat A$ is the formal symmetry for latter equation
and coefficients of that series can be rewritten in terms of
$u=v_1$ (i.e. invariant with respect shift in $v$) then the
similarity transformation yields formal symmetry for $u_t=Dg.$

{\it Proof.} Using the definition (\ref{genlin}) we find that
\[ f=D(g)\RA f_*=D\circ g_*\RA D_t-f_*=D(D_t-g_*)D^{-1}.  \]
Differentiating $\hat A=D^{-1}AD$ we find that $[D_t-g_*,\hat
A]=0$ and vice versa.

\qed

 This proposition allows one, in particular, to find the formal symmetry
indicated in Example 2.2. Namely, the Burgers equation has a form
$u_t=D(u_1+u^2)$ and in result of integration one obtains the
equation
\[v_t=v_{xx}+v_x^2,\quad u=v_x  \]
which equivalent to linear one
\[ v_t=v_{xx}+v_x^2,\quad w=e^v,\quad w_t=w_{xx}. \]
Starting with the formal symmetry $\ti A=D$ for the last equation
we find in notations of Proposition 2.9
\[ \hat A=e^{-v}De^v=D+v_1=D+u,\quad A=D\hat{A}D^{-1}=D+u+u_1D^{-1}.  \]

Closing this chapter we provide the list the most popular third
order evolutionary equations of the form
 \[      u_t=u_3+F(x,u,u_1,u_2) \]
 which posess
the formal symmetry.
\begin{eqnarray}
 u_t&=& u_3+P(u)u_1,\quad P'''=0, \label{KdV1}\\
 u_t&=& u_3-\frac12u^3_1 +(\alpha e^{2u} +\beta e^{-2u})u_1,\label{KdV2}\\
 u_t&=& u_3-\frac{3}{2u_1}(u^2_2+Q(u)),\quad Q^V=0.  \label{KN}
\end{eqnarray}

\section{Exercises for Chapter 5}
\paragraph{1.}
\paragraph{2.}
\paragraph{3.}
\paragraph{4.}

\backmatter
\printindex



\begin{thebibliography}{99}

\bibitem{shub} Shubin M.A. "Pseudodifferential operators and spectral
theory". Springer, 1987.

\bibitem{Olver} Olver P.J.
 {\bf Applications of Lie groups to Differential Equations.} 2nd Ed.,
 Graduate Texts in Mathematics, vol. 107. Springer-Verlag, New York
 (1993)

\bibitem{ibsh2} N.H.Ibragimov and A.B.Shabat,
         "On infinite dimension Lie-B\"{a}clund Algebras",
      {\em Funkts.  Anal. Prilozhen}, 14(4):79--80, 1980
\bibitem{sh1981} A.B. Shabat, R.I. Yamilov. "Exponential systems of type 1 and Cartan
matricies." Preprint. Bashkirskii Filial Akad. Nauk SSSR, Ufa
(1981)

\bibitem{zak1971} Zakharov V.E., Shabat A.B.
  "Exact theory of two-dimensional self-focusing and one-dimensional
   self-modulation of waves in nonlinear media",
   {\em Zh. Eksp. Teor. Fiz.}, v.61, No.1, (1971)
\bibitem{ritt} J.F. Ritt. {\bf Differential Algebra}.
American Math. Soc. Colloquim Publications, {\bf v}: 33, (1948).


\bibitem{Jaco1} C.G. Jacobi. "Vorlesungen
über Dynamik", Königsberg University 1842 - 1843 (edited by
Clebsch and published from Reimer, Berlin, 1884)

\bibitem{Liou1}
J. Liouville. "Note sur les équations de la dynamique", J. Math.
Pures Appl. 20 (1855), 137-138.


\bibitem{Neum} C. Neumann. "De problemate quodam
mechanico, quod ad primam integralium ultraellipticorum classem
revocatur", J. Reine Angew. Math. 56 (1859), 46-63.

\bibitem{Kirch} G. Kirchhoff. "Über die Bewegung eines Rotationskörper in einer Flüssigkeit", J.
reine. angew. Math. 71 (1870), 237-262.

\bibitem{Cleb} A. Clebsch. "Über die Bewegung eines festen Körper in einer Flüssigkeit", Math.
Ann. 3? (1871), 238-262.

\bibitem{Stek} V. Steklov (Stekloff). "Über die Bewegung eines festen Körper in einer
Flüssigkeit", Math. Ann. 42 (1893),
273-374.

\bibitem{Web} H. Weber.    "Anwendung der Thetafunctionen zweiter Veränderlichen
auf die Theorie der Bewegung eines festen Körper in einer
Flüssigkeit", Math. Ann. 14 (1878), 143-206.

\bibitem{Kow} S. Kowalevski.   "Sur le problème de la
rotation d'un corps solide autour d'un point fixe", Acta Math. 12
(1889), 177-232.


\bibitem{Stae} P. Stäckel. " Über die
Integralen der Hamilton-Jacobischen Differential Gleichung
mittelst Separation der Variable", Habilitationsschrift, Halle,
(1891)


\bibitem{Dar} G. Darboux. "Leçons sur la théorie générale des
surfaces", Gauthier-Villars, Paris, (1895)

\bibitem{Gar} R. Garnier. "Sur une classe de systè:mes différentiels abéliens
déduits de la théorie des équations linéaires", Rend. Circ. Mat.
Palermo 43 (1918-19), 155-191.


\bibitem{Drach1} J. Drach. "Détermination des cas de réduction de l'équation
différentielle $d2/dx2 = [\phi(x) + h ]y$", C.R. Acad. Sci. Paris
168 (1919), 47-50.

\bibitem{Drach2} J. Drach. "Sur l'intégration par quadratures
    de l'équation $d2/dx2 = [\phi(x) + h ]y$", C.R. Acad. Sci. Paris 168 (1919), 337-340.

\bibitem{Burch} J.L. Burchnall and T.W. Chaundy Commutative ordinary differential
operators, Proc. London Math. Soc. (2) 21 (1922), 420-440.

\bibitem{Gardner} C.S. Gardner, J.M. Greene, M.D. Kruskal and R.M. Miura. "Method for
solving the Korteweg-de Vries equation", Phys. Rev. Lett. 19
(1967), 1095-1097.

\bibitem{Lax} P.D. Lax.  "Integrals of nonlinear equations of evolutions and
solitary waves", Comm. Pure and Appl. Math. 21 (1968), 467-490.


\bibitem{ca} Calogero F. "Ground state a one-dimensional N-body system."
{\it J.Math.Phys. (USA)}, V.10, pp.2197-200 (1969)

\bibitem{ca1} Calogero F.
"Solution of the one-dimensional N-body problem with quadratic
and/or inversely quadratic pair potentials." {\it J. Math.
Phys.(USA)} V.12 , pp.419-436 (1971)
\bibitem{p3} Gromak V., Laine I., Shimomura S. {\bf Painlev\'e Differential
Equations in the complex plane.} {\it Walter de Grugter,} Berlin,
NY, 2002.

\bibitem{fu} Fujiwara T. "Synchronized Similar Triangles for Three-Body
Orbit with Zero Angular Momentum", Proc. (Sardinia), (2004)

\bibitem{has1962} Hasselman K. "On nonlinear energy transfer in gravity-wave
spectrum. Part 1. General theory." {\it J. Fluid Mech.}, V.12,
pp.481-500 (1962)

\bibitem{kar1994} Kartashova E.  "Clipping - a new investigation method
for PDEs in compact domains." {\it Theor. Math. Phys.}, V.99,
pp.675-680 (1994)

\bibitem{kar1998}  Kartashova E. "Wave resonances in systems with discrete
spectra." In book: (Ed.) Zakharov V.E. {\bf Nonlinear Waves and
Weak Turbulence}. {\it Advances in the Mathematical Sciences},
American Mathematical Society, Providence, RI, pp.95-129 (1998)

\bibitem{kar2005}  Kartashova E. "Diophantine equations of nonlinear
physics. Part 1." RISC preprint, {\bf to be published}, (2005)

\bibitem{sh1987} Mikhailov A. V., Shabat A.B., Yamilov.  R.I.
"The symmetry approach to the classification of non-linear
equations. Complete lists of integrable systems." {\it Uspechi
Mat. Nauk}, V.42 (4), pp.3-53 (1987) (English translation: {\it
Russian Math. Surveys}, V.42 (4), pp.1-63, 1987)

\bibitem{sh1991}  Mikhailov A.V., Shabat A.B., Sokolov V.V. "The Symmetry
Approach to Classification of Integrable Equations". In book:
(Ed.) V.E.Zakharov, {\bf What is Integrability?} Springer, Berlin,
pp. 115-184 (1991)

\bibitem{Olver} Olver P.J.
 {\bf Applications of Lie groups to Differential Equations.} 2nd Ed.,
 Graduate Texts in Mathematics, vol. 107. Springer-Verlag, New York
 (1993)

\bibitem{sh1975} Shabat A.B. "Inverse scattering problem for a system of
differential equations", {\em Funct. Anal. Appl. (English
translation)}, v.9, No.3, (1975).

\bibitem{sh1979} Zakharov V., Shabat A.
"A Scheme for integration the nonlinear equations of mathematical
physics by the method of the inverse scattering problem, II", {\em
Funct. Anal. Appl.} 13: 166-74 (1979)

\bibitem{zak1999} Zakharov V. "Statistical theory of gravity and capillary waves
on the surface of a finite-depth fluid." {\it Eur. J. Mech. B:
Fluids}, pp.327-344 (1999)


\bibitem{zak1971} Zakharov V.E., Shabat A.B.
  "Exact theory of two-dimensional self-focusing and one-dimensional
   self-modulation of waves in nonlinear media",
   {\em Zh. Eksp. Teor. Fiz.}, v.61, No.1, (1971)


\bibitem{zhi1979} Zhiber A.V., Shabat A.B.  "Klein-Gordon equations with a
non-trivial group", {\em Dokl. Akad. Nauk SSSR}, v.247, No.5,
(1979)

\bibitem{Ric} J. Riccati. "Opere", Treviso, 1758

\bibitem{Reid} W.T. Reid. {\bf Riccati differential equations}.
N.Y.-L., 1972

\bibitem{Codd} E.A.Coddington, N.Levinson {\bf Theory of ordinary differential equations}.
N.Y.-L., 1955.
\bibitem{ASY} V.E.Adler, A.B.Shabat, R.I. Yamilov, "Symmetry approach to the
integrability problem," {\it Theor. Math. Phys.}, Vol. 125(3),
pp.1603-1661 (2000).

\bibitem{sh70} A.B.Shabat, "Transparent potentials",{\it Dinamika
sploshnoi sredy}, Institute of Hydrodynamics, Novosibirsk,
        No.5, p.130-145, (1970).

\bibitem{Ad} V.E. Adler, "$N-$soliton solution of KdV equation,"
in {\em Asimtotic methods in Math. Phys.}, Ural Branch RAN, Ufa,
1989.

\bibitem{Adler} V.E. Adler. Private communication. (2005)

\bibitem{Gol} G.M.Goluzin, "Geometrical theory of functions of a complex
variable". Moscow, "Nauka", pp.628 [in Russian] (1966)

\bibitem{kodama} Yuji Kodama,
 "Young diagrams and N-soliton solutions of the KP
equation", J. Phys. A: Math. Gen. 37 (2004) 11169-11190,
arXiv:nlin.SI/0406033

\bibitem{Kov} Kovalevskii S.V. "Perepiska Kovalevskoi i
Mittag-Lefflera. Nauchnoe nasledie.", t.7, "Nauka", 1984

\bibitem{zabu} Zabusky, N. J. and Kruskal, M. D. "Interaction of Solitons in a
Collisionless Plasma and the Recurrence of Initial States." Phys.
Rev. Let. 15, 240-243, 1965

\bibitem{Novikov1} S. P. Novikov, " A periodic problem for the Korteweg-de Vries equations. I",
Funktsional Anal. i Prilozhen., 1974, v. 8, N 3, 54-66.


\bibitem{Dubr1} B. A. Dubrovin, V. B. Matveev and S. P. Novikov,
 "Nonlinear equations of KdV type, finite-zone linear operators and
 abelian varieties," {\em Russ.Math.Surv.}, {\bf 31}(1): 59--146, 1976.




\bibitem{shub} Shubin M.A. "Pseudodifferential operators and spectral
theory". Springer, 1987.
\bibitem{Olver} Olver P.J.
 {\bf Applications of Lie groups to Differential Equations.} 2nd Ed.,
 Graduate Texts in Mathematics, vol. 107. Springer-Verlag, New York
 (1993)

\bibitem{ibsh2} N.H.Ibragimov and A.B.Shabat,
         "On infinite dimension Lie-B\"{a}clund Algebras",
      {\em Funkts.  Anal. Prilozhen}, 14(4):79--80, 1980
\bibitem{sh1981} A.B. Shabat, R.I. Yamilov. "Exponential systems of type 1 and Cartan
matricies." Preprint. Bashkirskii Filial Akad. Nauk SSSR, Ufa
(1981)

\bibitem{zak1971} Zakharov V.E., Shabat A.B.
  "Exact theory of two-dimensional self-focusing and one-dimensional
   self-modulation of waves in nonlinear media",
   {\em Zh. Eksp. Teor. Fiz.}, v.61, No.1, (1971)
\bibitem{ritt} J.F. Ritt. {\bf Differential Algebra}.
American Math. Soc. Colloquim Publications, {\bf v}: 33, (1948).


\bibitem{shub} Shubin M.A. "Pseudodifferential operators and spectral
theory". Springer, 1987.
\bibitem{Olver} Olver P.J.
 {\bf Applications of Lie groups to Differential Equations.} 2nd Ed.,
 Graduate Texts in Mathematics, vol. 107. Springer-Verlag, New York
 (1993)

\bibitem{ibsh2} N.H.Ibragimov and A.B.Shabat,
         "On infinite dimension Lie-B\"{a}clund Algebras",
      {\em Funkts.  Anal. Prilozhen}, 14(4):79--80, 1980
\bibitem{sh1981} A.B. Shabat, R.I. Yamilov. "Exponential systems of type 1 and Cartan
matricies." Preprint. Bashkirskii Filial Akad. Nauk SSSR, Ufa
(1981)

\bibitem{zak1971} Zakharov V.E., Shabat A.B.
  "Exact theory of two-dimensional self-focusing and one-dimensional
   self-modulation of waves in nonlinear media",
   {\em Zh. Eksp. Teor. Fiz.}, v.61, No.1, (1971)
\bibitem{ritt} J.F. Ritt. {\bf Differential Algebra}.
American Math. Soc. Colloquim Publications, {\bf v}: 33, (1948).


\bibitem{svin} Svinolupov S. I., \\
"Second order evolution equations possessing symmetries,"\\
 {\em Russian Math. Surveys.}, {\bf 40}(5), 263--264, 1985.

\bibitem{bk2005} R. Beals, E. Kartashova. "Constructive
factorization of LPDO in two variables." {\it Theor. Math. Phys.},
{\bf to be published}, (2005)

\bibitem{sh1991}  Mikhailov A.V., Shabat A.B., Sokolov V.V. "The Symmetry
Approach to Classification of Integrable Equations". In book:
(Ed.) V.E.Zakharov, {\bf What is Integrability?} Springer, Berlin,
pp. 115-184 (1991)



\bibitem{bk2005} R. Beals, E. Kartashova. "Constructive
factorization of LPDO in two variables." {\it Theor. Math. Phys.},
{\bf to be published}, (2005)

\bibitem{sh1981} A.B. Shabat, R.I. Yamilov. "Exponential systems of type 1 and Cartan
matricies." Preprint. Bashkirskii Filial Akad. Nauk SSSR, Ufa
(1981)

\bibitem{lez}  A.N. Leznov, M.P. Saveliev. {\bf Group-theoretical methods
for integration on non-linear dynamical systems} (Russian),
Moscow, Nauka (1985). English version: Progress in Physics, 15.
Birkhäuser Verlag, Basel, pp. xviii+290pp (1992)

\bibitem{bianchi} L. Bianchi, Lezioni di geometria differenziale, Zanichelli,
Bologna, 1924

\bibitem{ziz} G. Tzitzeica G., Sur un theoreme de M. Darboux.
Comptes Rendu de l'Academie des Aciences 150 (1910), pp.955-956;
971-974

\bibitem{sha1} A.V.Zhiber, A.B.Shabat  "Klein-Gordon equations with a
non-trivial group", {\em Dokl. Akad. Nauk SSSR}, v.247, No.5,
(1979)

\bibitem{sha2} A.V.Mikhailov, A.B.Shabat and R.I.Yamilov  "The
      Symmetry Approach to the Classification of Non-linear Equations.
      Complete Lists of Integrable Systems",
       {\em Uspekhi Mat. Nauk}, 42(4): 3--53, 1987

\bibitem{bob} A.I.Bobenko "Surfaces in Terms of 2 by 2 Matrices.
Old and New Integrable Cases". In: A.P. Fordy, J.C. Wood (eds.)
"Harmonic Maps and Integrable Systems", Vieweg,


Braunschweig/Wiesbaden 1994, pp. 81-127

\bibitem{AKNS74} M.J. Ablowitz, D.J. Kaup, A.C. Newell, and H. Segur.
     \newblock  The inverse scattering transform --- Fourier analysis for nonlinear problems.
     \newblock  \emph{Stud. Appl. Math.} \textbf{53}: 249 (1974).

\bibitem{Bo76} O. I. Bogoyavlensky.
     \newblock   On perturbations of the Toda lattice.
     \newblock     \emph{ Commun. Math. Phys.} \textbf{51}: 201--209 (1976).

\bibitem{BK09_1}
M. D. Bustamante and E. Kartashova.
     \newblock Dynamics of nonlinear resonances in Hamiltonian systems.
     \newblock \emph{EPL}, \textbf{85}:  14004-1--5 (2009).

\bibitem{Cal71} F. Calogero.
     \newblock   Solution of the one-dimensional N-body problem with quadratic and/or inversely quadratic pair
potentials,
     \newblock  \emph{J. Math. Phys.} \textbf{12}:  419--436 (1971).

\bibitem{CaMa74} F. Calogero and C. Marchioro.
     \newblock     Exact solution of a one-dimensional three-body scattering problem with two-body
and/or three body inverse square potential.
     \newblock  \emph{J. Math. Phys.} \textbf{15}: 1425--1430 (1974).

\bibitem{Fla74a} H. Flaschka.
     \newblock  The Toda lattice.
     \newblock       \emph{Phys. Rev. B} \textbf{9}: 1924 (1974).

\bibitem{Fla74b} H. Flaschka.
     \newblock       The Toda lattice II.
     \newblock  \emph{Progr. Theor. Phys.} \textbf{51}: 703--716 (1974).

\bibitem{CUP} Kartashova, E.
\textbf{Nonlinear Resonance Analysis}
     \newblock (Cambridge University Press, 2010).

\bibitem{Ma74} S.V. Manakov.
     \newblock  Complete integrability and stochastization of discrete dynamical systems.
          \newblock  \emph{Soviet Phys. JETP} \textbf{40}: 269--274 (1974).

\bibitem{M075}  J. Moser.
     \newblock  Three integrable Hamiltonian systems connected with
isospectral deformations.
     \newblock  \emph{Adv. Math.} \textbf{16}: 197--220 (1975).

\bibitem{M080}  J. Moser.
     \newblock  Geometry of quadrics and spectral theory.
     \newblock  In ``Chern Symposium, Berkeley, 1979'' (Springer-Verlag,
1980) pp. 147--188.

\bibitem{OP76} M.A. Olshanetsky and A.M. Perelomov.
     \newblock  Completely Integrable Hamiltonian Systems Connected with Semisimple Lie Algebras.
     \newblock        \emph{Inventions Math.} \textbf{37}: 93-108 (1976).

\bibitem{Su72} B. Sutherland.
     \newblock  Exact results for a quantum many-body problem in one-dimension. II.
     \newblock       \emph{Phys. Rev. A} \textbf{5}: 1372--1376 (1972).

\bibitem{Toda67a} M. Toda.
     \newblock Vibration of a chain with a non-linear interaction.
     \newblock \emph{J. Phys. Soc. Japan} \textbf{22}:  431--436 (1967).

\bibitem{Toda67b} M. Toda.
     \newblock     Wave propagation in anharmonic lattice.
     \newblock \emph{J. Phys. Soc. Japan} \textbf{23}:  501--596 (1967).

\bibitem{Ver68a}
F. Verheest.
     \newblock Proof of integrability for five-wave interactions in a case with unequal coupling constants.
     \newblock \emph{Phys. A: Math. Gen.} \textbf{21} (1988), L545--49.

\bibitem{Ver68b}
F. Verheest.
     \newblock Integrability of restricted multiple three-wave interactions. II. Coupling constants with ratios 1 and 2.
     \newblock \emph{J. Math. Phys.} \textbf{29} (1988), 2197--201.

\bibitem{ZSh74} V.E. Zakharov and A.B. Shabat.
     \newblock A scheme for integrating the nonlinear equations of mathematical
physics by the method of the inverse scattering problem, I.
     \newblock \emph{Funct. Anal. Appl.} \textbf{8} 226-235 (1974).


\end{thebibliography}
\end{document}